\documentclass[useAMS,usenatbib]{./aa}
\usepackage{txfonts}
\usepackage{natbib}
\bibpunct{(}{)}{;}{a}{}{,}
\usepackage[dvips]{graphicx}

\begin{document}
\title{Scaling Relations and Mass Calibration of the X-ray
Luminous Galaxy Clusters at redshift $\sim0.2$: XMM-Newton Observations
\thanks{This work is based on observations
made with the XMM-Newton, an ESA science mission with instruments and
contributions directly funded by ESA member states and the USA
(NASA).}}
\author{Y.-Y. Zhang\inst{1},
A. Finoguenov\inst{1},
H. B\"ohringer\inst{1},
J.-P. Kneib\inst{2,3},
G. P. Smith\inst{4},
O. Czoske\inst{5,6},
and G. Soucail\inst{7}
}
\institute{Max-Planck-Institut f\"ur extraterrestrische Physik, Giessenbachstra\ss e, 85748 Garching, Germany
\and OAMP, Laboratoire d'Astrophysique de Marseille, traverse du Siphon, 13012 Marseille, France
\and Caltech-Astronomy, MC105-24, Pasadena, CA 91125, USA
\and School of Physics and Astronomy, University of Birmingham, Edgbaston, Birmingham, B152TT, UK
\and AIFA, Universit\"at Bonn, Auf dem H\"ugel 71, 53121 Bonn, Germany
\and Kapteyn Astronomical Institute, PO Box 800, 9700AV Groningen, Netherlands
\and Observatoire Midi-Pyrenees, Laboratorire d'Astrophysique,
UMR 5572, 14 Avenue E. Belin, 31400 Toulouse, France
}
\authorrunning{Zhang et al.}
\titlerunning{Scaling Relations and Mass Calibration of the X-ray Luminous Galaxy Clusters at z=0.2}
\date{Received 16/10/06 / Accepted 20/02/07}

\offprints{Y.-Y. Zhang}

\abstract{We present the X-ray properties and scaling relations of
a flux-limited morphology-unbiased sample of 12 X-ray luminous galaxy
clusters at redshift around 0.2 based on XMM-Newton observations. The
scaled radial profiles are characterized by a self-similar behavior at
radii outside the cluster cores ($>0.2 r_{500}$) for the temperature
($T\propto r^{-0.36}$), surface brightness, entropy ($S \propto
r^{1.01}$), gas mass and total mass. The cluster cores contribute up
to 70\% of the bolometric X-ray luminosity. The X-ray scaling
relations and their scatter are sensitive to the presence of the cool
cores. Using the X-ray luminosity corrected for the cluster central
region and the temperature measured excluding the cluster central
region, the normalization agrees to better than 10\% for the cool core
clusters and non-cool core clusters, irrelevant to the cluster
morphology. No evolution of the X-ray scaling relations was observed
comparing this sample to the nearby and more distant samples. With the
current observations, the cluster temperature and luminosity can be
used as reliable mass indicators with the mass scatter within
20\%. Mass discrepancies remain between X-ray and lensing and lead to
larger scatter in the scaling relations using the lensing masses
(e.g. $\sim 40$\% for the luminosity--mass relation) than using the
X-ray masses ($<20$\%) due to the possible reasons discussed.

}

\maketitle

\section{Introduction}

Both the gravitational growth of density fluctuations and the
expansion history of the Universe can be used to constrain the
cosmology.

The gravitational growth of the fluctuation can be measured by,
e.g. the evolution of the galaxy cluster mass function (e.g. Schuecker
et al. 2003). The most massive clusters show the cleanest results in
comparing theory with observations. The mass function of luminous
galaxy clusters probes the cosmic evolution of large-scale structure
(LSS) and is thus an extremely effective test of cosmological
models. It is sensitive to the matter density, $\Omega_{\rm m}$, and
the amplitude of the cosmic power spectrum on cluster scales,
$\sigma_8$ (e.g. Schuecker et al. 2003).

The expansion history of the Universe can be used to determine the
metric and thus to constrain the cosmological parameters. Typical
examples are Supernova Ia (SN Ia, e.g. Astier et al. 2006) and galaxy
clusters combining the Sunyaev-Zeldovich effect (Sunyaev \& Zeldovich
1972, the SZ effect) and thermal Bremsstrahlung X-ray approaches
(Molnar et al. 2002), in which the luminosity distance and volume are
used to determine the metric, respectively.

The multi-pole structure of the cosmic microwave background (CMB)
anisotropy power spectrum depends on the normalized overall amount of
dark matter (DM) in the Universe. The WMAP three year results imply
values of the parameters $\Omega_{\rm m} h^2$, $\Omega_{\rm b} h^2$,
$h$, $n_{\rm s}$, and $\sigma_8$ of $0.127^{+0.007}_{-0.013}$,
$0.0223^{+0.0007}_{-0.0009}$, $0.73^{+0.03}_{-0.03}$,
$0.951^{+0.015}_{-0.019}$, and $0.74^{+0.05}_{-0.06}$ (Spergel et al.
2006). Combining the CMB approach with the other cosmological tests,
e.g. galaxy cluster surveys, the large-scale galaxy distribution
and/or SN~Ia, is sensitive to the effects of dark energy, the density
of which is characterized by the parameter $\Omega_{\Lambda}$ and its
time evolution by the equation of state parameter $w(z)$ (e.g.
Vikhlinin et al. 2002; Allen et al. 2004; Chen
\& Ratra 2004; Hu et al. 2006).

The construction of the mass function of galaxy clusters for large
cosmological cluster samples is based on the cluster X-ray
temperature/luminosity estimates and the mass--observable scaling
relations (Reiprich \& B\"ohringer 2002). A better understanding of
the mass--observable scaling relations and their scatter is of prime
importance for galaxy clusters as a unique means to study the LSS and
thus to constrain the cosmological parameters (Voit 2005). Precise
mass estimates and mass--observable scaling relations can potentially
push the current measurement uncertainty (10--30\%) of the
cosmological tests to higher precision (Smith et al. 2003, 2005;
Bardeau et al. 2006).

The mass distributions in galaxy clusters can be probed in a variety
of ways using: (i) the X-ray gas density and temperature distributions
under the assumptions of hydrostatic equilibrium and spherical
symmetry, (ii) the velocity dispersion and spatial distribution of
cluster galaxies, (iii) the distortion of background galaxies caused
by the gravitational lensing effect, and (iv) the SZ effect. For the
relaxed low-redshift clusters, the uncertainties are small in the
masses measured from the precise intra-cluster medium (ICM)
temperature and density profiles up to large radii (e.g. $r_{500}$)
using X-ray observations of XMM-Newton and Chandra (Allen et al. 2004;
Pointecouteau et al. 2005; Arnaud et al. 2005; Vikhlinin et
al. 2006a). Particularly, the X-ray results for the relaxed clusters
have achieved a very high level of precision ($<10$\%) in Arnaud et
al. (2005) and Vikhlinin et al. (2006a). The methods to measure
cluster mass using the velocity dispersion of cluster galaxies and
gravitational lensing are improving (Girardi et al. 1998; Smith et
al. 2003, 2005; Bardeau et al. 2006). Up to now, it is technically
more difficult for the SZ effect to resolve radial temperature
distributions in galaxy clusters to obtain precise cluster masses but
the future holds promise (Zhang \& Wu 2003).

Some interpretations of the cluster dynamics, such as spherical
symmetry and hydrostatic equilibrium, are required to obtain the
cluster masses for the X-ray approach (also for the SZ
effect). Substructure is observed in galaxy clusters and the frequency
of its occurrence in ROSAT observations has for example been estimated
to be of the order of about $52\pm 7$\% (Schuecker et al. 2001). Using
high quality XMM-Newton data, Finoguenov et al. (2005) found that
substructures (on $>10$\% level) can be observed in all the clusters
in the REFLEX-DXL sample at $z\sim 0.3$. Based on the strong lensing
results using high resolution HST data, Smith et al. (2005) measured
the substructure fractions and identified the clusters with
multi-modal DM morphologies. The up-coming XMM-Newton/Chandra results
of comprehensive samples without exclusion of the mergers (HIFLUGS,
Reiprich \& B\"ohringer 2002, Reiprich 2006; LP, B\"ohringer et
al. 2007), would be promising to investigate the mass measurement
accuracy and the scaling relations for a morphology-unbiased sample at
low redshifts.

Independent of the X-ray method, gravitational lensing provides a
direct measure of projected cluster masses irrespective to their
dynamical state. Comparison of ground-based lensing data including
estimated arc redshifts with ASCA/ROSAT X-ray data show that the
masses between X-ray and lensing are closely consistent for relaxed
clusters (Allen 1998). However, detailed investigation with high
quality (HST/XMM-Newton/Chandra) data reveals that for a few
apparently "relaxed" clusters in X-rays there remain large mass
discrepancies between X-ray and lensing, especially on small scales,
e.g.  CL0024$+$17 (Kneib et al. 2003; Ota et al. 2004; Zhang et
al. 2005a) and Abell~2218 (Soucail et al. 2004; Smith et al. 2005;
Pratt et al. 2005). Stanek et al. (2006) illustrate the interplay
between parameters and sources of systematic error in cosmological
applications using galaxy clusters and stress the need for independent
calibration of the $L$--$M$ relation such as gravitational weak
lensing and X-ray approaches. It is extremely important to investigate
cluster dynamics and to calibrate the scaling relations of a set of
morphology-unbiased cluster samples at different redshifts to improve
the understanding of intrinsic scatter and to control the sources of
systematic errors.

Recently two morphology-unbiased flux-limited samples at medium
distant redshifts were constructed. One (Smith et al. 2005; Bardeau et
al. 2005, 2006; hereafter the pilot LoCuSS sample) is in the redshift
range $0.176<z<0.253$ with $L~\geq~8.0 \times 10^{44}~{\rm
erg~s^{-1}}$ based on the XBACS in Ebeling et al. (1996). The other
(Zhang et al. 2004a, 2006; Finoguenov 2005; B\"ohringer et al. 2006;
the REFLEX-DXL sample) is in the redshift range $0.258<z<0.308$ with
$L~\geq~5.9 \times 10^{44}~{\rm erg~s^{-1}}$ based on the REFLEX
survey in B\"ohringer et al. (2004). The selection effect is well
understood for both samples (Smith et al. 2005; B\"ohringer et
al. 2006), consisting of massive galaxy clusters spanning the full
range of cluster morphologies (Bardeau et al. 2006; Zhang et
al. 2006). Using the same method described in Zhang et al. (2006) for
the REFLEX-DXL sample, we present here the studies of the pilot LoCuSS
sample of 12 clusters (Abell~2219 is flared) based on high quality
XMM-Newton observation. Smith et al. (2005) performed the strong
lensing studies for most clusters in this sample except for Abell~1689
(Broadhurst et al. 2005a; Halkola et al. 2006) and Abell~2667 (Covone
et al. 2006). The strong lensing studies of Abell~2218 can also be
found in Kneib et al. (1996, 2004). Weak lensing analysis using HST,
CFH12k and Subaru can be found in Bardeau et al. (2005, 2006) and
Broadhurst et al. (2005b, for Abell~1689). We note that Abell~773 has
no weak lensing mass estimates because no CFH12k data were taken for
this cluster. We will compare the X-ray masses with the lensing masses
(e.g. Kneib et al. 1996, 2004; Smith et al. 2003, 2005; Bardeau et
al. 2006) to discuss possible mass discrepancies.

The main goals of this work are: (1) to derive precise cluster
mass and gas mass fraction, (2) to investigate the self-similarity
of the scaled profiles of the X-ray properties, (3) to seek for
proper definition of the global quantities for the X-ray scaling
relations with the least scatter, (4) to study the evolution of
the X-ray scaling relations combining the pilot LoCuSS sample with
the nearby and more distant samples, and (5) to compare the X-ray
and lensing masses and to understand the scatter in the
luminosity--mass scaling relations using the lensing masses.

The data reduction is described in Sect.~\ref{s:method} and the X-ray
properties of the ICM of the sample in Sect.~\ref{s:result}.  We
investigate the self-similarity of the scaled profiles of the X-ray
properties in Sect.~\ref{s:prof} and the X-ray scaling relations in
Sect.~\ref{s:scale}. In Sect.~\ref{s:discu} we compare the X-ray and
lensing masses, discuss the peculiarities in individual clusters, and
address the possible explanation for the scatter in the relation
between the lensing mass and X-ray luminosity. We our the conclusions
in Sect.~\ref{s:conclusion}.  Unless explicitly stated otherwise, we
adopt a flat $\Lambda$CDM cosmology with the density parameter
$\Omega_{\rm m}=0.3$ and the Hubble constant $H_{\rm
0}=70$~km~s$^{-1}$~Mpc$^{-1}$. We adopt the solar abundance table of
Anders \& Grevesse (1989).  Confidence intervals correspond to the
68\% confidence level. We apply the Orthogonal Distance Regression
method (ODRPACK~2.01\footnote{http://www.netlib.org/odrpack/ and
references here}, e.g. Boggs et al. 1987) taking into account
measurement errors on both variables to determine the parameters and
their error bars of the fitting throughout this paper. We use Monte
Carlo simulations for the uncertainty propagation on all quantities of
interest.

\section{Data reduction}
\label{s:method}

\subsection{Data preparation}

All 13 clusters of the pilot LoCuSS sample at $z \sim 0.2$ were
observed by XMM-Newton. However, the XMM-Newton observations of
Abell~2219 are seriously contaminated by soft proton flares and are
therefore excluded in this work. In total, 12 galaxy clusters observed
by XMM-Newton were uniformly analyzed using the same method as
described in Zhang et al. (2006). We use the XMMSAS v6.5.0 software
for the data reduction. In Table~\ref{t:obs} ({\it See the electronic
edition of the Journal}) we present a detailed XMM-Newton log of the
observations for all clusters.

All observations were performed with medium or thin filter for three
detectors. The MOS data were taken in Full Frame (FF) mode except for
MOS1 of Abell~1835. The pn data were taken in Extended Full Frame
(EFF) mode or FF mode. For pn, the fractions of the out-of-time (OOT)
effect are 2.32\% and 6.30\% for the EFF mode and FF mode,
respectively. An OOT event file is created and used to statistically
remove the OOT effect. 

Above 10~keV (12~keV), there is little X-ray emission from
clusters detected with MOS (pn) due to the low telescope
efficiency at these energies. The particle background therefore
dominates. The light curve in the energy range 10--12~keV
(12--14~keV) for MOS (pn), binned in 100~s intervals, is used to
monitor the particle background and to excise periods of high
particle flux. Since episodes of ``soft proton flares'' (De Luca
\& Molendi 2004) were detected in the soft band, the light curve
of the 0.3--10~keV band, binned in 10~s intervals, is used to monitor
and to excise the soft proton flares. A 10~s interval bin size is
chosen for the soft band to ensure a similar good photon statistic as
for the hard band. The average and variance of the count rate have
been interactively determined for each light curve from the count rate
histogram. Good time intervals (GTIs) are those intervals with count
rates below the threshold, which is defined as 2-$\sigma$ above the
quiet average. The GTIs of both the hard band and the soft band are
used to screen the data. The background observations are screened by
the GTIs of the background data, which are produced using exactly the
same thresholds as for the corresponding target observations. Settings
of $FLAG=0$ and $PATTERN<13$ ($PATTERN<5$) for MOS (pn) are used in
the screening process.

An ``edetect\_chain'' command was used to detect point-like
sources. Point sources were subtracted before the further data
reduction.

\subsection{Background subtraction}

The background consists of several components exhibiting different
spectral and temporal characteristics (e.g. De Luca \& Molendi 2001;
Lumb et al. 2002; Read \& Ponman 2003). The background components can
approximately be divided into two groups (e.g.  Zhang et
al. 2004a). One contains the background components showing significant
vignetting, e.g. the cosmic X-ray background (CXB). The other contains
the components with very little or no vignetting,
e.g. particle-induced background.

Suitable background observations guarantee similar background
components showing vignetting as for the target in the same detector
coordinates. We choose the blank sky accumulations in the Chandra Deep
Field South (CDFS) as background. The background observations were
processed in the same way as the cluster observations. The CDFS
observations used the thin filter for all detectors and the FF/EFF
mode for MOS/pn. The medium filter was used for some target
observations. Using the medium filter the background is different from
the background using the thin filter at energies below
0.7~keV. Therefore we performed all the analysis at energies above
0.7~keV, in which the difference of the background is negligible. One
can subtract the background extracted in the same detector coordinates
as for the target. The cluster emission covers the inner part of the
field of view (FOV), $R \le 8^{\prime}$, and leaves the outer region
to exam the background. The difference between the background in the
target and background observations is taken into account as a residual
background in the background subtraction. We applied such a
double-background subtraction method specially developed for the
medium distant clusters in Zhang et al. (2006) for spectral imaging
analysis. More details about this method can be found in Zhang et
al. (2006). Note independently a similar type of double background
subtraction was earlier described in Pratt et al. (2001) and
formalised in Arnaud et al. (2002). Here we only briefly described our
double-background subtraction method (Zhang et al. 2006) as follows.

\subsubsection{Spectral analysis}
\label{s:bkgspe}

The spectra are extracted using the annuli within the truncation
radius (see $r_{\rm t}$ in Sect.~\ref{s:sx} and Table~\ref{t:global})
corresponding to a $S/N$ of 3 of the observational surface brightness
profile. The width of each annulus is selected to achieve precise
temperature estimates. In order to obtain temperature measurements
with uncertainties of $\sim 15$\%, we used the criterion of $\sim
2000$ net counts per bin in the 2--7~keV band\footnote{The 2--7~keV
band is sensitive to the determination of the temperatures for massive
clusters which have cluster temperatures higher than 3~keV.}.

For a given target spectrum ($S_{\rm a}$) extracted from the annulus
region with an area of $A_{\rm a}$ in the target observations, the
background spectrum $B_{\rm a}$ is extracted in the background
observations in the same detector coordinates as for the target
spectrum. Hereafter we call the count rate ratio of the target and
background observations limited to 10--12~keV (12--14~keV) for MOS
(pn) as $c_{\rm AB}$. The first-order background spectrum is $c_{\rm
AB} B_{\rm a}$. Because the cluster emission covers radii smaller than
$8^{\prime}$ as shown in Sect.~\ref{s:sx} (also see the truncation
radius $r_{\rm t}$ in Table~\ref{t:global}), the second-order
background spectrum can be prepared as follows. A spectrum ($S_{\rm
o}$) is extracted from the outer region
(e.g. $8^{\prime}<R<8.33^{\prime}$) with an area of $A_{\rm o}$ in the
target observations and its background spectrum ($B_{\rm o}$) in the
same detector coordinates but in the background observations. The
second-order background spectrum is $\frac{A_{\rm a}}{A_{\rm o}}
(S_{\rm o}-c_{\rm AB} B_{\rm o})$. Assuming that the second-order
background spectrum consists of the vignetted components such as CXB,
the cluster spectrum is then given by $S_{\rm a}-c_{\rm AB} B_{\rm
a}-\frac{A_{\rm a}}{A_{\rm o}}(S_{\rm o}-c_{\rm AB} B_{\rm o})$. We
applied the second-order background spectrum with and without
vignetting correction in the spectral analysis, and found that the
difference in temperature and metallicity measurements increases with
radius only up to 3\% and 17\%, respectively. Therefore the influence
on the spectral measurements due to possible non-vignetted components
such as the particle-induced background in the second-order background
is negligible.

We performed spectral fitting of the MOS and pn data simultaneously
with the XSPEC v11.3.1 software. Both the response matrix file (rmf)
and auxiliary response file (arf) are used to recover the correct
spectral shape and normalization of the cluster emission
component. The following are usually taken into account for the rmf
and arf, (i) a pure redistribution matrix giving the probability that
a photon of energy $E$, once detected, will be measured in data
channel PI, (ii) the quantum efficiency (without any filter, which, in
XMM-Newton calibration, is called closed filter position) of the CCD
detector, (iii) filter transmission, and (iv) geometric factors such
as bad pixel corrections and gap corrections (e.g. around 4\% for
MOS). The vignetting correction to effective area for off-axis
observations can be accounted for in the event lists by a weight
column created by ``evigweight''. All spectra are extracted
considering the vignetting correction by the weight column in the
event list produced by ``evigweight''. The on-axis rmf and arf are
co-created to account for (i) to (iv). We fixed the redshift to the
peak value of the cluster galaxy redshift histogram (e.g. Smith et al.
2005; Bardeau et al. 2006; Covone et al. 2006) and the Galactic
absorption to the weighted value in Dickey \& Lockman (1990). A
combined model ``wabs$*$mekal'' is then used with the on-axis arf and
rmf in XSPEC for the fitting.

\subsubsection{Image analysis}

The 0.7--2~keV band is used to derive the surface brightness
profiles. This ensures an almost temperature-independent X-ray
emission coefficient over the expected temperature range. The
vignetting correction to effective area is accounted for by the weight
column in the event lists created by ``evigweight''.  Geometric
factors such as bad pixel corrections are accounted for in the
exposure maps. The width of the radial bins is $2^{\prime \prime}$. An
azimuthally averaged surface brightness profile of the CDFS is derived
in the same detector coordinates as for the target. The count rate
ratios of the target and CDFS in the 10--12~keV band and 12--14~keV
band for MOS and pn, respectively, are used to scale the CDFS surface
brightness. The residual background in each annulus of the surface
brightness is the count rate in the 0.7--2~keV band of the area scaled
residual spectrum obtained in the spectral analysis. Both the scaled
CDFS surface brightness profile and the residual background are
subtracted from the target surface brightness profile. The background
subtracted and vignetting corrected surface brightness profiles for
three detectors are added into a single profile, and re-binned to
reach a significance level of at least 3-$\sigma$ in each annulus. The
truncation radii $r_{\rm t}$ are listed in Table~\ref{t:global}. For
most clusters, the particle-induced background varies by less than
10\% comparing the background observations to the target
observations. Therefore the dispersion of the re-normalization of the
background observations is typically 10\%. We take into account a 10\%
uncertainty of the scaled CDFS background and residual background.

\subsection{PSF and de-projection}
\label{s:deproj}

In the imaging analysis, we correct the PSF effect by fitting the
observational surface brightness profile with a surface brightness
profile model convolved with the empirical PSF matrices (Ghizzardi
2001).

Using the XMM-Newton point-spread function (PSF) calibrations in
Ghizzardi (2001) we estimated the redistribution fraction of the
flux. We found 20\% for bins with width about $0.5^{\prime}$ and less
than 10\% for bins with width greater than $1^{\prime}$ neglecting
energy dependent effects. We thus require an annulus width greater
than $0.5^{\prime}$ in the spectral analysis. For such distant massive
clusters, the PSF effect is only important within $0.5^{\prime}$,
which corresponds $\le 0.1r_{\rm 500}$ and introduces an uncertainty
to the final results of the temperature profiles. This has to be
investigated using deeper exposures with better photon statistic. The
PSF blurring can not be completely considered for the spectral
analysis the same way as for the image analysis because of the limited
photon statistic.

The projected temperature is the observed temperature from a
particular annulus, containing in projection the foreground and
background temperature measurements. Under the assumption of spherical
symmetry, the gas temperature in each spherical shell is derived by
de-projecting the projected spectra. In this procedure, the inner
shells contribute no flux to the outer annuli. The projected spectrum
in the outermost annulus is thus equal to the spectrum in the
outermost shell. The projected spectrum in the neighboring inner
annulus has contributions from all the spectra in the shell at the
radius of this annulus and in the outer shells (e.g. Suto et
al. 1998). We de-projected the spectra as described in Jia et
al. (2004) and performed the spectral fitting of the de-projected
spectra in XSPEC to measure the radial temperature and metallicity
profiles.

\section{X-ray properties}
\label{s:result}

The primary parameters of all 13 galaxy clusters are given in
Table~\ref{t:global}.

\subsection{Density contrast}

To determine the global cluster parameters, one needs a fiducial outer
radius that was defined as follows. The mean cluster density
contrast is the average density with respect to the critical density,
\begin{equation}
\Delta(<r)= \frac{3 M(<r)}{4 \pi r^3 \rho_{\rm c}(z)} \;.
\label{e:del}
\end{equation}
The critical density at redshift $z$ is $\rho_{\rm c}(z)=\rho_{\rm c0}
E^2(z)$, where $E^2(z)=\Omega_{\rm
m}(1+z)^3+\Omega_{\Lambda}+(1-\Omega_{\rm
m}-\Omega_{\Lambda})(1+z)^2$. $r_{\Delta}$ is the radius within which
the density contrast is $\Delta$. $M_{\Delta}$ is the total mass
within $r_{\Delta}$. For $\Delta=500$, $r_{500}$ is the radius within
which the density contrast is 500 and $M_{500}$ is the total mass
within $r_{500}$. For $\Delta=2500$, $r_{2500}$ is the radius within
which the density contrast is 2500 and $M_{2500}$ is the total mass
within $r_{2500}$.

\subsection{Temperature profiles}
\label{s:kt}

\begin{figure*}
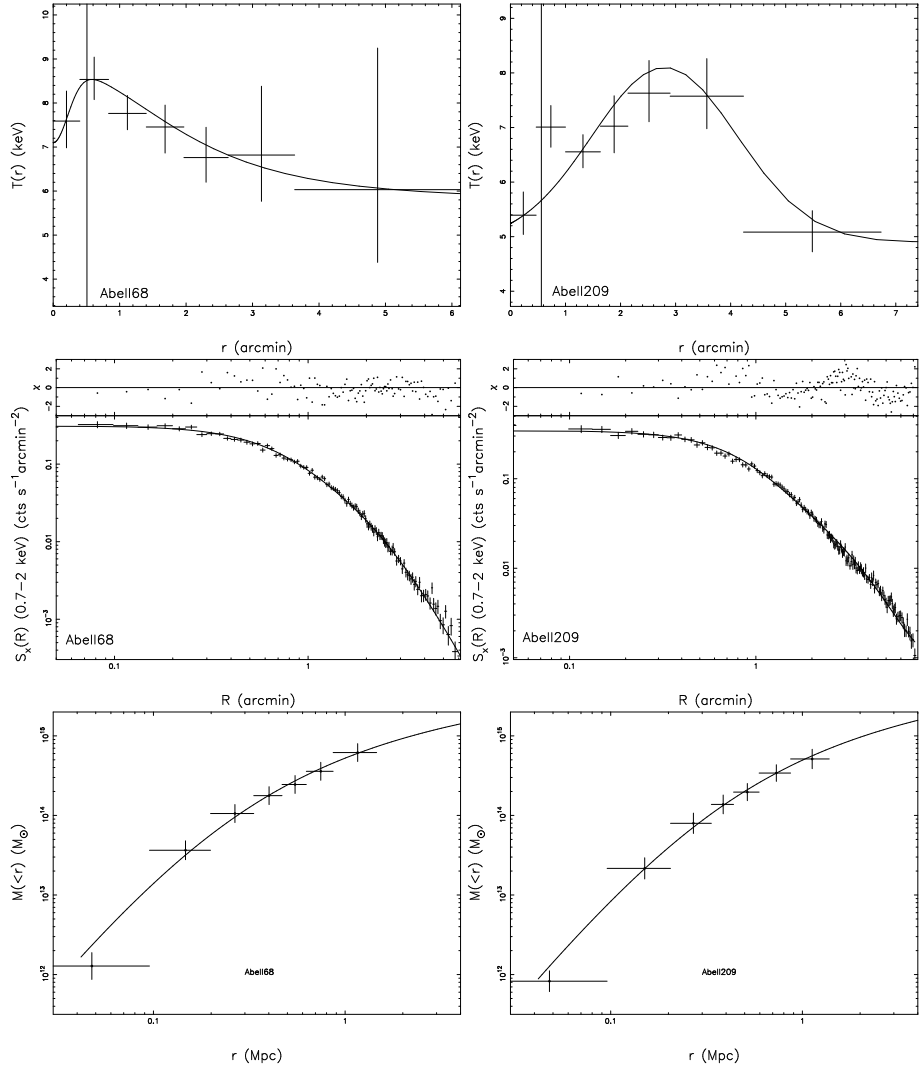

\begin{center}
\includegraphics[angle=270,width=6cm]{6567f1a.ps}
\includegraphics[angle=270,width=6cm]{6567f1b.ps}

\includegraphics[angle=270,width=6cm]{6567f1c.ps}
\includegraphics[angle=270,width=6cm]{6567f1d.ps}

\includegraphics[angle=270,width=6cm]{6567f1e.ps}
\includegraphics[angle=270,width=6cm]{6567f1f.ps}

\end{center}
\caption{De-projected temperature profiles (upper panels), surface
brightness profiles (middle panels) and mass profiles (lower
panels). The temperature profiles are approximated by the
parameterization $T(r)=T_3
\exp[-(r-T_1)^2/T_2]+T_6(1+r^2/T_4^2)^{-T_5}+T_7$ crossing all the
data points (solid). The vertical line denotes $0.1 r_{500}$. The
surface brightness profiles are parameterized by double-$\beta$
model. The observed mass profiles are fitted by generalized NFW
model. \label{f:ktcom1}}
\end{figure*}

\begin{figure*}
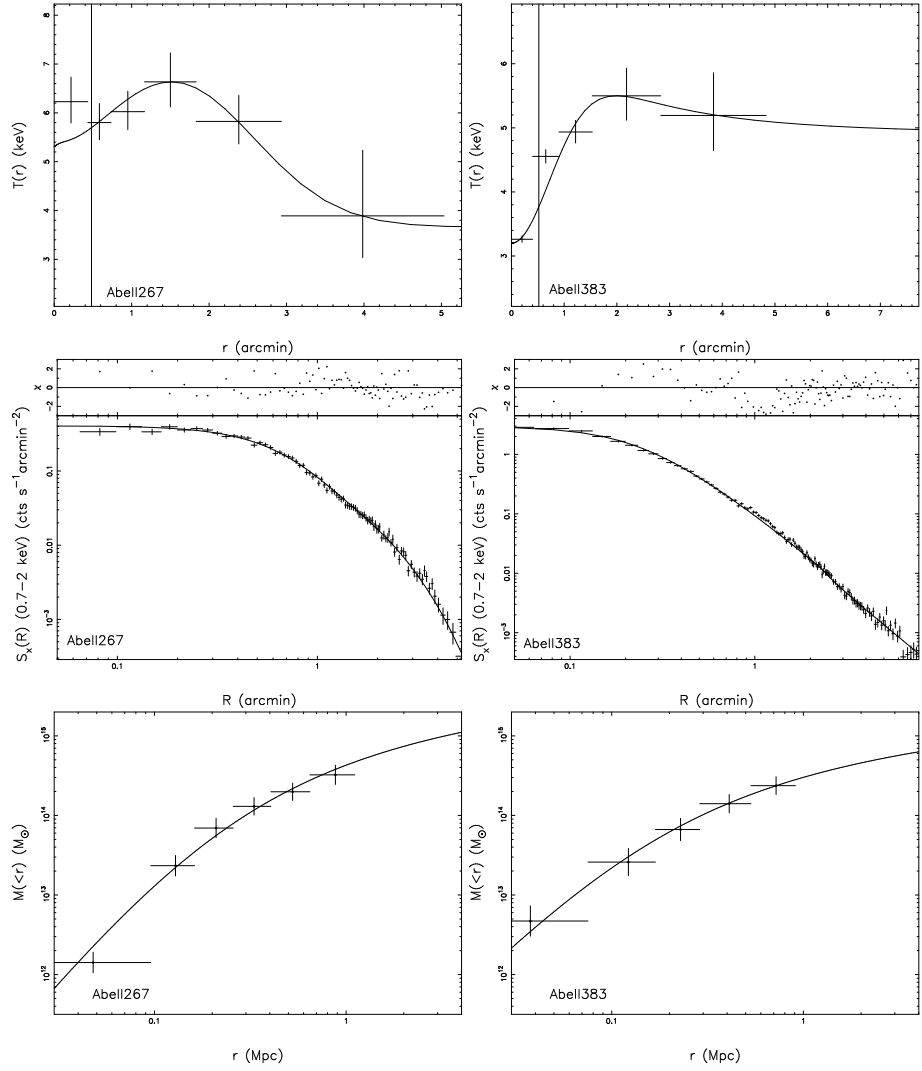

\begin{center}

\includegraphics[angle=270,width=6cm]{6567fb1a.ps}
\includegraphics[angle=270,width=6cm]{6567fb1b.ps}

\includegraphics[angle=270,width=6cm]{6567fb1c.ps}
\includegraphics[angle=270,width=6cm]{6567fb1d.ps}

\includegraphics[angle=270,width=6cm]{6567fb1e.ps}
\includegraphics[angle=270,width=6cm]{6567fb1f.ps}

\end{center}
\caption{See the caption in Fig.~\ref{f:ktcom1}.
\label{f:ktcom2}}
\end{figure*}

\begin{figure*}
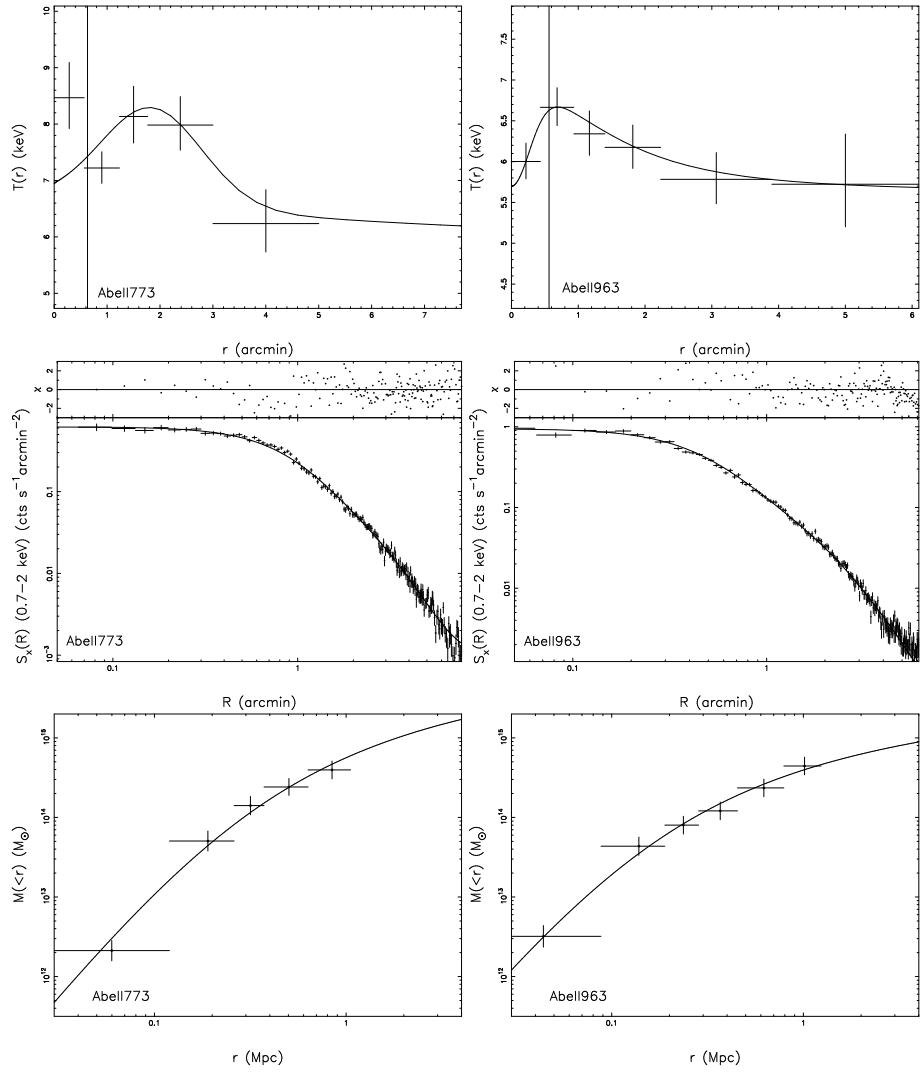

\begin{center}
\includegraphics[angle=270,width=6cm]{6567fb2a.ps}
\includegraphics[angle=270,width=6cm]{6567fb2b.ps}

\includegraphics[angle=270,width=6cm]{6567fb2c.ps}
\includegraphics[angle=270,width=6cm]{6567fb2d.ps}

\includegraphics[angle=270,width=6cm]{6567fb2e.ps}
\includegraphics[angle=270,width=6cm]{6567fb2f.ps}
\end{center}
\caption{See the caption in Fig.~\ref{f:ktcom1}.
\label{f:ktcom3}}
\end{figure*}

\begin{figure*}
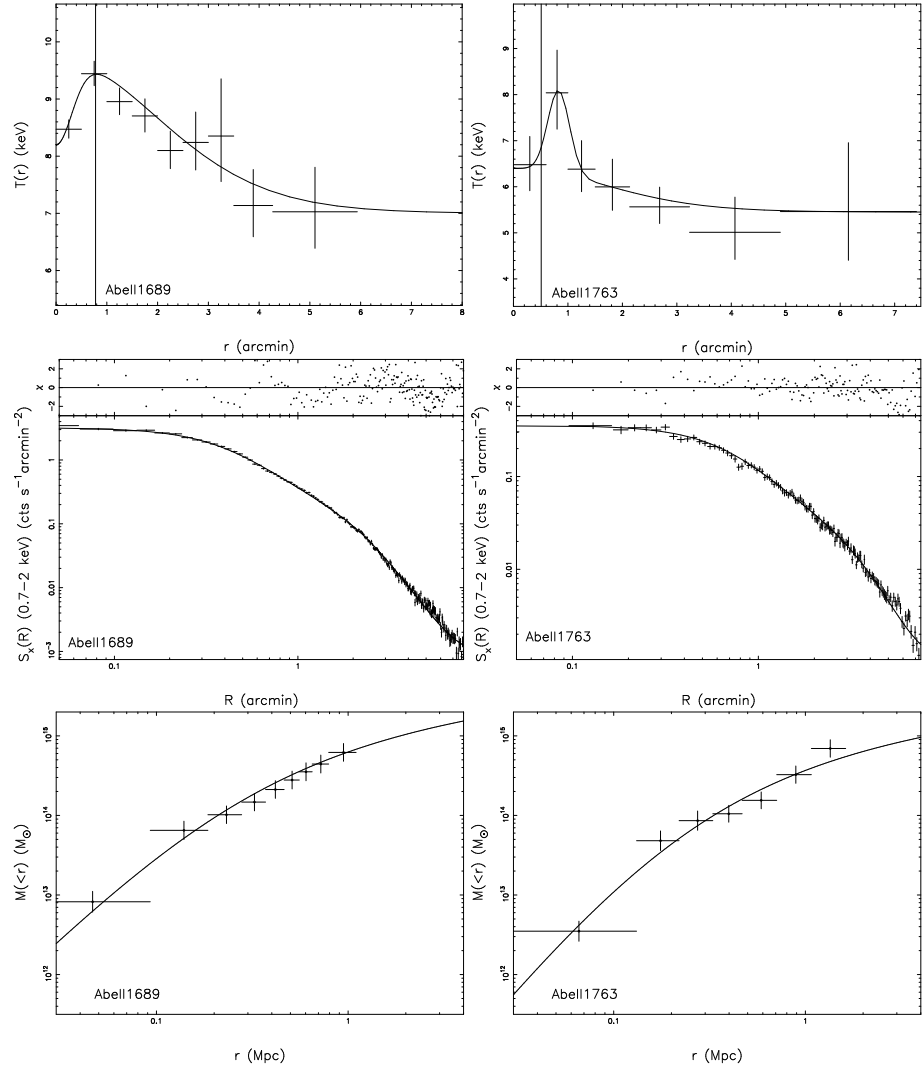

\begin{center}
\includegraphics[angle=270,width=6cm]{6567fb3a.ps}
\includegraphics[angle=270,width=6cm]{6567fb3b.ps}

\includegraphics[angle=270,width=6cm]{6567fb3c.ps}
\includegraphics[angle=270,width=6cm]{6567fb3d.ps}

\includegraphics[angle=270,width=6cm]{6567fb3e.ps}
\includegraphics[angle=270,width=6cm]{6567fb3f.ps}
\end{center}
\caption{See the caption in Fig.~\ref{f:ktcom1}.
\label{f:ktcom4}}
\end{figure*}

\begin{figure*}
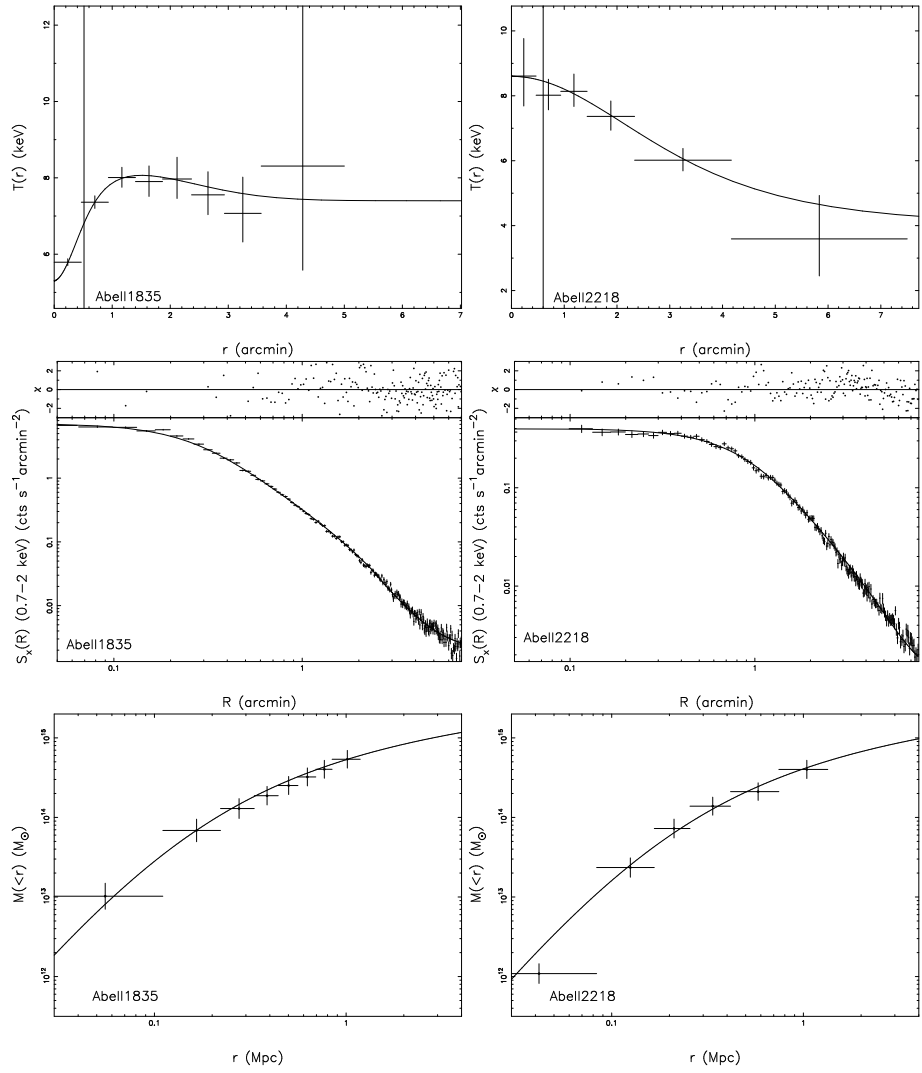

\begin{center}
\includegraphics[angle=270,width=6cm]{6567fb4a.ps}
\includegraphics[angle=270,width=6cm]{6567fb4b.ps}

\includegraphics[angle=270,width=6cm]{6567fb4c.ps}
\includegraphics[angle=270,width=6cm]{6567fb4d.ps}

\includegraphics[angle=270,width=6cm]{6567fb4e.ps}
\includegraphics[angle=270,width=6cm]{6567fb4f.ps}
\end{center}
\caption{See the caption in Fig.~\ref{f:ktcom1}.
\label{f:ktcom5}}
\end{figure*}

\begin{figure*}
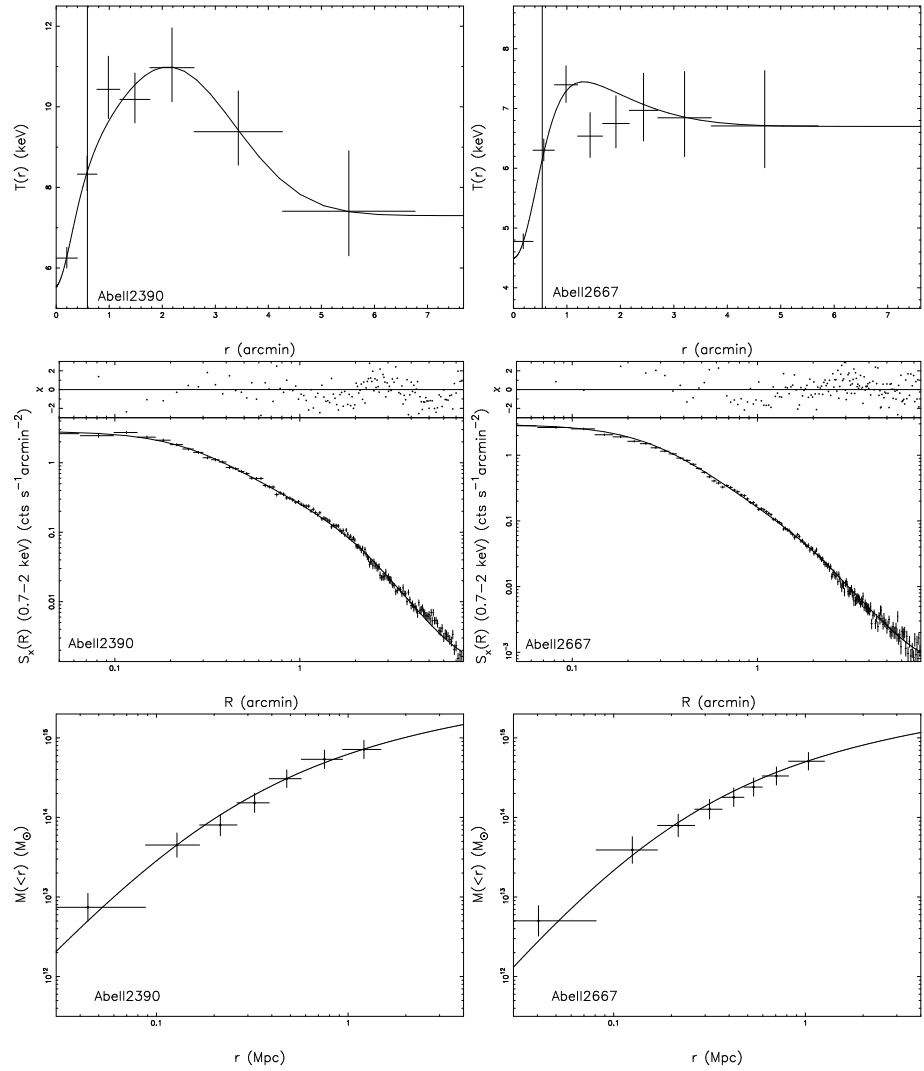

\begin{center}
\includegraphics[angle=270,width=6cm]{6567fb5a.ps}
\includegraphics[angle=270,width=6cm]{6567fb5b.ps}

\includegraphics[angle=270,width=6cm]{6567fb5c.ps}
\includegraphics[angle=270,width=6cm]{6567fb5d.ps}

\includegraphics[angle=270,width=6cm]{6567fb5e.ps}
\includegraphics[angle=270,width=6cm]{6567fb5f.ps}
\end{center}
\caption{See the caption in Fig.~\ref{f:ktcom1}.
 \label{f:ktcom6}}
\end{figure*}

Temperature profiles probe the thermodynamical history of galaxy
clusters. XMM-Newton (also Chandra), in contrast to earlier
telescopes, has a less energy-dependent, smaller PSF, more reliable to
study cluster temperature profiles. We de-projected the spectra as
described in Sect.~\ref{s:deproj} and performed the spectral fitting
in XSPEC to obtain the radial measurements of the temperature and
metallicity. In the left panels of Fig.~\ref{f:ktcom1} ({\it Also see
Figs.~\ref{f:ktcom2}--\ref{f:ktcom6} in the electronic edition of the
Journal for the whole sample}) we show the radial temperature
profiles. The temperature profiles are approximated by
\begin{equation}
T(r)=T_3
\exp[-(r-T_1)^2/T_2]+T_6(1+r^2/T_4^2)^{-T_5}+T_7\; ,
\label{e:t}
\end{equation}
where $T_{i}$, $i=1,...,7$, are simply for parameterization without
physical meaning.

A cool core cluster (CCC) often shows both a peaked surface brightness
profile and a steep temperature drop towards the cluster center. The
classical mass deposition rate is derived by the radiative cooling
model linking to the X-ray luminosity and thus the surface brightness
profile (e.g. Peres et al. 1998). Chen et al. (2007) defined CCCs by
two criteria, (1) scaled mass deposition rate
$\dot{M}/M_{500}>10^{-13}/$yr, and (2) mass deposition rate
$\dot{M}>0.01 M_{\odot}$/yr. They found that 60\% of the HIFLUGS
sample are CCCs. With high resolution Chandra observations Vikhlinin
et al. (2006b) defined CCCs by the cuspness of the surface brightness
profile ($\alpha =\frac{d \log \rho_{\rm g}}{d \log r} > 0.5$ at
$r=0.04 R_{500}$), and found that 2/3 of the clusters at $z<0.5$ and
15\% at $z>0.5$ are CCCs, respectively. At $z>0.5$, they found no
pronounced CCCs with $\alpha > 0.7$.

For Abell~383, Abell~1835, Abell~2390 and Abell~2667, the temperatures
drop to the values less than 70\% of the maximum temperatures towards
the cluster center. There is no strict division between CCCs and
non-CCCs. We empirically define the CCC as the cluster whose
temperature drop is greater than 30\% of the peak temperature towards
the cluster center. Hereafter these 4 clusters are called CCCs in the
sample. As we show later, the 4 CCCs show shorter central cooling
time, larger cooling radii (scaled to $r_{500}$) and lower central
entropies compared to the non-CCCs. Using our definition of the CCC,
the fractions of the CCCs are 33\% and 15\%, respectively, for the
samples at $z\sim 0.2$ in this work and $z
\sim 0.3$ in Zhang et al. (2006). This could 
indicate the evolution of the cool cores for massive clusters between
$z\sim 0.2$ and $z\sim 0.3$ that the CCC fraction decreases with
increasing redshift. This evolution could have already become
important at $z\sim 0.3$.

However, the Poisson noise is quite significant for such small
samples. Using re-randomization of the Monte Carlo simulated
temperature profiles, the CCC fraction is 25--40\% for the pilot
LoCuSS sample, and 15--30\% for the REFLEX-DXL sample,
respectively. The difference in the fraction of the CCCs is actually
not significant between the samples at redshift $\sim 0.2$ and
0.3. The evolution of the cool cores is thus not justified by the low
significance of our result. Furthermore, the bootstrap effect
accounting for missing clusters due to (1) the selection effect of the
sample, and (2) the flares the XMM-Newton observational run, can only
be recovered by mock catalogs in simulations. Additionally, the
detection of the temperature drop in the cluster center becomes
difficult for distant clusters, which can enhance the phenomenon that
the CCC fraction decreases with the increasing redshift. A less
resolution-dependent way to investigate the evolution of the cool
cores is to use the luminosity based on the flux images, which will be
addressed in Sect.~\ref{s:lxcoolcore}.

\subsection{Surface brightness}
\label{s:sx}

A $\beta$ model (e.g. Cavaliere \& Fusco-Femiano 1976; Jones \& Forman
1984) is often used to describe electron density profiles in
clusters. To obtain an acceptable fit for all clusters in this sample,
we adopt a double-$\beta$ model of the electron number density
profile, $n_{\rm e}(r)=n_{\rm e01}(1+r^2/r_{\rm
c1}^2)^{-3\beta/2}+n_{\rm e02}(1+r^2/r_{\rm c2}^2)^{-3\beta/2}$, where
$n_{\rm e0}=n_{\rm e01}+n_{\rm e02}$ is the central electron number
density (see Table~\ref{t:catalog2}), $\beta$ the slope parameter, and
$r_{\rm c1}$ and $r_{\rm c2}$ the core radii of the inner and outer
components, respectively (e.g. Bonamente et al. 2006). The soft band
(e.g. 0.7--2~keV) X-ray surface brightness profile model $S(R)$, in
which $R$ is the projected radius, is linked to the radial profile of
the ICM electron number density $n_{\rm e}(r)$ as an integral
performed along the line-of-sight for hot clusters ($T >2$~keV),
%6
\begin{equation}
S_{\rm X}(R)\propto\int_{R}^{\infty} n_{\rm e}^2 d\ell.
\end{equation}
We fitted the observed surface brightness profile by this integral
convolved with the PSF matrices (middle panels in Fig.~\ref{f:ktcom1},
{\it also see Figs.~\ref{f:ktcom2}--\ref{f:ktcom6} in the electronic
edition of the Journal for the whole sample}) and obtained the
parameters of the double-$\beta$ model of the electron density
profile. The fit was performed within the truncation radius ($r_{\rm
t}$, see Table~\ref{t:global}) corresponding to a $S/N$ of 3 of the
observational surface brightness profile. The truncation radii $r_{\rm
t}$ are above $r_{500}$ for all 12 clusters. The cluster cores,
referring to the central $\beta$ model of the double-$\beta$ model,
span a broad range up to 0.2~$r_{500}$. The cluster cores for the CCCs
are resolved with the current observations.

\subsection{Mass distribution}
\label{s:massdis}

We assume that the ICM is in hydrostatic equilibrium within the
gravitational potential dominated by DM and its distribution appears
spherical symmetry. The ICM can then be used to trace the cluster
potential. The cluster mass is calculated from the X-ray measured ICM
density and temperature distributions,
\begin{equation}
\frac{1}{\mu m_{\rm p} n_{\rm e}(r)}\frac{d[n_{\rm e}(r) T(r)]}{dr}=
  -\frac{GM(<r)}{r^2}~,
\label{e:hyd}
\end{equation}
where $\mu=0.62$ is the mean molecular weight per hydrogen atom.
Following the method in Zhang et al. (2006), we use a set of input
parameters of the approximation functions, in which $\beta$, $n_{\rm
e0i}$, $r_{\rm ci}$ ($i=1,2$) represent the electron number density
profile $n_{\rm e}(r)$ and $T_{\rm i}$ ($i=1,...,7$) represent the
temperature profile $T(r)$, respectively, to compute the mean cluster
mass. The mass uncertainties are propagated using the uncertainties of
the electron number density and temperature measurements by Monte
Carlo simulations as follows. For each cluster, we simulated
electron density and temperature profiles of a sample of 100 clusters
using the observed electron density and temperature profiles and their
errors. The mass profiles and other properties of the 100 simulated
clusters were calculated to derive the error bars.

The observed mass profile was used to estimate $M_{500}$ and $r_{500}$
(see Table~\ref{t:catalog2}). The NFW model (e.g.  Navarro et
al. 1997, NFW) does not provide an acceptable fit for the observed
mass profiles. We adopt the acceptable fit of a generalized NFW model
(e.g. Hernquist 1990; Zhao 1996; Moore et al. 1999, Navarro et
al. 2004), $\rho_{\rm DM}(r)=\rho_{\rm s} (r/r_{\rm
s})^{-\alpha}(1+r/r_{\rm s})^{\alpha-3}$, where $\rho_{\rm s}$ and
$r_{\rm s}$ are the the characteristic density and scale of the halo,
respectively (see Table~\ref{t:mcom}). Note $\alpha=0$ could be an
artificial fact in fitting the central data points of the mass profile
due to low spatial resolution and/or significant PSF effect. In the
right panels of Fig.~\ref{f:ktcom1} ({\it Also see
Figs.~\ref{f:ktcom2}--\ref{f:ktcom6} in the electronic edition of the
Journal for the whole sample}), we present the observed mass profiles
and their best generalized NFW model fits. We note that the data
points in the mass profiles are not completely independent, which can
introduce uncertainties in the fitting.

\subsection{Gas mass fraction distribution}

The gas mass fraction is an important parameter for cluster physics,
e.g. heating and cooling processes, and cosmological applications
using galaxy clusters (e.g. Vikhlinin et al. 2002; Allen et
al. 2004). The gas mass fraction distribution is defined as $f_{\rm
gas}(<r)=M_{\rm gas}(<r)/M(<r)$. The gas mass, total mass and gas mass
fraction at $r_{500}$ are given in Table~\ref{t:catalog2}. We obtained
an average gas mass fraction of $0.13 \pm 0.01$ at $r_{\rm 500}$ for
the pilot LoCuSS sample. This result agrees with the average gas mass
fraction of $0.116 \pm 0.007$ at $r_{\rm 500}$ for the REFLEX-DXL
sample at $z
\sim 0.3$ in Zhang et al. (2006). The average gas mass fraction at
$r_{2500}$ of the pilot LoCuSS sample is $0.107 \pm 0.005$, in good
agreement with the value of the REFLEX-DXL sample in Zhang et
al. (2006) giving $0.100\pm 0.004$, and the value in Allen et
al. (2002) based on Chandra observations of 7 clusters giving $f_{\rm
gas}\sim 0.105$--$0.138h^{-3/2}_{70}$.

As expected, the gas mass fraction distributions of all the clusters
are lower than the universal baryon fraction, $f_{\rm b}=\Omega_{\rm
b}/\Omega_{\rm m}=0.176 \pm 0.019$, based on the WMAP three year
results in Spergel et al. (2006). This is because that the baryons in
galaxy clusters reside mostly in hot gas together with a small
fraction of stars as implied in simulations (15\%~$f_{\rm gas}$,
e.g. Eke et al. 1998, Kravtsov et al. 2005; 5--7\%~$f_{\rm gas}$,
e.g. Eke et al. 2005). In principle, $\Omega_{\rm m}$ can be
determined from the baryon fraction, $f_{\rm b}=f_{\rm gas}+f_{\rm
gal}$, in which a contribution from stars in galaxies is given by
$f_{\rm gal}=0.02\pm 0.01 h_{50}^{-1}$ (White et al. 1993). The gas
mass fractions of the pilot LoCuSS sample support a low matter density
Universe as also shown in recent studies (e.g. Allen et al. 2002;
Ettori et al. 2003; Vikhlinin et al. 2003).

\subsection{Global temperature and metallicity}

Vikhlinin et al. (2005) used the volume average of the radial
temperature profile in a certain radial range as the global
temperature. We seek well defined global temperatures for the pilot
LoCuSS clusters as follows. We calculated the volume average of the
radial temperature profile in the radial range of $0.1 r_{500}<r<0.5
r_{500}$ and $0.2 r_{500}<r<0.5 r_{500}$, respectively. We also
measured the spectral temperatures ($T^{\rm spec}$) and metallicities
($Z$) in the annuli of $R<2/3r_{\rm t}$, $0.1 r_{500}<R<0.5 r_{500}$
and $0.2 r_{500}<R<0.5 r_{500}$, respectively (see
Table~\ref{t:tglobal}). The spectral temperature agrees with the
volume average of the radial temperature profile better in the annulus
of 0.2--0.5$r_{500}$ than in the annulus of 0.1--0.5$r_{500}$ because
the measurements limited to 0.2--0.5$r_{500}$ are less affected by the
cool cores for the CCCs.

We found an average of $0.29 \pm 0.02$~$Z_{\odot}$ for the global
metallicities of the sample measured in the $<2/3r_{\rm t}$
regions. For the 8 non-CCCs of this sample the average metallicity,
$0.26\pm 0.02 Z_{\odot}$, agrees with the value of $0.24 \pm
0.03$~$Z_{\odot}$ for the REFLEX-DXL sample (Zhang et al. 2006), and
$0.21^{+0.10}_{-0.05} Z_{\odot}$ for 18 distant ($0.3<z<1.3$) clusters
in Tozzi et al. (2003) within the measurement uncertainties. The
average metallicity of the CCCs of the sample, $0.36\pm 0.03
Z_{\odot}$, agrees with the value of $0.38 \pm 0.07Z_{\odot}$ for 21
CCCs in Allen \& Fabian (1998). 

\section{Self-similarity of the scaled profiles of the X-ray properties}
\label{s:prof}

Simulations (e.g. Navarro et al. 1997, 2004) suggest a
self-similar structure for galaxy clusters in hierarchical
structure formation scenarios. The scaled profiles of the X-ray
properties and their scatter can be used to quantify the
structural variations. This is a probe to test the regularity of
galaxy clusters and to understand their formation and evolution.
The accuracy of the determination of the scaling relations,
limited by how precise the cluster mass and other global
quantities can be estimated, is of prime importance for the
cosmological applications of clusters of galaxies.

Because the observational truncation radii ($r_{\rm t}$) in the
surface brightness profiles are above $r_{500}$ for all clusters but
below $r_{200}$ for most clusters in the sample, we use $r_{500}$ for
radial scaling.

The following redshift evolution corrections (e.g. Zhang et al. 2006)
are usually used to account for the dependence on the evolution of the
cosmological parameters,

$S_{\rm X} \cdot E^{-3}(z) \; (\Delta_{c,z}/\Delta_{c,0})^{-1.5}
\propto f(T)$,

$S \cdot E^{4/3}(z) \; (\Delta_{c,z}/\Delta_{c,0})^{2/3} \propto f(T)$,

$L \cdot E^{-1}(z) \; (\Delta_{c,z}/\Delta_{c,0})^{-0.5} \propto f(T)$,

$M \cdot E(z) \; (\Delta_{c,z}/\Delta_{c,0})^{0.5} \propto f(T)$,

$M_{\rm gas} \cdot E(z) \; (\Delta_{c,z}/\Delta_{c,0})^{0.5} \propto f(T)$,

$r \cdot E(z) \; (\Delta_{c,z}/\Delta_{c,0})^{0.5} \propto f(T)$,

where $\Delta_{\rm c,z}=18\pi^2+82(\Omega_{\rm m,z}-1)-39(\Omega_{\rm
m,z}-1)^2$ is the analytic approximation derived from the top-hat
spherical collapse model for a flat Universe (Bryan \& Norman 1998)
and $\Omega_{\rm m,z}$ the cosmic density parameter at redshift
$z$. The function $f(T)$ denotes the best fitting power law
parameterization.

We investigated the self-similarity of the scaled profiles of the
X-ray properties for this sample as follows.

\subsection{Scaled temperature profiles}

\begin{figure*}
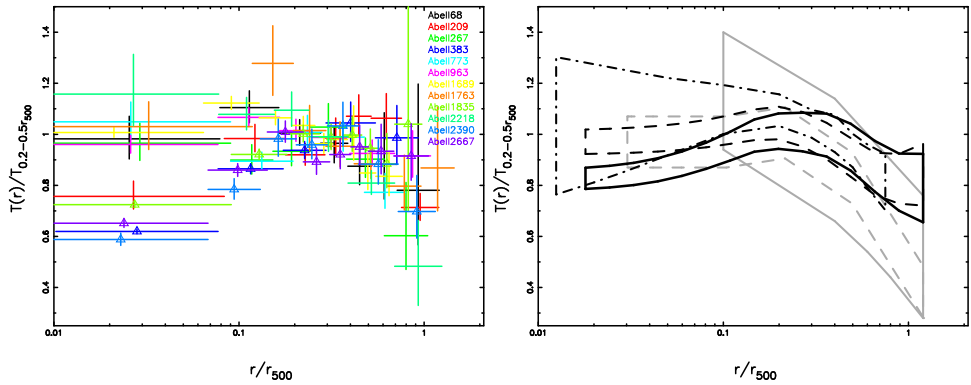

\begin{center}
\includegraphics[width=5cm,angle=270]{6567f2a.ps}
\includegraphics[width=5cm,angle=270]{6567f2b.ps}
\end{center}
\caption{{\it Left:} Scaled radial temperature profiles for the
pilot LoCuSS sample. The CCCs are in triangles. {\it See the
electronic edition of the Journal for a color version.} {\it
Right:} An average temperature profile of the pilot LoCuSS sample
(black, solid) compared to the temperature profile ranges in
Markevitch et al. (1998, grey, solid), Vikhlinin et al. (2005,
grey, dashed), Zhang et al. (2006, black, dashed) and Pratt et al.
(2007, black, dash-dotted). \label{f:scalet} }
\end{figure*}

We scaled the radial temperature profiles by $T_{0.2-0.5r_{500}}$ and
$r_{500}$ (Fig.~\ref{f:scalet}). Within $0.2r_{\rm 500}$, we observed
a temperature drop to at least 70\% of the maximum value towards the
cluster center for 4 clusters (defined as the CCCs) and an almost
constant temperature distribution for the non-CCCs. Previous
observations have shown that temperature measurements on scales below
$0.2 r_{500}$ tend to show peculiarities linked to the cluster
dynamical history. For example, the temperatures of merger clusters
can be boosted (Smith et al. 2005). However, the boosting is serious
mainly in the cluster cores. Using global temperatures excluding the
$<0.2 r_{500}$ regions can thus (1) tight the scaled profiles of the
X-ray properties, (2) minimize the scatter in the X-ray scaling
relations and (3) reach a better agreement in the normalization
between the X-ray scaling relations for the CCCs and non-CCCs.

An average temperature profile region is derived by averaging the
1-$\sigma$ boundary of the scaled radial temperature profiles. The
average temperature profile for the whole sample gives $T(r)
\propto r^{0.21 \pm 0.04}$, and for the CCC subsample $T(r) \propto
r^{0.38 \pm 0.04}$, respectively, in the $r<0.2 r_{500}$ region. Note
the segregation could be more pronounced considering the spatial
resolution and PFS effects. This temperature behavior in the cluster
cores for the CCC subsample is very similar to the behavior for the
nearby CCC sample in Sanderson et al. (2006) giving $T(r) \propto
r^{0.4}$ based on Chandra observations. In the outskirts ($0.2
r_{500}<r<r_{500}$), the whole sample shows a self-similar behavior
giving $T(r) \propto r^{-0.36 \pm 0.18}$ with scatter within 20\%. A
temperature profile decreasing down to 80\% of the maximum value with
an intrinsic scatter of $\sim 20$\% has been observed at about $r_{\rm
500}$ for the average of the sample.

Studies of the cluster temperature distributions (e.g. Markevitch et
al. 1998; De Grandi \& Molendi 2002; Vikhlinin et al. 2005; Zhang et
al. 2004a, 2006; Pratt et al. 2007) indicate a universal temperature
profile with a significant decline beyond an isothermal center. The
average temperature profile of the pilot LoCuSS sample is consistent
with the average profiles from ASCA in Markevitch et al. (1998),
BeppoSAX in De Grandi \& Molendi (2002) and XMM-Newton in Zhang et
al. (2004a, 2006) and Pratt et al. (2007), but is slightly less steep
than the profile from Chandra (using an assumed uncertainty of 20\% of
the averaged temperature profile as an approximate illustration) in
Vikhlinin et al. (2005). A similarly universal temperature profile is
indicated in outskirts of clusters by simulations (e.g. Borgani et
al. 2004; Borgani 2004).

\subsection{Scaled metallicity profiles}

\begin{figure*}
\begin{center}
\includegraphics[width=5cm,angle=270]{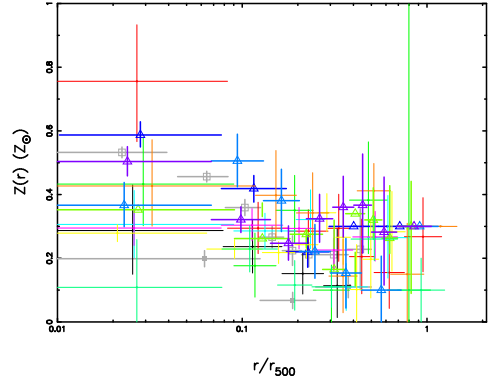}
\end{center}
\caption{Scaled metallicity profiles. The values in the last bins
for some clusters are fixed to 0.3~$Z_{\odot}$ in the spectral
fitting which have no error bars. The CCCs are in triangles. The
colors have the same meaning as those in Fig.~\ref{f:scalet}. Open
and filled boxes denote the average of the scaled metallicity
profiles of the CCCs and non-CCCs, respectively, in De Grandi \&
Molendi (2002, grey). \label{f:met} }
\end{figure*}

The metallicity profiles are shown in Fig.~\ref{f:met} with their
radii scaled to $r_{500}$. We observed metallicity peaks towards
the cluster centers within 0.2~$r_{500}$ for the CCCs. There is no
evident evolutionary effect comparing the pilot LoCuSS sample to
the nearby CCCs in De Grandi \& Molendi (2002) within the measure
uncertainties.

\subsection{Scaled surface brightness profiles}

\begin{figure*}
\begin{center}
\includegraphics[width=5cm,angle=270]{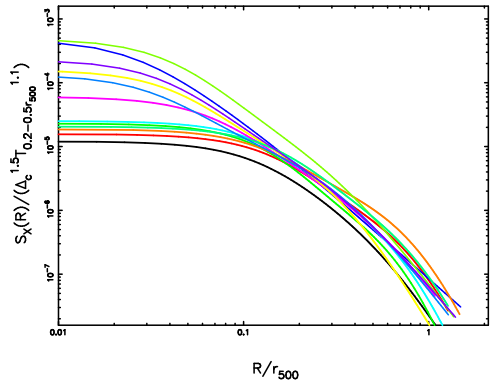}
\end{center}
\caption{Surface brightness profile fits scaled according to the
empirical scaling, $S_{\rm X} \propto T^{1.1}$. The colors have the
same meaning as those in Fig.~\ref{f:scalet}. \label{f:scalesx} }
\end{figure*}

In the standard self-similar model, gas mass scales with mass and thus
temperature as $M_{\rm gas}\propto M \propto T^{1.5}$, which gives
$S_{\rm X} \propto T^{0.5}$ (e.g. Arnaud et al. 2002). The $S_{\rm
X}$--$T$ scaling based on the empirical $M_{\rm gas}$--$T$ relation
provides the least scatter in the scaled surface brightness profiles
(e.g. Neumann \& Arnaud 2001; Arnaud et al. 2002). We thus applied the
current empirical relation $M_{\rm gas}\propto T^{1.8}$ (e.g. Mohr et
al. 1999; Vikhlinin et al. 1999; Castillo-Morales \& Schindler 2003)
corresponding to the scaling $S_{\rm X} \propto T^{1.1}$ to scale the
surface brightness profiles (Fig.~\ref{f:scalesx}).  We found a less
scattered self-similar behavior at $R>0.2 r_{500}$ for the scaled
surface brightness profiles compared to the profiles scaled by $S_{\rm
X} \propto T^{0.5}$.

The bolometric X-ray luminosity (here we use the 0.01--100~keV band)
is given by $L_{\rm bol} \propto \int \widetilde{\Lambda}(r) n^2_{\rm
e}(r) dV$, practically an integral of the X-ray surface brightness to
2.5$r_{500}$ ($L_{\rm bol}^{\rm incc}$ in Table\ref{t:catalog2}). The
value 2.5$r_{500}$ is used because that the extrapolated virial radii
are about 2.2--2.6$r_{500}$. We note the luminosity only varies within
3\% setting the truncation radius to the value from $r_{500}$ to $2.5
r_{500}$.

The surface brightness profiles show small core radii for the CCCs and
large core radii for the non-CCCs. Similar to the REFLEX-DXL sample
(Zhang et al. 2006) the core radii populate a broad range of values up
to 0.2~$r_{500}$. The X-ray luminosity is sensitive to the presence of
the cool core. It can thus be used to probe the evolution of the cool
cores. We present the bolometric luminosity including and excluding
the $R<0.2 r_{500}$ region ($L_{\rm bol}^{\rm incc}$ and $L_{\rm
bol}^{\rm excc}$) in Table~\ref{t:global}. The fractions of the X-ray
luminosity attributed to the $R<0.2 r_{500}$ region span a broad range
up to 70\%. This introduces large uncertainties in the use of the
luminosity $L_{\rm bol}^{\rm incc}$ as a mass indicator. The
normalization varies with the fraction of the CCCs in the sample and
the significance of their cool cores by a factor of 30--70\%. The
X-ray luminosity excluding the $<0.2 r_{500}$ region ($L_{\rm
bol}^{\rm excc}$) is much less dependent on the presence of the cool
core. However, to better re-produce the cluster luminosity and the
normalization of the luminosity--temperature/mass relation, we use the
X-ray luminosity corrected for the cool core by assuming\footnote{Note
this assumption was used only in the calculation for $L_{\rm bol}^{\rm
corr}$, not for any other quantity.}  $S_{\rm X}(R)=S_{\rm X}(0.2
r_{500})$ for the $R<0.2 r_{500}$ region ($L_{\rm bol}^{\rm corr}$ in
Table~\ref{t:global}). Using $L_{\rm bol}^{\rm corr}$ for the X-ray
scaling relations we obtained reduced scatter and consistent
normalization for the CCCs and non-CCCs as shown later. Note the
luminosity is lower (by up to 10\%) assuming a constant luminosity in
the cluster core instead of a beta model, especially for the
pronounced CCCs. However, the luminosity corrected for the cool core
by the beta model still introduces relatively significant scater
dominated by the CCCs due to the correlation between the core radius
and the slope parameter.

\subsection{Scaled cooling time profiles}
\label{s:tc}

The cooling time is derived by the total energy of the gas divided by
the energy loss rate
%11
\begin{equation}
t_{\rm c} = \frac{2.5 n_{\rm g}T}{n_{\rm e}^2\widetilde{\Lambda}}
\label{e:tcool}
\end{equation}
where $\widetilde{\Lambda}$, $n_{\rm g}$, $n_{\rm e}$ and $T$ are the
radiative cooling function, gas number density, electron number
density and temperature, respectively. We compute the upper limit of
the age of the cluster as an integral from the cluster redshift $z$ up
to $z=100$. Cooling regions are those showing cooling time less than
the upper limit of the cluster age. The boundary radius of such a
region is called the cooling radius. The cooling radius is zero when
the cooling time in the cluster center is larger than the upper limit
of the cluster age. The cooling time at the resolved inner most radii
of the surface brightness profiles and cooling radii are given in
Table~\ref{t:catalog2}.

\begin{figure*}
\begin{center}
\includegraphics[width=5cm,angle=270]{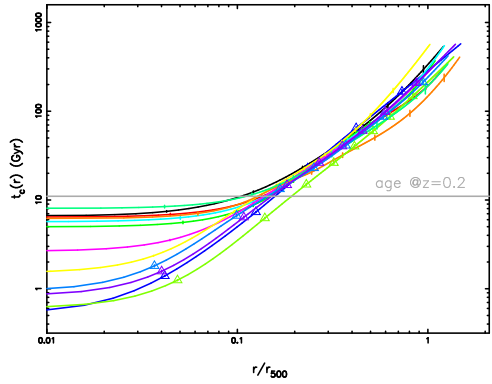}
\end{center}
\caption{Cooling time profiles. The CCCs are in triangles. The
horizontal line denotes the age of the Universe at $z \sim 0.2$.
The colors have the same meaning as those in Fig.~\ref{f:scalet}.
\label{f:tc} }
\end{figure*}

In Fig.~\ref{f:tc}, we show the cooling time profiles with their radii
scaled to $r_{500}$. The CCCs show larger cooling radii in unit of
$r_{500}$. In the cluster centers, the cooling time is all shorter
than the upper limit of the cluster age. The CCCs show much shorter
cooling time than for the non-CCCs. The scaled cooling time profiles
show a self-similar behavior above $0.2 r_{500}$ for the CCC and
non-CCC subsamples, respectively. The best fit power law above $0.2
r_{500}$ gives $t_{\rm c}(r) \propto r^{1.61\pm 0.01}$ for the whole
sample. For the 4 CCCs, the best fit power law above $0.2 r_{500}$
gives $t_{\rm c}(r) \propto r^{1.70\pm 0.01}$. For the non-CCCs, the
best fit power law above $0.2 r_{500}$ is more flat, $t_{\rm c}(r)
\propto r^{1.54\pm 0.01}$. The central cooling time is similar to the
work in Bauer et al. (2005) using the Chandra data for the same
clusters towards the most inner bin resolved. For the pilot
LoCuSS sample, the slope of the cooling time profile within
0.2$r_{500}$ becomes steeper with increasing significance of the cool
core. This was also observed for the nearby clusters based on Chandra
data in Sanderson et al. (2006) and Dunn \& Fabian (2006).

The cooling time is calculated towards the cluster center to the inner
most bin that the surface brightness profile is resolved.  There can
be large uncertainties in the cooling time at the inner most bins
where the temperature measurements are not resolved. We note that the
spatial resolution is important to calculate the proper cooling time
in the cluster center. When the cluster core is not well resolved, the
value calculated from the measured temperature and electron number
density gives the upper limit of the cooling time, specially for the
CCCs showing cusped surface brightness profiles.

\subsection{Scaled entropy profiles}
\label{s:en}

The entropy is the key to the understanding of the thermodynamical
history since it results from shock heating of the gas during cluster
formation. The observed entropy is generally defined as $S=T n_{\rm
e}^{-2/3}$ for clusters (e.g. Ponman et al. 1999). Radiative cooling
can raise the entropy of the ICM (e.g Pearce et al. 2000) or produce a
deficit below the scaling law (e.g. Lloyd-Davies et al. 2000). For all
12 galaxy clusters, the temperature data points are resolved below
$0.1r_{500}$. The entropy at $0.1r_{200}$, $S_{\rm 0.1 r_{200}}$, can
thus be used as an indicator of the central entropy.

\begin{figure*}
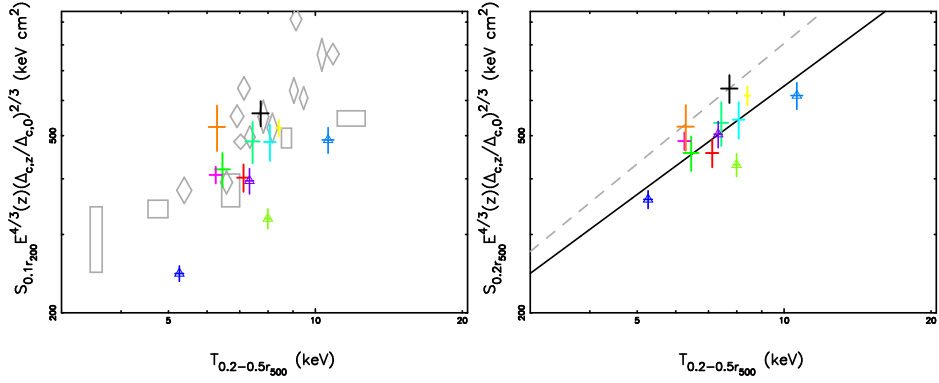

\begin{center}
\includegraphics[width=5cm,angle=270]{6567f6a.ps}
\includegraphics[width=5cm,angle=270]{6567f6b.ps}
\end{center}
\caption{Entropy at $0.1 r_{\rm 200}$ (left) and 
$0.2 r_{\rm 500}$ (right) vs. temperature for the pilot LoCuSS sample
(crosses) in which the CCCs are in triangles. Nearby clusters in
Ponman et al. (2003, boxes) and the REFLEX-DXL sample (diamonds) are
shown for comparison. The lines denote the best fits for the pilot
LoCuSS sample (solid) and the REFLEX-DXL sample in Zhang et al. (2006,
dashed). The colors have the same meaning as those in
Fig.~\ref{f:scalet}.
\label{f:cores}
}
\end{figure*}

According to the standard self-similar model the entropy scales as $S
\propto T$ (e.g. Ponman et al. 1999). We investigated the
entropy--temperature relation ($S$--$T$) using $S_{\rm 0.1 r_{200}}$
as the central entropy and $T_{0.2-0.5r_{500}}$ as the cluster
temperature. $S_{\rm 0.1 r_{200}}$ indicates the physics on core
scales and $T_{0.2-0.5r_{500}}$ on global scales. Therefore the CCCs
show significantly lower central entropies compared to the $S$--$T$
scaling law. This can be due to that lower temperature systems show
more pronounced cool cores corresponding to lower central
entropies. The entropy at $0.1 r_{200}$ can thus be used as a
mechanical educt of the non-gravitational process which introduces
large scatter of the $S_{\rm 0.1 r_{200}}$--$T$ relation (left panel
of Fig.~\ref{f:cores}).

\begin{figure*}
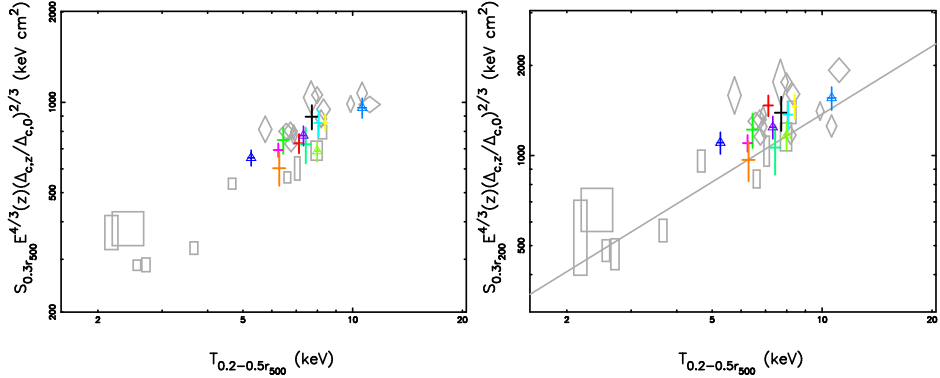

\begin{center}
\includegraphics[width=5cm,angle=270]{6567f7a.ps}
\includegraphics[width=5cm,angle=270]{6567f7b.ps}
\end{center}
\caption{Entropy at $0.3 r_{\rm 500}$ (left) and 
$0.3 r_{\rm 200}$ (right) vs. temperature for the pilot LoCuSS sample
(crosses), the REFLEX-DXL sample (diamonds) and the sample in Pratt et
al. (2006, boxes). The line denote the best fit in Pratt et al. (2006,
including only the $E(z)$ correction). The pilot LoCuSS CCCs are in
triangles. The colors have the same meaning as those in
Fig.~\ref{f:scalet}.
\label{f:cores2}
}
\end{figure*}

Ponman et al. (2003) suggested to scale the entropy as $S \propto
T^{0.65}$ based on the observations for nearby clusters, with which
the pilot LoCuSS sample agrees (see the left panel in
Fig.~\ref{f:cores}). $0.1 r_{200}$ is slightly smaller than $0.2
r_{500}$ for the LoCuSS sample. To avoid the extrapolation to
calculate $r_{200}$, we also used the entropies at $0.2 r_{500}$,
$S_{\rm 0.2 r_{500}}$. The scatter is reduced by almost a factor of 2
for the $S$--$T$ relation using $S_{\rm 0.2 r_{500}}$ instead of
$S_{\rm 0.1 r_{200}}$ because that the entropy profiles are more
self-similar beyond $0.2 r_{500}$ than in the cluster cores. We show
the $S$--$T$ relation using $S_{\rm 0.2 r_{500}}$ in the right panel
of Fig.~\ref{f:cores} and Table~\ref{t:mtx_lite}. Within the error,
the $S_{\rm 0.2 r_{500}}$--$T$ relations for the pilot LoCuSS sample
and the REFLEX-DXL sample cannot rule out the standard self-similar
model $S \propto T$ while also being consistent with Ponman et
al. (2003). The CCCs ($S_{\rm 0.2 r_{500}} \propto T^{0.71 \pm 0.21}$)
agree better with the empirically determined scaling (Ponman et
al. 2003) $S\propto T^{0.65}$. As shown in Fig.~\ref{f:cores2}, the
further away from the cluster center, the less scatter there is in the
$S$--$T$ relation (also see Pratt et al. 2006). There is no noticable
evolution in the $S$--$T$ relation (e.g. using $S_{\rm 0.3 r_{500}}$
and $S_{\rm 0.3 r_{200}}$) comparing the pilot LoCuSS sample to the
REFLEX-DXL sample ($z \sim 0.3$, Zhang et al. 2006) and the nearby
relaxed cluster sample in Pratt et al. (2006). With the redshift
correction, the combined fit of the pilot LoCuSS sample, the
REFLEX-DXL sample and the nearby relaxed cluster sample in Pratt et
al. (2006) gives $S_{\rm 0.3 r_{500}} \propto T^{0.93 \pm 0.06}$ and
$S_{\rm 0.3 r_{200}} \propto T^{0.84 \pm 0.08}$, respectively. Note
there is the segregation between the CCCs and non-CCCs in which the
CCCs show low normalization of the $S$--$T$ relation. With the
combined data, at the high mass end the clusters are un-biased to CCCs
and at the low mass end the clusters are biased to CCCs (relaxed
clusters in Pratt et al. 2006). The slope of the $S$--$T$ relation for
the combined data is thus steeper than the slope for the CCCs or
non-CCCs. This is less significant for the $S$--$T$ relation using the
entropies at larger radii, e.g. $S_{\rm 0.3 r_{200}}$.

\begin{figure*}
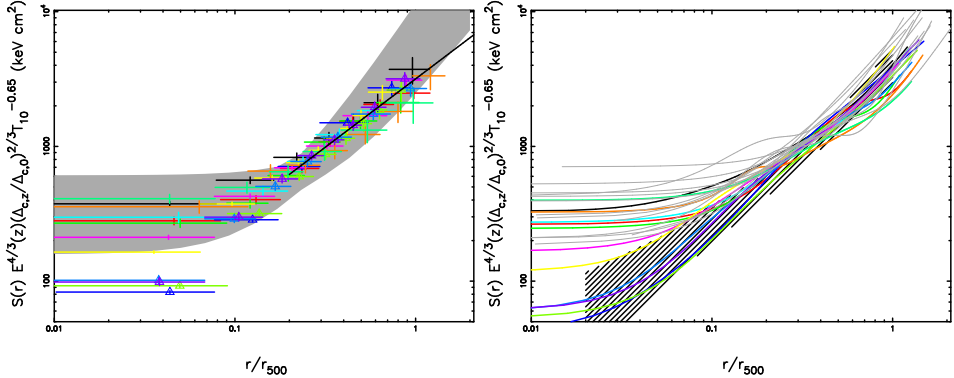

\begin{center}
\includegraphics[width=5cm,angle=270]{6567f8a.ps}
\includegraphics[width=5cm,angle=270]{6567f8b.ps}
\end{center}
\caption{{\it Left:} Scaled entropy profiles for the pilot LoCuSS
sample and the combined best fit. The shadow denotes the range of the
entropy profiles of the nearby clusters in the same temperature range
in Ponmen et al. (2003). {\it Right:} Scaled entropy profile fits for
the pilot LoCuSS sample compared to the REFLEX-DXL sample in Zhang et
al. (2006, grey, thin) and the sample in Pratt et al. (2006, black,
hatched). The CCCs are in triangles. $T_{10}$ denotes
$T_{0.2-0.5r_{500}}/10$~keV. The colors have the same meaning as those
in Fig.~\ref{f:scalet}. \label{f:en} }
\end{figure*}

We scaled the radial entropy profiles using the empirically scaling
(Ponman et al. 2003) $S\propto T^{0.65}$ and $r_{500}$. As shown in
Fig.~\ref{f:en}, the scaled entropy profiles for the pilot LoCuSS
sample are self-similar above $0.2 r_{500}$ and show the least scatter
around 0.2--0.3$r_{500}$. However, we found the redshift correction is
required to obtain an agreement on global radial scales for the pilot
LoCuSS sample, Birmingham-CfA sample (Ponman et al. 2003), REFLEX-DXL
sample (Zhang et al. 2006) and the cluster sample in Pratt et
al. (2006).

After the redshift correction the combined entropy profiles of the
pilot LoCuSS sample give the best fit $S(r)\propto r^{1.01\pm 0.04}$
above $0.2r_{\rm 500}$. A similar power law was found as $S\propto
r^{0.97}$ by Ettori et al. (2002), $S\propto r^{0.95}$ by Piffaretti
et al. (2005), and $S(r)\propto r^{1.00\pm 0.07}$ for the REFLEX-DXL
sample ($>0.2 r_{500}$). The spherical accretion shock model predicts
$S\propto r^{1.1}$ (e.g. Tozzi \& Norman 2001; Kay 2004). The combined
fit of the entropy profiles for the 4 CCCs in the pilot LoCuSS sample
gives $S\propto r^{1.10\pm 0.05}$, similar to the prediction of the
spherical accretion shock model and consistent with the relaxed nearby
clusters in Pratt et al. (2006) giving $S\propto r^{1.08}$. Since the
clusters appear more ``relaxed'' in Pratt et al (2006) the entropy
profiles of their sample are very consistent with the CCCs in the
pilot LoCuSS sample. The combined fit of the entropy profiles for
the 8 non-CCCs in the pilot LoCuSS sample gives $S\propto r^{0.97\pm
0.05}$.

\begin{figure*}
\begin{center}
\includegraphics[angle=0,width=6.5cm]{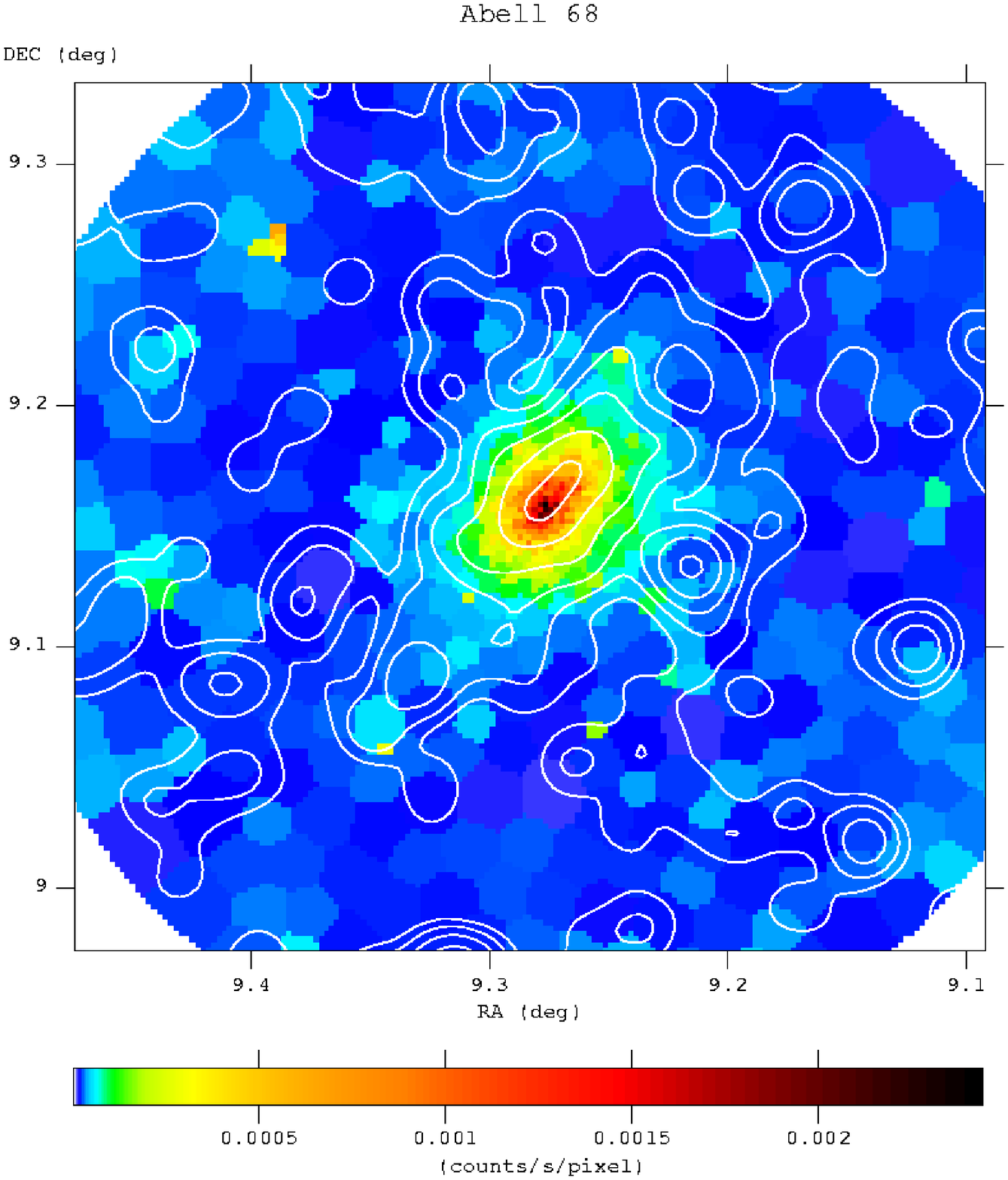}
\includegraphics[angle=0,width=6.5cm]{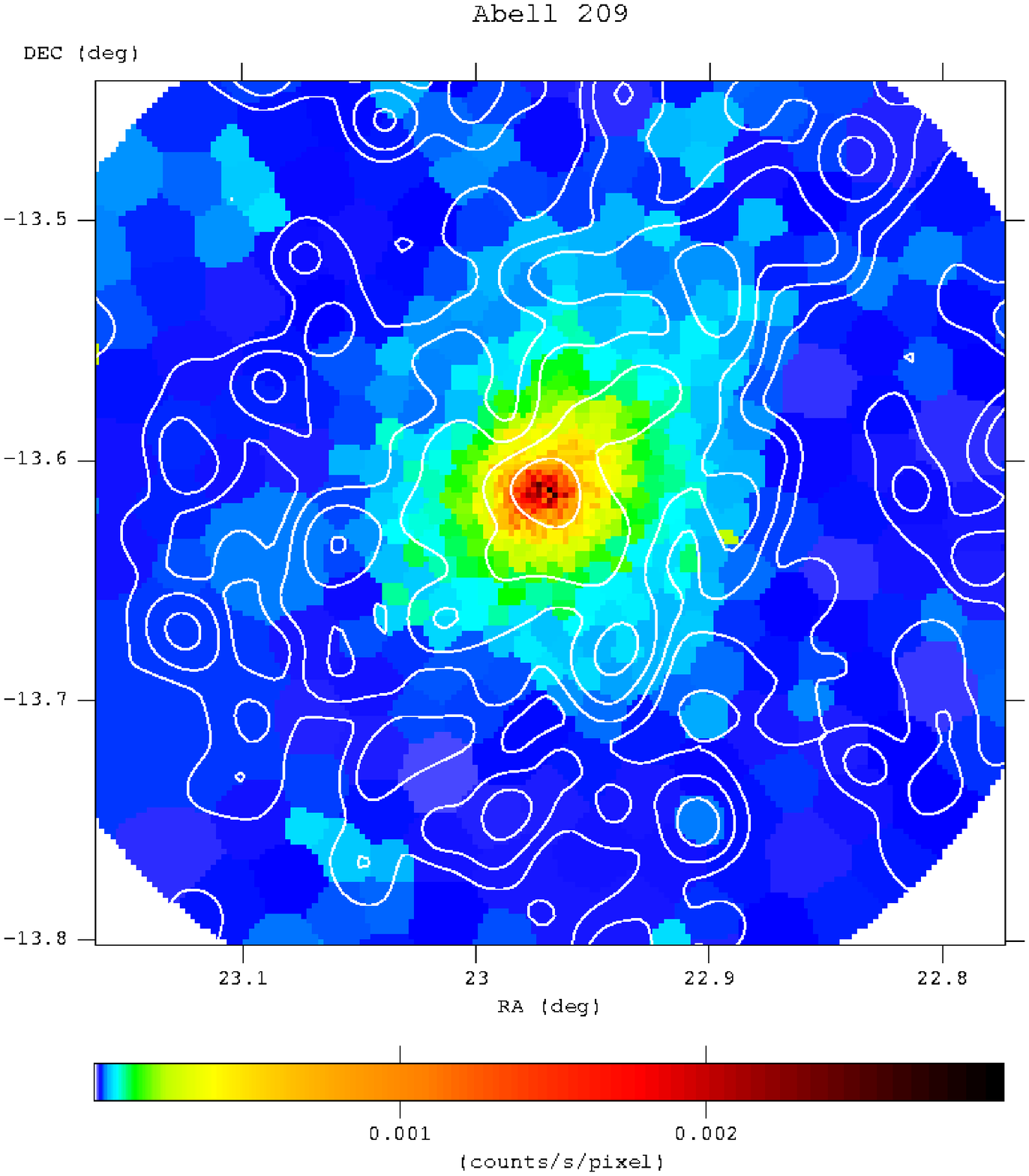}

\includegraphics[angle=0,width=6.5cm]{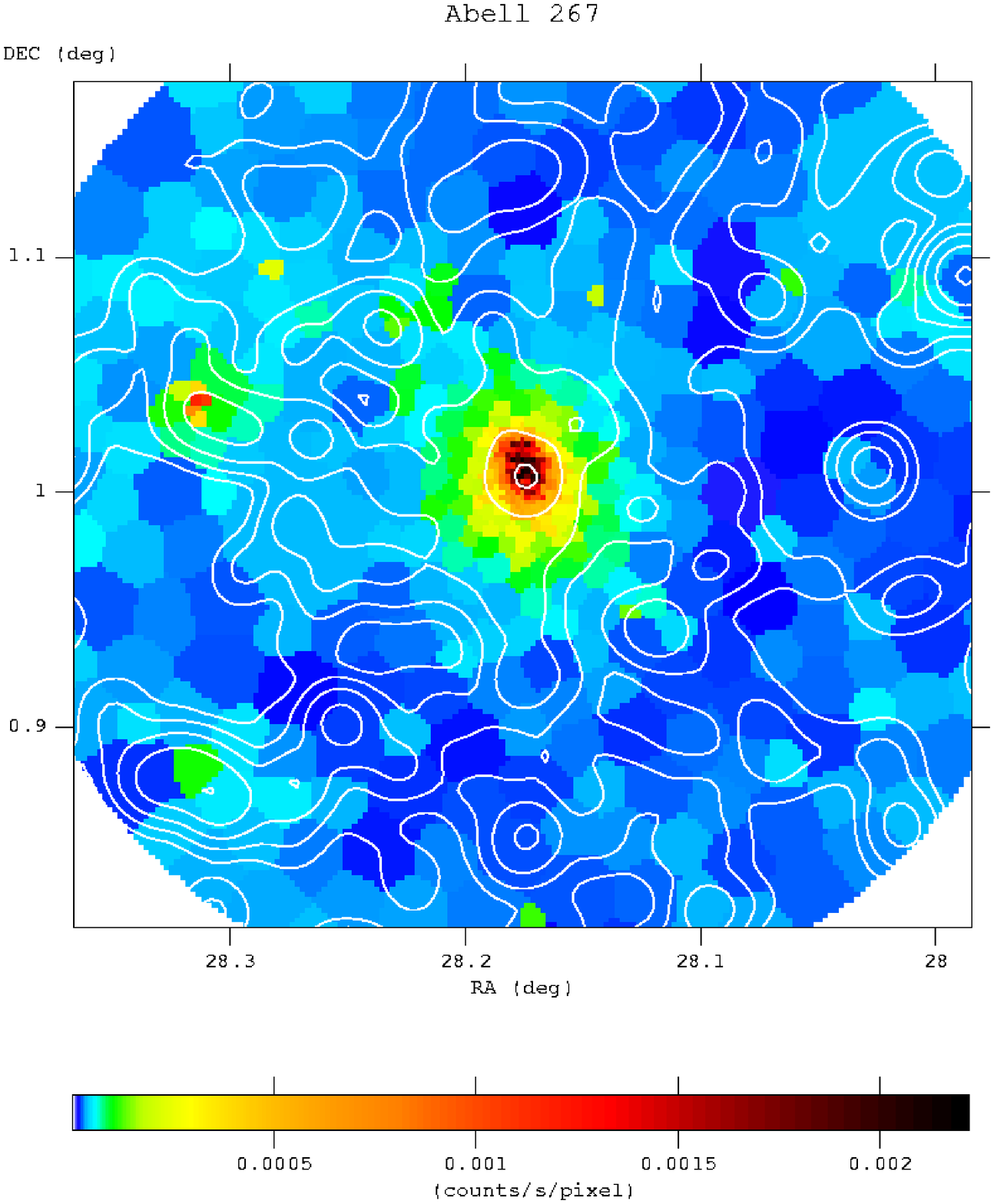}
\includegraphics[angle=0,width=6.5cm]{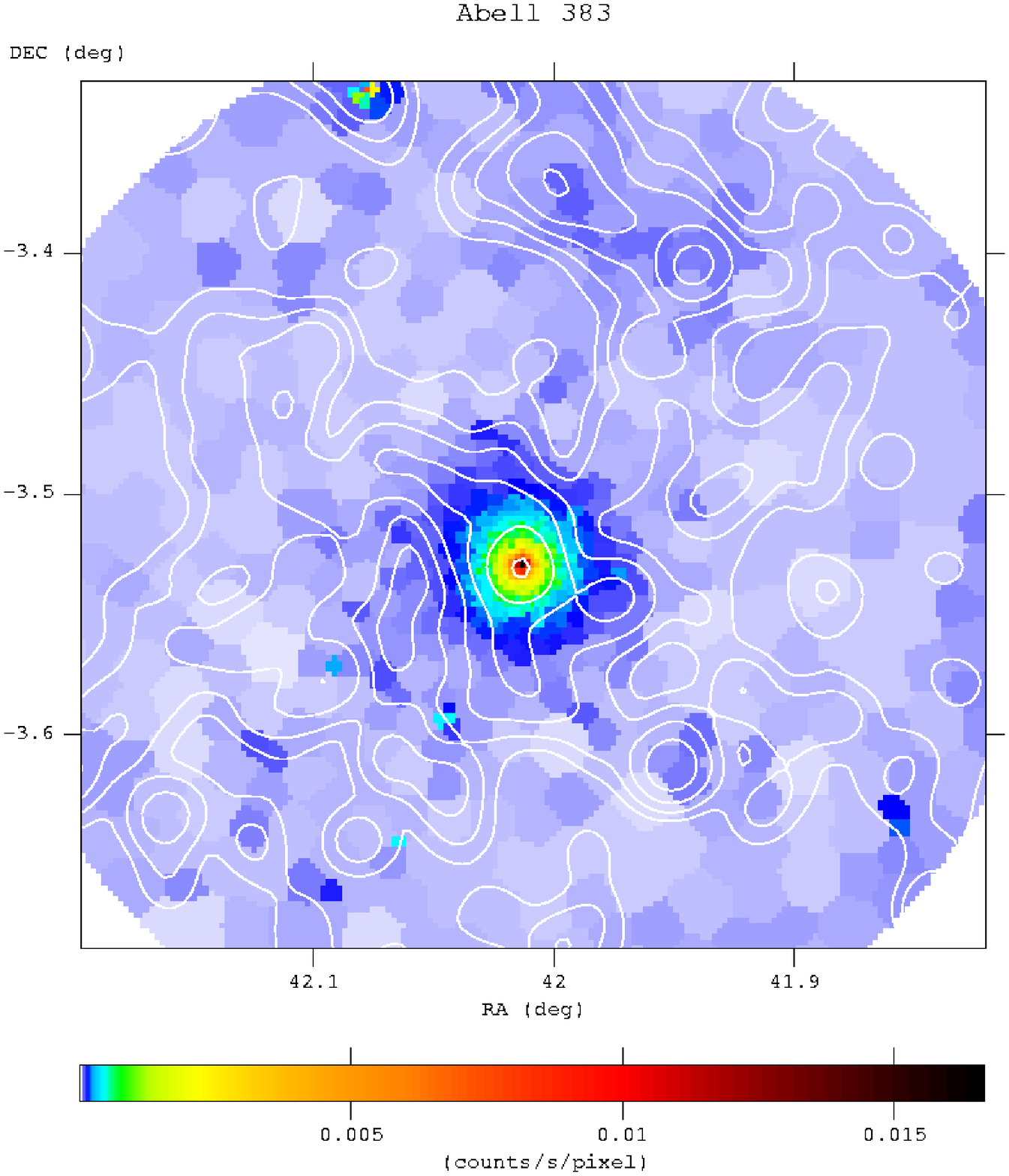}
\end{center}
\caption{Flat fielded XMM-Newton MOS1 flux images in the 0.7--2~keV
band superposed with the CFH12k optical luminosity weighted galaxy
density contours.
\label{f:imga} }
\end{figure*}

\begin{figure*}
\begin{center}
\includegraphics[angle=0,width=6.5cm]{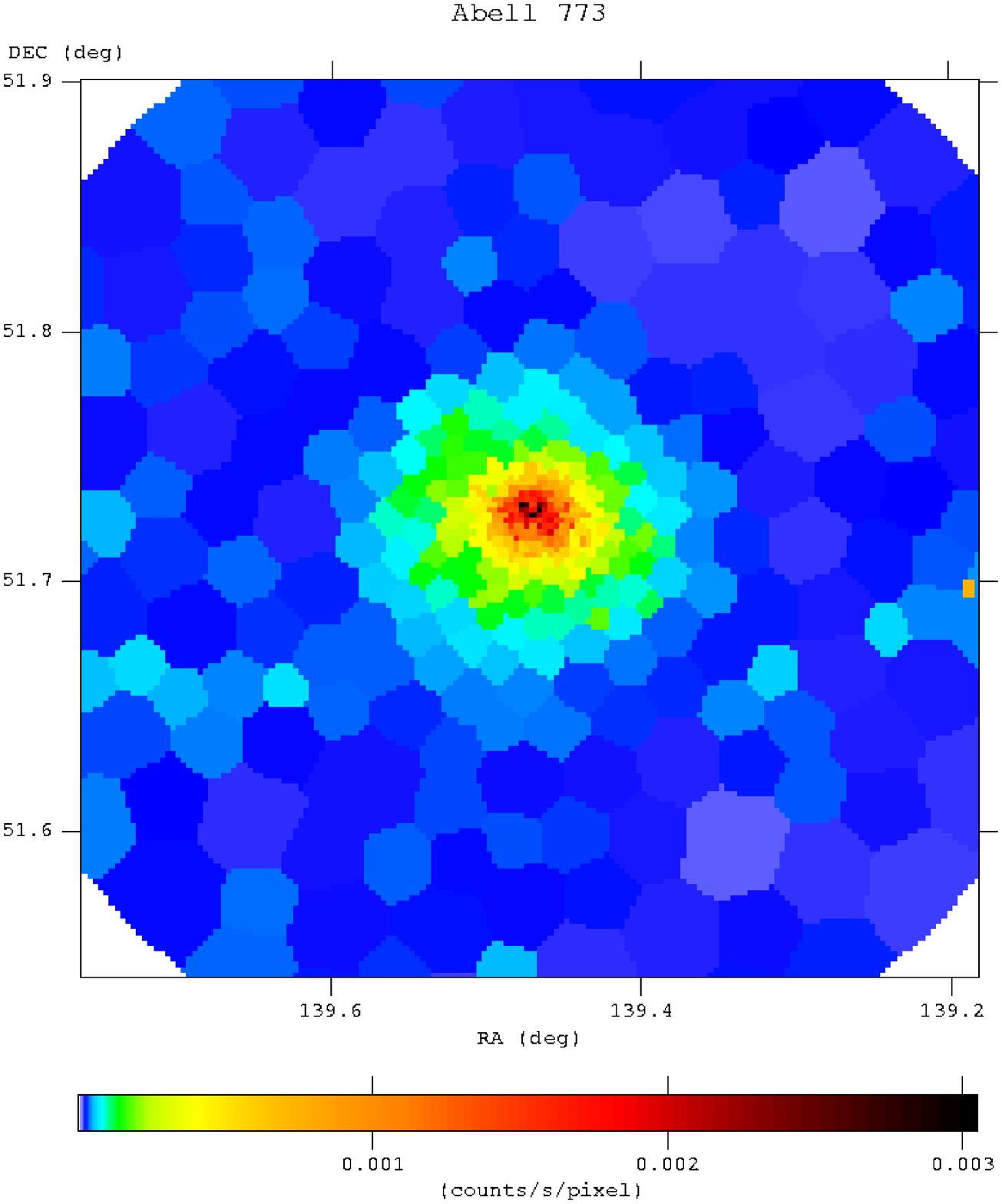}
\includegraphics[angle=0,width=6.5cm]{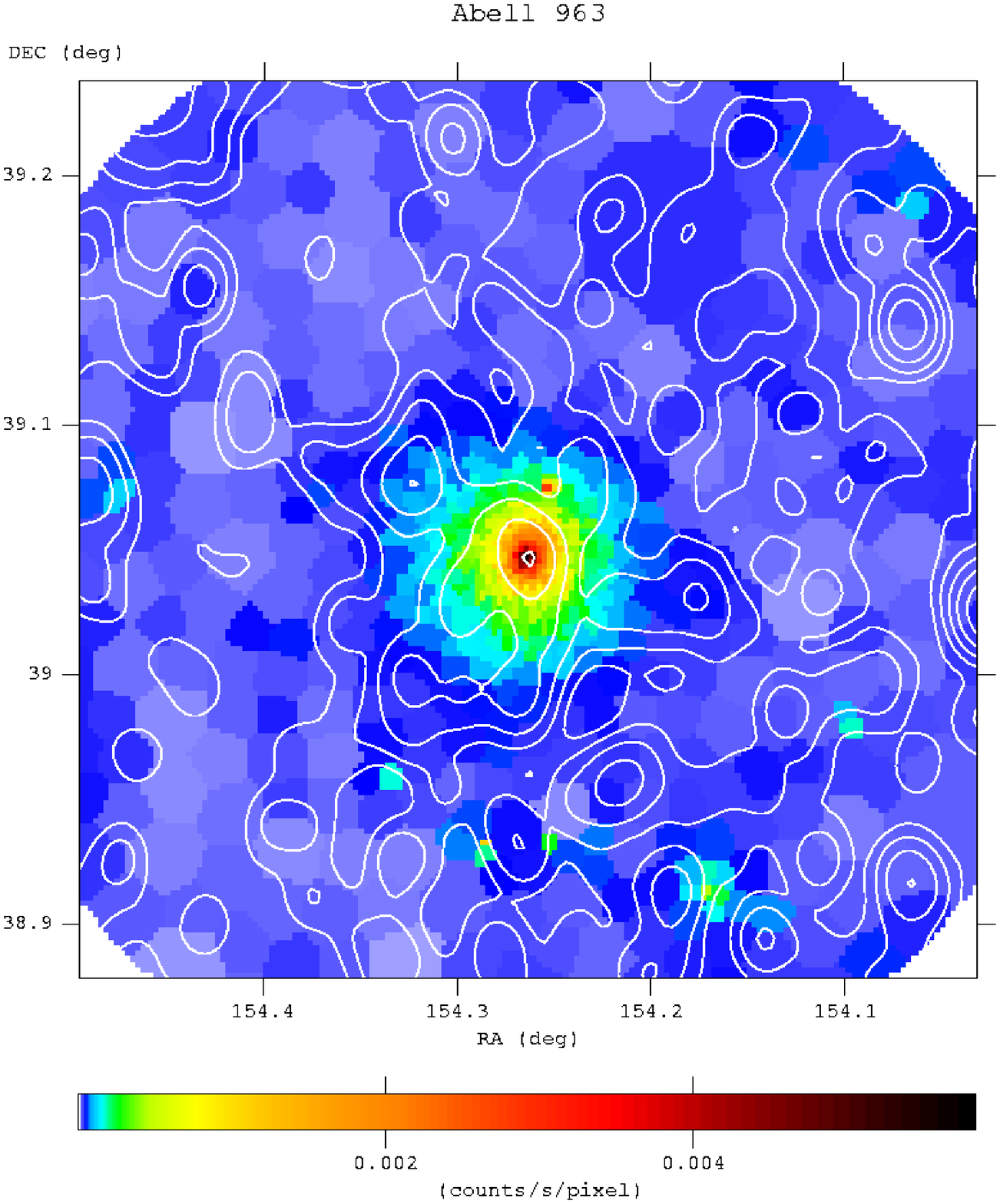}

\includegraphics[angle=0,width=6.5cm]{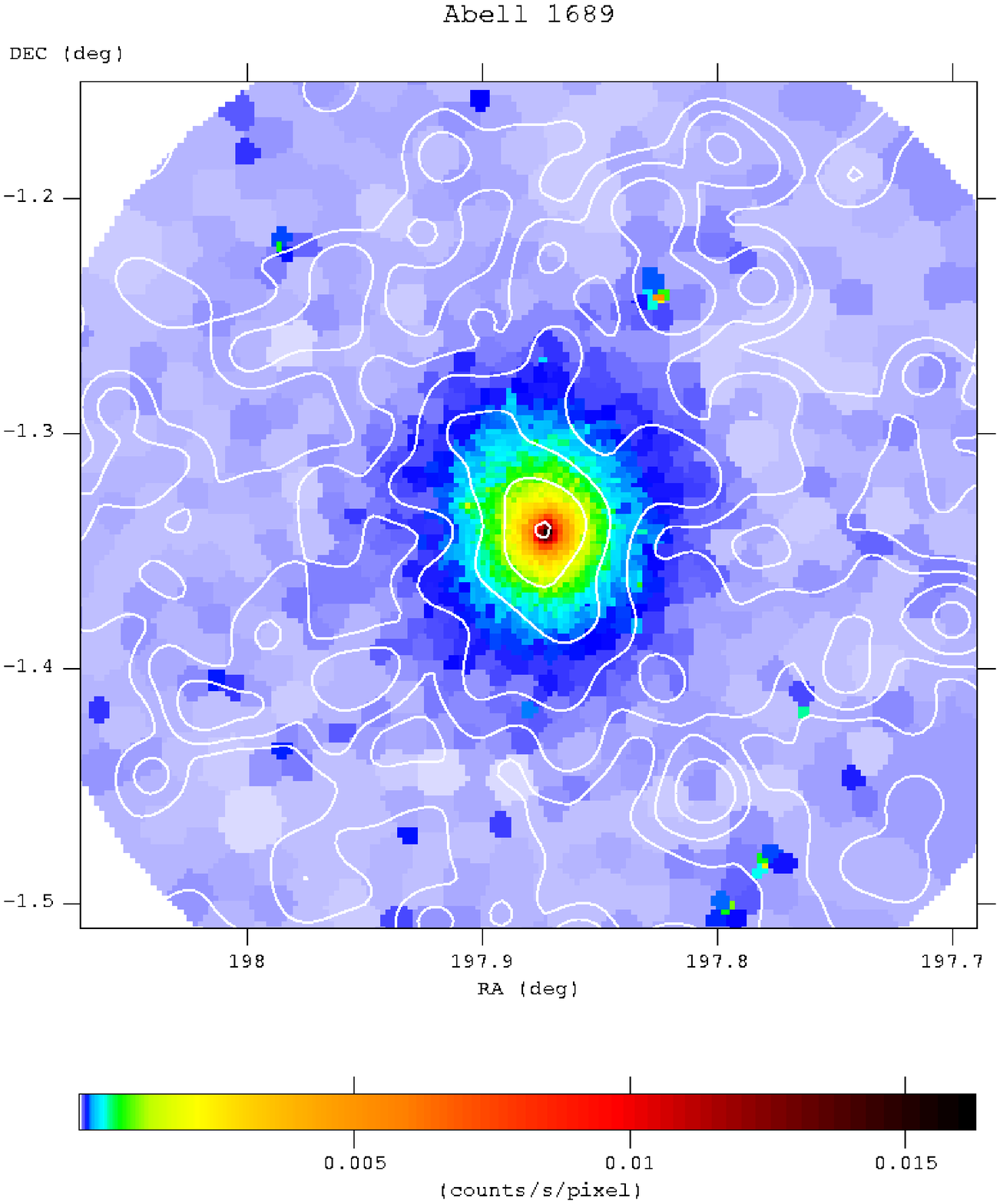}
\includegraphics[angle=0,width=6.5cm]{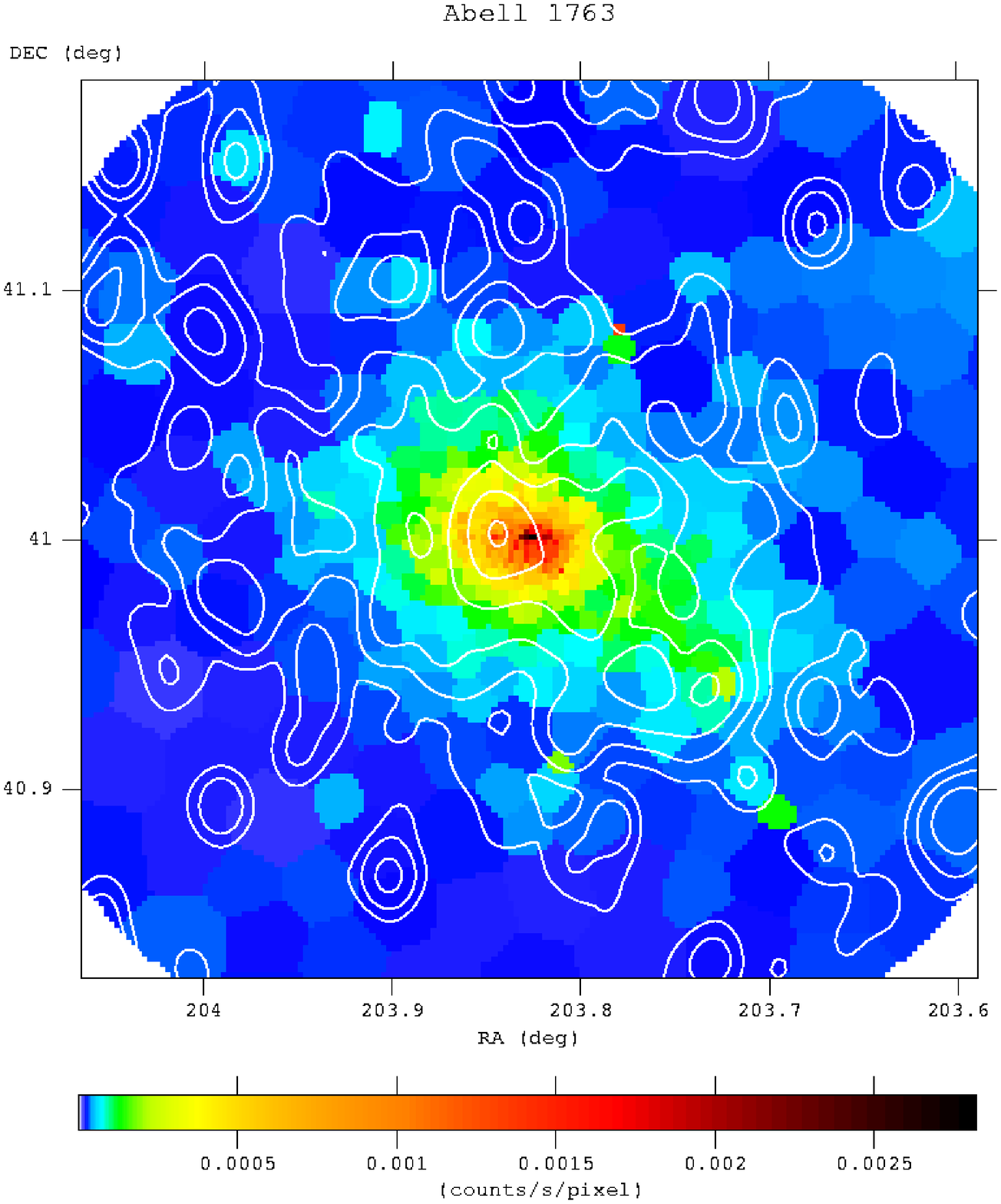}
\end{center}

\caption{See the caption in Fig.~\ref{f:imga}. Abell~773 has no CFH12k data.
\label{f:imgb} }
\end{figure*}

\begin{figure*}
\begin{center}
\includegraphics[angle=0,width=6.5cm]{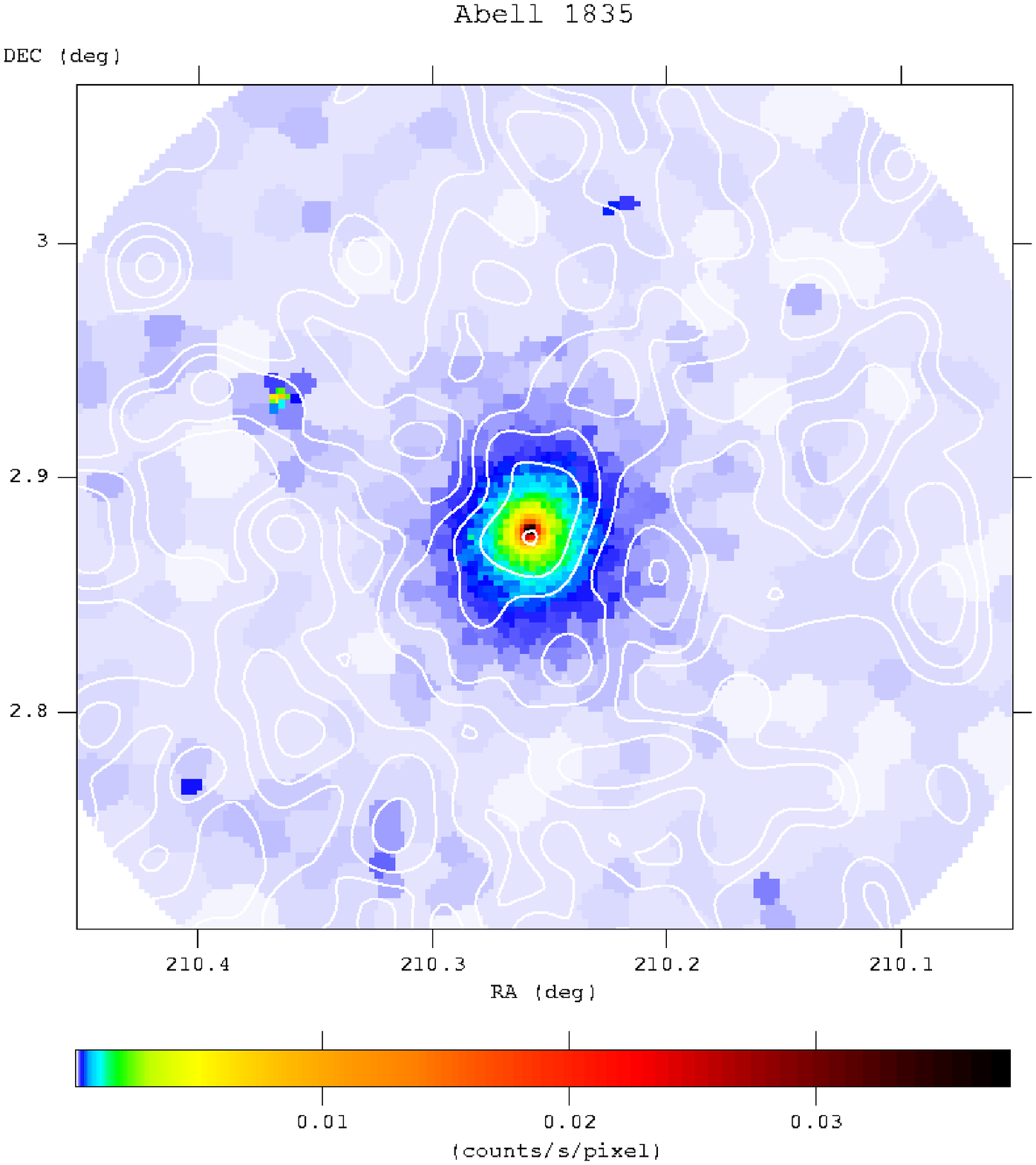}
\includegraphics[angle=0,width=6.5cm]{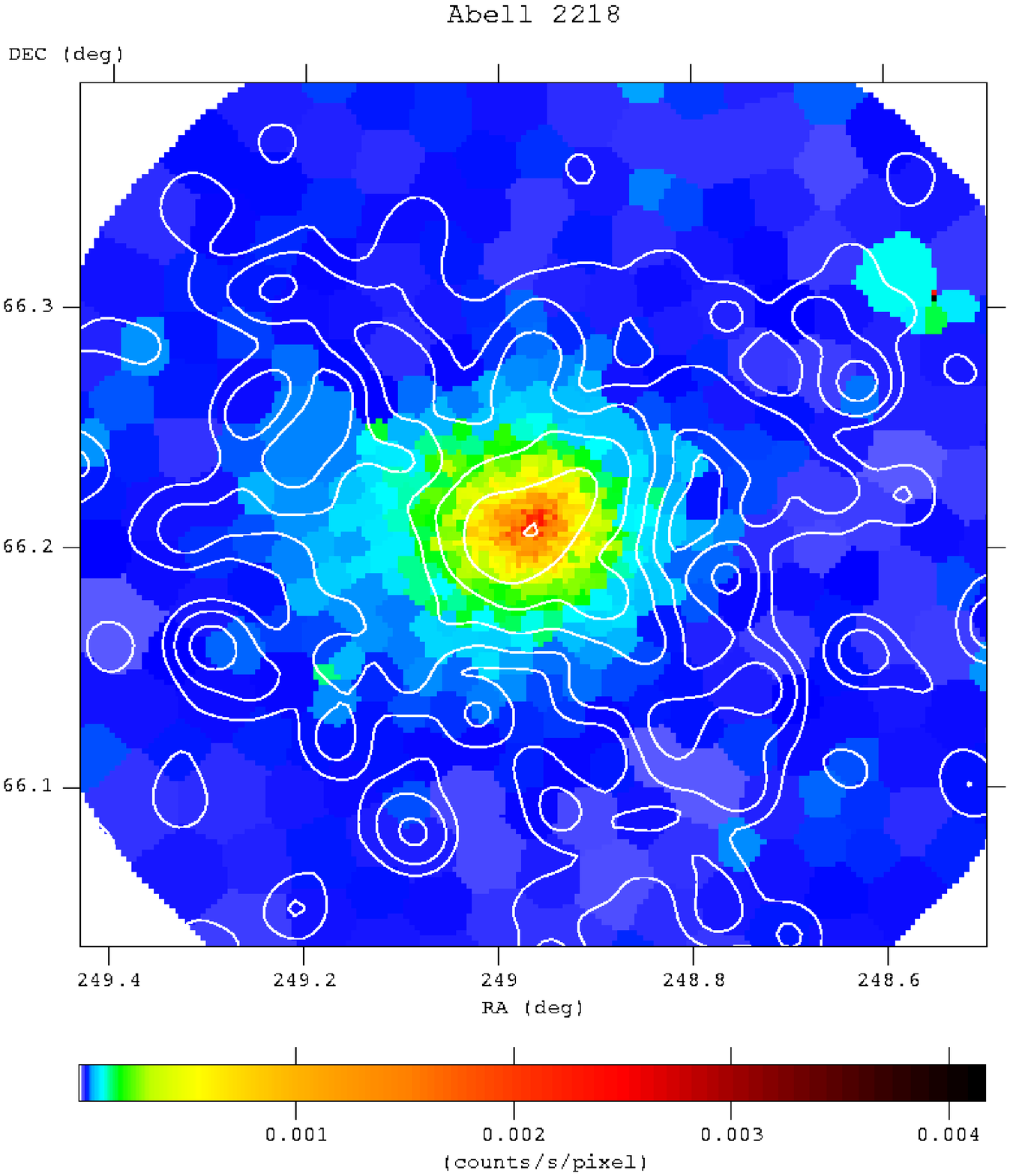}

\includegraphics[angle=0,width=6.5cm]{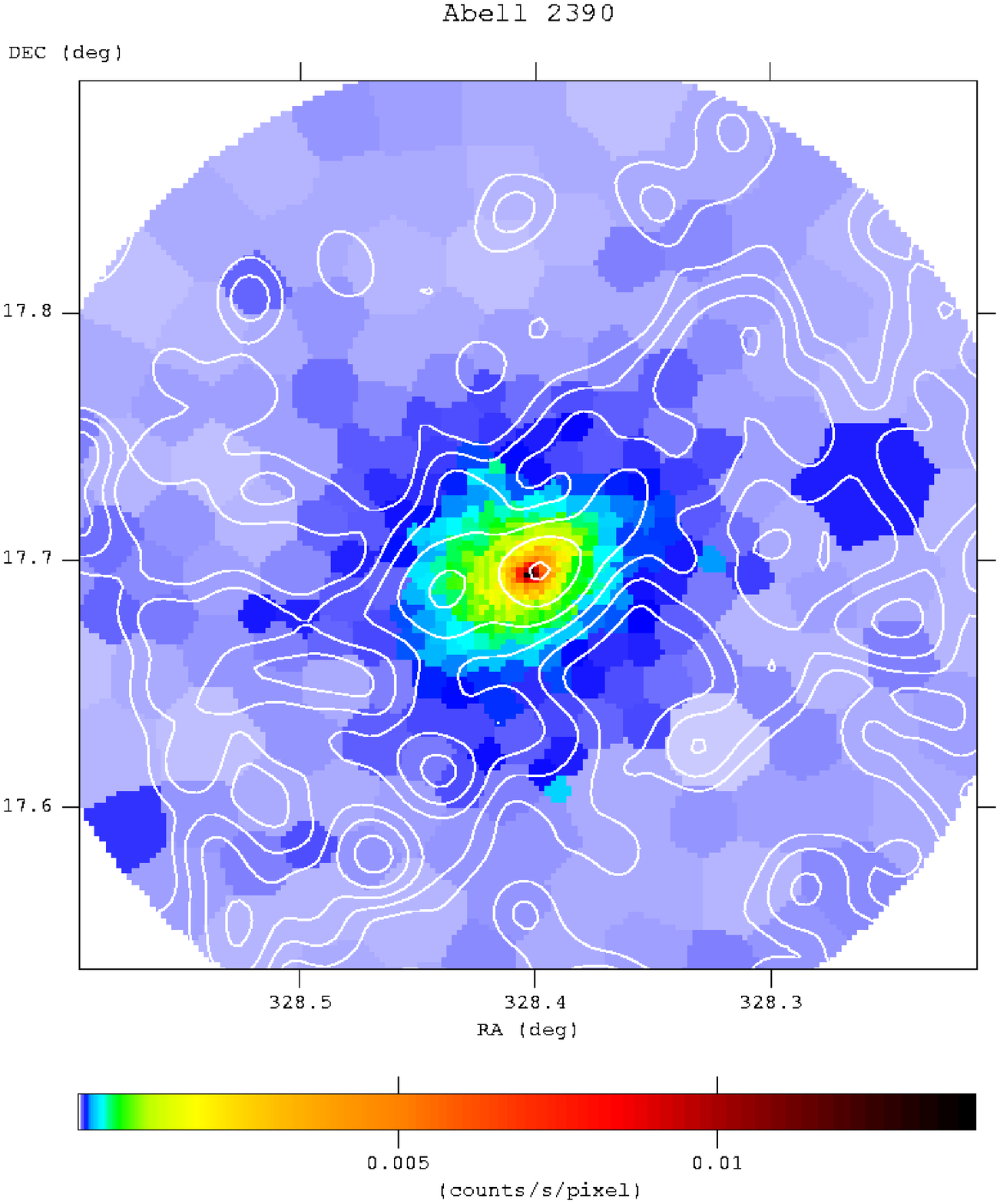}
\includegraphics[angle=0,width=6.5cm]{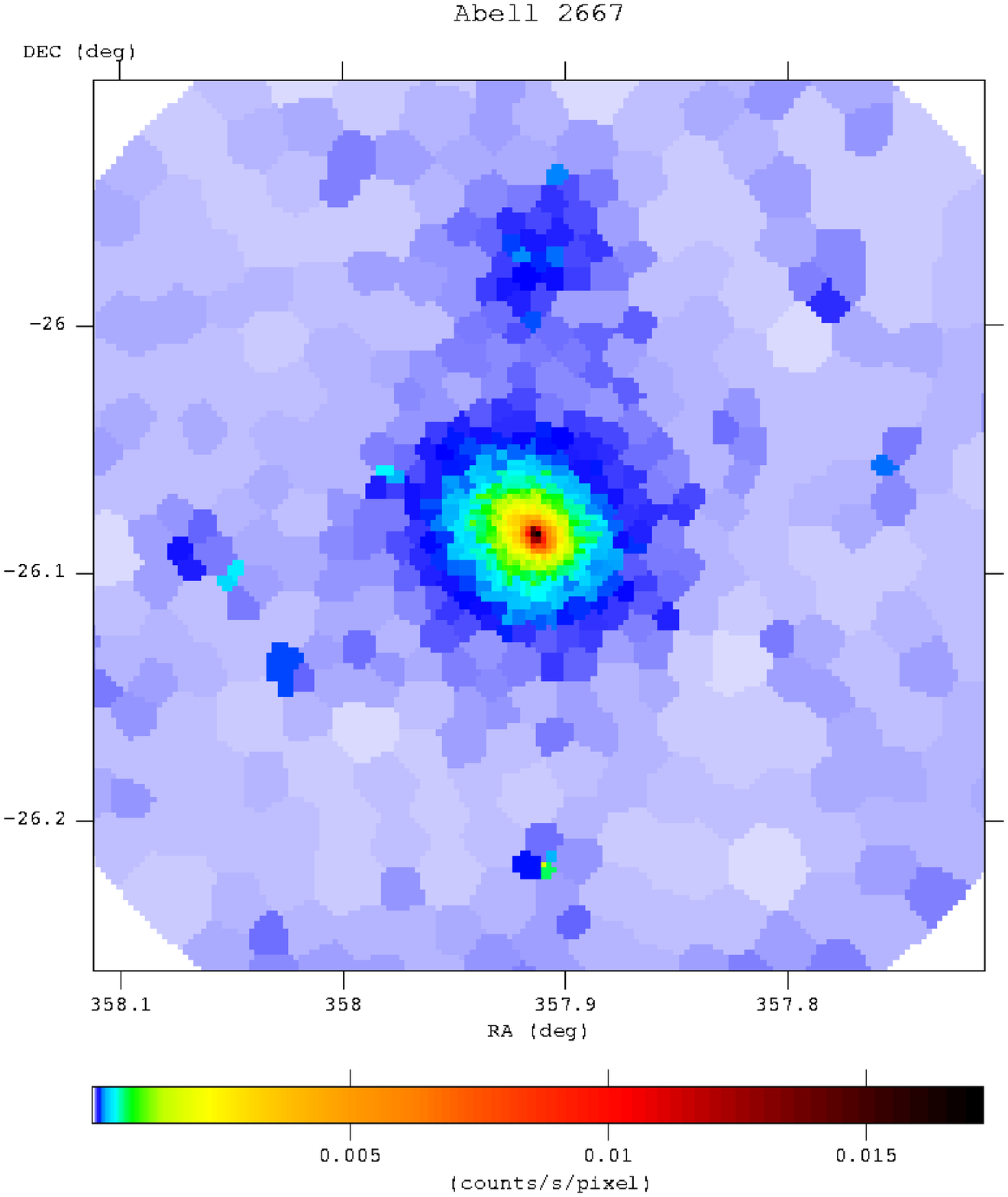}
\end{center}
\caption{See the caption in Fig.~\ref{f:imga}. Abell~2667 has no CFH12k data.
\label{f:imgc} }
\end{figure*}

The different slopes of the entropy profiles for the CCCs and non-CCCs
may indicate the different phases or different origins of the clusters
with respect to the cluster morphologies. Observations suggest an
evolution of the cluster morphology such that mergers happen more
frequently in high redshift galaxy clusters (e.g. Jeltema et al. 2005;
Hashimoto et al. 2006). The non-CCCs are dynamically young and thus
show the evidence of the recent mergers judging from their X-ray flux
images such as elongation and extended low surface brightness features
(e.g. Abell~1763 in the pilot LoCuSS sample). Their entropy profiles
can be flattened by mergers and thus become less steep than the
prediction of the spherical accretion shock model. There is an
evidence of AGN activities in some clusters (e.g. Abell~1763,
Hardcastle \& Sakelliou 2004), in which the central AGN could also
affect the entropy profile. On the contrary, the CCCs appear
``relaxed'', and their entropy profiles agree with the prediction of
the spherical accretion shock model. On the other hand, the different
slopes of the entropy profiles of the CCCs and non-CCCs can also be
due to geometry. The non-CCCs appear more elongated due to the recent
mergers while the CCCs more symmetric as shown in the XMM-Newton flux
images (Fig.~\ref{f:imga}, {\it also see
Figs.~\ref{f:imgb}--\ref{f:imgc} in the electronic edition of the
Journal for the whole sample}). The flattening of the entropy profiles
of the non-CCCs can also be a geometric effect due to the azimuthal
average.

\subsection{Scaled total mass and gas mass profiles}

\begin{figure*}
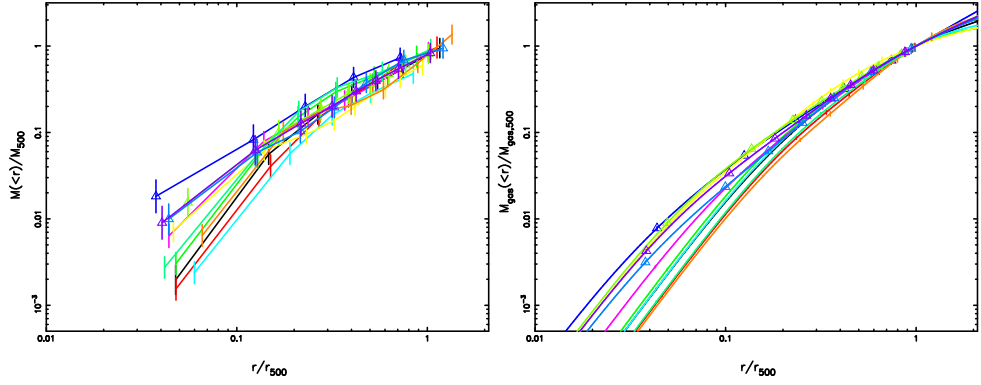

\begin{center}
\includegraphics[width=5cm,angle=270]{6567f9a.ps}
\includegraphics[width=5cm,angle=270]{6567f9b.ps}
\end{center}
\caption{Scaled total mass profiles (left) and gas mass profiles
(right). The CCCs are in triangles. The colors have the same meaning
as those in Fig.~\ref{f:scalet}. The error bars (a few per cent) of
the gas mass profiles are too small to be seen in this
plot. \label{f:scaleym} }
\end{figure*}

The mass profiles were scaled with respect to $M_{\rm 500}$ and
$r_{\rm 500}$, respectively (Fig.~\ref{f:scaleym}). We found the least
scatter at radii above $0.2 r_{\rm 500}$. In the inner regions ($<0.2
r_{\rm 500}$), the mass profiles vary significantly with the cluster
central dynamics (Fig.~\ref{f:scaleym}). As found in the strong
lensing studies based on high quality HST data (Smith et al. 2005),
the cluster cores can be very complicated with multi-clump DM
halos. The peculiarities in the cluster cores for the individual
clusters may explain the large scatter of the mass profiles on core
scales where the ICM does not always trace DM. This scatter may also
suggest the different phases of the clusters. 

Similar to the mass profiles, the scaled gas mass profiles appear
self-similar at radii above $0.2 r_{\rm 500}$ but show less scatter (a
few per cent) than for the scaled mass profiles.

\section{X-ray scaling relations}
\label{s:scale}

To use the mass function of the cluster sample to constrain
cosmological parameters, it is important to calibrate the scaling
relations between the X-ray luminosity, temperature and gravitational
mass, the fundamental cluster properties including also velocity
dispersion. Massive clusters, selected in a narrow redshift range are
important to constrain the normalization and to understand the scatter
in the scaling relations. Both REFLEX-DXL and pilot LoCuSS samples are
such flux-limited, morphology unbiased samples. Comparing the X-ray
scaling relations of such samples to samples in other narrow
redshift bins can constrain the evolution of the X-ray scaling
relations.

The scaling relations are generally parameterized by a power law
($Y=Y_{0} X^{\gamma}$). Note the pilot LoCuSS clusters are in a narrow
temperature range, which has the disadvantage to constrain the slope
parameter. To study the normalization segregation of the scaling
relations for the CCCs and non-CCCs, we fit the normalization with
fixed slope parameters as often used in previous studies for the pilot
LoCuSS sample and the pilot LoCuSS non-CCCs. For each relation, we
also performed the fitting with both normalization and slope free.
The fitting slopes generally appear consistent with the fixed slopes
within the errors. The scatter describes the dispersion between the
observational data and the best fit. We list the best fit power law
and the scatter in Table~\ref{t:mtx_lite}. For the pilot LoCuSS sample
the cluster masses are uniformly determined from high quality
XMM-Newton data. This guarantees the minimum systematic error due to
the analysis method.

\subsection{Cool cores and the X-ray scaling relations}
\label{s:lxcoolcore}

In the cluster cores we observed broad scatter of the X-ray properties
with respect to the cluster morphologies (e.g. CCCs and non-CCCs) for
the scaled profiles of the X-ray quantities (e.g. temperature,
metallicity, surface brightness, cooling time, entropy, total mass and
gas mass). This is most probably due to the effects of different
physical processes rather than statistical fluctuations in the
measurements (Zhang et al. 2004b, 2005b; Finoguenov et al. 2005). The
CCCs and non-CCCs can be the clusters at their different phases.

The X-ray luminosity integrated over the whole cluster region
introduces significant normalization segregation between CCCs and
non-CCCs, and large scatter of the X-ray scaling relations dominated
by non-CCCs. It is thus worth to take a closer look how the cool cores
affect the scaling relations and their scatter as follows.

For $S_{0.1 r_{200}}$--$T$ relation, including the CCCs in the sample
introduces not only a steeper slope but also a lower normalization
($\sim 30$\%).

Using the X-ray luminosity including the cool core ($L^{\rm incc}$),
the normalization of the $L$--$T$ and $L$--$M$ relations excluding the
4 CCCs is reduced by 40\%. If the temperature is measured also
including the cool core, the normalization segregation will be even
larger (by a few per cent) for the $L$--$T$ relation. Using the
luminosity corrected for the $< 0.2 r_{500}$ region and temperature
excluding the $< 0.2 r_{500}$ region, the normalization of the scaling
relations agrees to better than 10\% for the CCCs and non-CCCs. In
such a way the scaling relations among X-ray luminosity, cluster mass
and temperature, and their scatter are insensitive to the exclusion of
CCCs for the pilot LoCuSS sample. The above results show the following
picture. CCCs usually appear more ``relaxed'', and are thus considered
to provide reliable X-ray mass measurements (e.g. Pierpaoli et
al. 2001; Allen et al. 2004). At the same time, the properties of the
cool cores exhibit the largest scatter. Therefore, the CCCs were
usually selected and the cool cores were often excised or corrected to
study the X-ray scaling relations.

\begin{figure*}
\begin{center}
\includegraphics[width=5cm,angle=270]{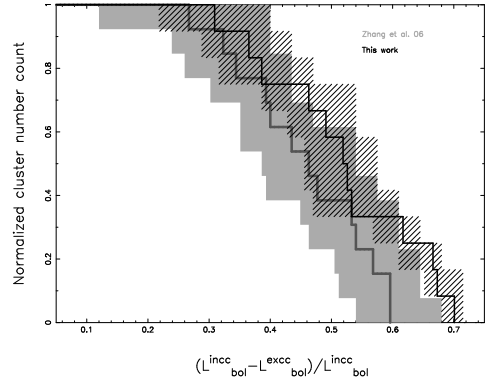}
\end{center}
\caption{Normalized cumulative cluster number count as a function of
the fraction of the total bolometric luminosity attributed by the
$<0.2 r_{500}$ region for the pilot LoCuSS sample at $z \sim 0.2$
compared to the REFLEX-DXL sample at $z \sim 0.3$ in Zhang et
al. (2006).
\label{f:fraclcc}}
\end{figure*}

With the bolometric X-ray luminosity including and excluding the $<
0.2 r_{500}$ region (Table~\ref{t:catalog2}), we show the normalized
cumulative cluster number count as a function of the fraction of the
luminosity attributed by the cluster core ($< 0.2 r_{500}$) in
Fig.~\ref{f:fraclcc}. Up to 70\% of the bolometric X-ray luminosity is
contributed by the cluster core ($< 0.2 r_{500}$). The evolution of
the cool cores is not significant as also explained using the
temperature drop towards the cluster center in Sect.~\ref{s:kt}.

As mentioned above the global temperature was calculated excluding the
$<0.2 r_{500}$ region and the X-ray luminosity was corrected for the
$<0.2 r_{500}$ region, respectively. In such a way the scatter in the
X-ray scaling relations is minimized ($<20$\%) and the normalization
of the X-ray scaling relations is insensitive to the exclusion of the
CCCs ($<10$\%).

\subsection{Mass--temperature relation}

\begin{figure*}
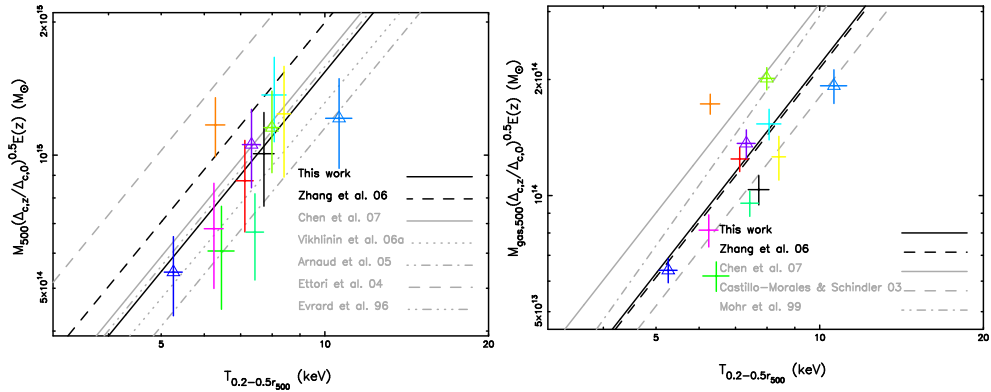

\begin{center}
\includegraphics[angle=270,width=6.5cm]{6567f11a.ps}
\includegraphics[angle=270,width=6.5cm]{6567f11b.ps}
\end{center}
\caption{Mass temperature relation (left) and gas mass temperature
relation (right). The CCCs are in triangles. The colors have the same
meaning as those in Fig.~\ref{f:scalet}. The best fit power law of the
pilot LoCuSS sample and REFLEX-DXL sample used the fixed slope of 1.5 for
the $M$--$T$ relation and 1.8 for the $M_{\rm gas}$--$T$ relation. 
\label{f:mt}}
\end{figure*}

Given the virial theorem ($T \propto M_{500}/r_{500}$) and the
spherical collapse model ($M_{500} \propto r_{500}^3$) one obtains
$M_{500} \propto T^{1.5}$. This scaling was also observed for nearby
clusters (e.g. Finoguenov et al. 2001; Reiprich \& B\"ohringer 2002;
Arnaud et al. 2005). To compare the pilot LoCuSS sample to the
existing nearby and more distant samples, we present the
$M_{500}$--$T$ relation fixing the slope parameter to the often used
value (1.5) in Fig.~\ref{f:mt}. The scatter in the cluster mass in the
$M_{500}$--$T$ relation is within 20\% for the pilot LoCuSS sample.

Evrard et al. (1996) simulated ROSAT observations of 58 nearby
clusters ($z \sim 0.04$, 1--10~keV), for which the normalization of
the $M$--$T$ relation agrees with the pilot LoCuSS sample. Chen et al
(2007) investigated ROSAT and ASCA observations of a flux-limited,
morphology-unbiased sample of 106 nearby clusters ($z<0.15$,
1--15~keV, extended HIFLUGCS). The scaling relation of the massive
non-CCCs ($T>3$~keV) in their sample shows an excellent agreement with
the pilot LoCuSS sample. This also indicates that the cool core
correction for the pilot LoCuSS sample introduces less bias into the
normalization towards the CCCs. Note the early work related to the
HIFLUGCS sample (Finoguenov et al. 2001; Popesso et al. 2005) gives
similar scaling relations as in Chen et al. (2007) for the whole
HIFLUGCS sample. Vikhlinin et al (2006a) derived the $M$--$T$ relation
for 13 low-redshift clusters ($z<0.23$, 0.7--9~keV) using Chandra
observations, which normalization agrees with ours within the scatter
(20\%) for the pilot LoCuSS sample. The scaling in Arnaud et
al. (2005) is based on XMM-Newton observations of 6 relaxed nearby
clusters ($z<0.15$, 2--9~keV, $T>3.5$~keV). The temperatures used in
their work were determined in the 0.1--0.5$r_{200}$ region, which also
excludes the cluster cores as we did using the 0.2--0.5$r_{500}$
region. The temperature profiles in Arnaud et al. (2005) are almost
flat in the $0.5 r_{500}$--$0.5 r_{200}$ regions and their values are
close to the maximum temperatures. The measured global temperature is
thus higher when the $0.5 r_{500}$--$0.5 r_{200}$ region is used as in
Arnaud et al. (2005) than the global temperature measured excluding
the $0.5 r_{500}$--$0.5 r_{200}$. We, also in Zhang et al. (2006),
measured the global temperatures using the region only up to
$0.5r_{500}$ determined by the photon statistic. Therefore the
normalization of the $M$--$T$ scaling relations in this work and Zhang
et al. (2006) is slightly higher than in Arnaud et al. (2005).

The $M$--$T$ relation in Ettori et al. (2004) is based on Chandra
observations of 28 high redshift clusters ($0.4<z<1.3$, 3--11~keV). It
shows a higher normalization compared to the above published relations
we mentioned. This could be due to the different way to define the
density contrast used in the redshift evolution correction and the
method to calculate the global temperatures in Ettori et
al. (2004). We used the redshift dependent density contrast (Bryan \&
Norman 1998) in the redshift evolution correction. We note that in
this work the cool cores are excluded in calculating the global
temperatures. In Ettori et al. (2004) the central regions including
cluster cool cores were used to measure the global temperatures, which
tends to give lower temperatures for given cluster masses for the
clusters showing cool cores. Therefore the normalization of the
$M$--$T$ relation in Ettori et al. (2004) can be higher. The scaling
relation in Zhang et al.  (2006) is based a flux-limited
morphology-unbiased sample of 13 medium distant clusters
($0.26<z<0.31$, $T>5$~keV, REFLEX-DXL) observed by
XMM-Newton. Compared to the sample in this work the REFLEX-DXL sample
shows a slightly higher normalization but well within the mass scatter
($\sim 20$\%) for the pilot LoCuSS sample.

No evident evolution is found for the $M$--$T$ relation comparing the
pilot LoCuSS sample to the nearby and more distant samples within the
scatter.

\subsection{Gas mass--temperature relation}

Assuming the gravitational effect dominates in galaxy clusters,
the gas should follow the collapse of the DM giving $M_{\rm gas,500}
\propto M_{500}$ and thus $M_{\rm gas,500} \propto T^{1.5}$
(e.g. Arnaud 2005). However, the non-gravitational effects become
important for the ICM such that the shape of the ICM density profile
depends on the cluster temperature (e.g. Ponman et al. 1999) and the
observed slope of the $M_{\rm gas,500}$--$T$ relation becomes steep
($\sim 1.8$, e.g. Mohr et al. 1999; Vikhlinin et al. 1999). The slope
of the gas mass--temperature relation for the pilot LoCuSS sample is
consistent with the previously observed value 1.8 within the
error. Therefore, we present the $M_{\rm gas,500}$--$T$ relation for
the pilot LoCuSS sample with the fixed slope of 1.8 to compare with
the recently published results in Fig.~\ref{f:mt}. The scatter of the
gas mass for the $M_{\rm gas,500}$--$T$ relation is within 15\% for
the pilot LoCuSS sample. Note the weighted (unweighted) fitting slope
of the $M_{\rm gas,500}$--$T$ relation is $2.8\pm0.8$ ($1.6\pm0.2$).

The $M_{\rm gas,500}$--$T$ relation in Mohr et al. (1999) is based on
a flux-limited sample of 45 nearby clusters spanning a temperature
range of 2--10~keV observed by ROSAT and ASCA. Castillo-Morales \&
Schindler (2003) investigated a sample of 10 nearby clusters
($0.03<z<0.09$, 4.7--9.4~keV) also observed by ROSAT and ASCA. The
$M_{\rm gas,500}$--$T$ relation for the pilot LoCuSS sample agrees
with the published relations for (i) the nearby samples in e.g. Mohr
et al. (1999), Castillo-Morales \& Schindler (2003), and Chen et
al. (2007, the non-CCCs with $T>3$~keV in HIFLUGCS), and (ii) the more
distant sample in Zhang et al. (2006, REFLEX-DXL).

\subsection{Luminosity--temperature relation}

\begin{figure*}
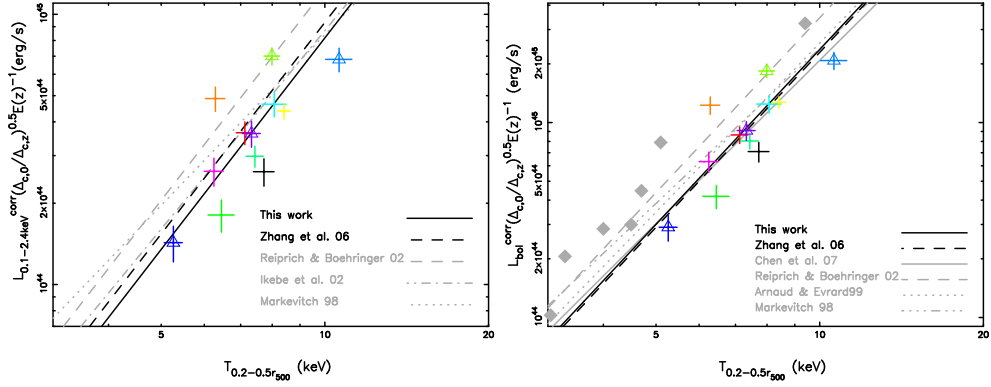

\begin{center}
\includegraphics[angle=270,width=6.5cm]{6567f12a.ps}
\includegraphics[angle=270,width=6.5cm]{6567f12b.ps}
\end{center}
\caption{X-ray luminosity in the 0.1--2.4~keV band vs. temperature
(left) and bolometric luminosity vs. temperature (right). The CCCs are
in triangles. The clusters in Kotov \& Vikhlinin (2005) are in
diamonds. The colors have the same meaning as those in
Fig.~\ref{f:scalet}. The best fit power law of the pilot LoCuSS sample
and REFLEX-DXL sample used the fixed slope of 2.6 for the $L_{\rm
0.1-2.4keV}^{\rm corr}$--$T$ relation and 2.98 for the $L_{\rm
bol}^{\rm corr}$--$T$ relation. \label{f:l0124t} }
\end{figure*}

In the standard self-similar model $M_{\rm gas,500} \propto
M_{500} \propto T^{1.5}$ and $M_{500} \propto r_{500}^3$ give $L
\propto M_{\rm gas}^2 T^{0.5}/r_{500}^3 \propto T^2$. However, 
the $L$--$T$ relation deviates from the $L\propto T^2$ scaling due to
the dependence of the ICM distribution on the cluster temperature
(e.g. Neumann \& Arnaud 2001) and becomes steeper. Given the empirical
scaling $M_{\rm gas,500} \propto T^{1.8}$, one gets $L \propto
T^{2.6}$ as often found in observations. We thus fixed the slope to
the often observed values (e.g. 2.6 for $L_{\rm 0.1-2.4keV}^{\rm
corr}$--$T$ and 2.98 for $L_{\rm bol}^{\rm corr}$--$T$, e.g. observed
in Reiprich
\& B\"ohringer 2002), and compared the findings for the pilot LoCuSS
sample to the recent results in Fig.~\ref{f:l0124t}. For the pilot
LoCuSS sample, the scatter of the luminosity for the $L$--$T$
relations is within 15\%.

The $L$--$T$ relation has been intensively studied for the nearby
cluster samples using the luminosity based on ROSAT observations and
the temperature based on ASCA observations, for example, in (i)
Markevitch (1998, 30 clusters, $0.04<z<0.09$, $T>3.5$~keV), (ii)
Evrard \& Arnaud (1999, 24 clusters, $z<0.37$, $T>2$~keV), (iii)
Reiprich \& B\"ohringer (2002, in which the flux-limited
morphology-unbiased sample HIFLUGCS was initially constructed and
investigated, note the fits are for the whole HIFLUGCS sample
including groups), (iv) Ikebe et al. (2002, a flux-limited sample of
62 clusters, $z<0.16$, 1--10~keV), and (v) Chen et al. (2007,
HIFLUGCS).

The $L$--$T$ relation deviates from the standard self-similar
prediction $L \propto T^2$, but shows no evolution comparing the pilot
LoCuSS sample to the nearby samples in Markevitch (1998) and Arnaud
\& Evrard (1999) using representative non-CCCs. As flux-limited
morphology-unbiased samples, the $L^{\rm corr}_{\rm 0.1-2.4keV}$--$T$
relation for the pilot LoCuSS sample agrees with the HIFLUGCS sample
(e.g. Reiprich \& B\"ohringer 2002, including groups; Chen et
al. 2007, non-CCCs with $T>3$~keV), the sample in Ikebe et al. (2002)
and the REFLEX-DXL sample (e.g. Zhang et al. 2006). Therefore no
evident evolution is observed for the $L$--$T$ relation up to redshift
0.3.

Kotov \& Vikhlinin (2005) applied an alternative redshift
evolution for the $L$--$T$ relation of 10 distant non-CCCs
($0.4<z<0.7$, $T>3.5$~keV), giving $L\propto T^{2.64}(1+z)^{1.8}$. We
applied the redshift correction to the 7 clusters available in Kotov
\& Vikhlinin (2005) also observed by XMM-Newton and found an agreement
with the pilot LoCuSS sample within the scatter. We note that in Kotov
\& Vikhlinin (2005) the temperature was measured in the 70~kpc--$r_{500}$ 
region and luminosity in the 70--1400~kpc region, in which a small
central region was excluded in calculating the temperature/luminosity.
This tends to give slightly lower/higher temperature/luminosity and
thus higher normalization of the $L$--$T$ relation compared to the
pilot LoCuSS sample. On the other hand, Kotov \& Vikhlinin (2005) used
the same aperture for the correction which gives a more significant
correction of the cool core for a less massive system and could thus
provide lower normalization of the $L$--$T$ relation for less massive
systems.

\subsection{Luminosity--mass relation}

\begin{figure*}
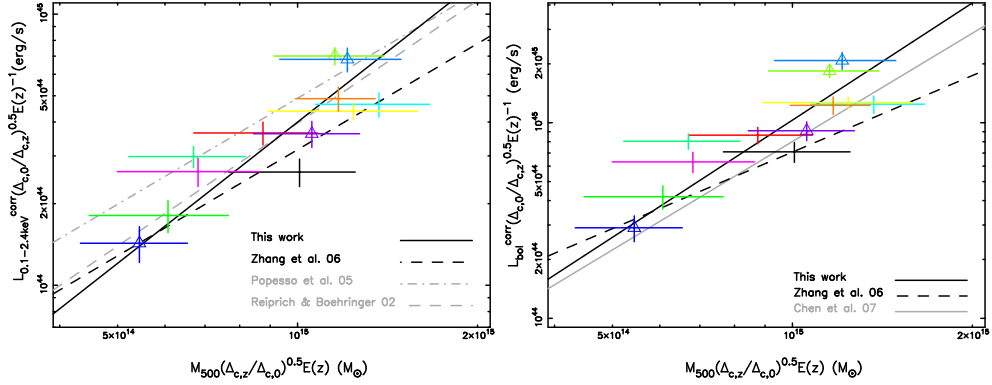

\begin{center}
\includegraphics[angle=270,width=6.5cm]{6567f13a.ps}
\includegraphics[angle=270,width=6.5cm]{6567f13b.ps}
\end{center}
\caption{X-ray luminosity in the 0.1--2.4~keV band vs. total mass
(left) and bolometric luminosity vs. cluster total mass (right).  The
CCCs are in triangles. The colors have the same meaning as those in
Fig.~\ref{f:scalet}. The best fit power law of the pilot LoCuSS sample
used the fixed slope of 1.73 for the $L_{\rm 0.1-2.4keV}^{\rm
corr}$--$M$ relation and 1.99 for the $L_{\rm bol}^{\rm corr}$--$M$
relation. The fixed slope of 1.3 was used for the REFLEX-DXL sample in
Zhang et al. (2006). \label{f:lm}}
\end{figure*}

With $M \propto T^{1.5}$ and $L \propto T^2$ in the standard
self-similar model one derives $L \propto M^{1.33}$. With the observed
scaling relations in e.g. Reiprich \& B\"ohringer (2002), $L_{\rm
0.1-2.4keV} \propto T^{2.6}$ gives $L_{\rm 0.1-2.4keV} \propto
M^{1.73}$ and $L_{\rm bol}
\propto T^{2.98}$ gives $L_{\rm bol} \propto M^{1.99}$,
respectively. The fitting of the pilot sample favors the observed
scaling relations. Therefore, we fixed the slope to 1.73 for $L_{\rm
0.1-2.4keV}^{\rm corr}$--$T$ and 1.99 for $L_{\rm bol}^{\rm
corr}$--$T$, and compared the pilot LoCuSS sample to the recent
results in Fig.~\ref{f:lm}. The scatter of the luminosity and mass for
the $L$--$T$ relations is within 15\% and 20\%, respectively.

The pilot LoCuSS sample agrees with the nearby sample HIFLUGCS
(Reiprich \& B\"ohringer 2002; Popesso et al. 2005; Chen et al.  2007,
non-CCCs with $T>3$~keV) and the more distant sample REFLEX-DXL (Zhang
et al. 2006) also as flux-limited morphology-unbiased
samples. Therefore we observed no evident evolution of the
luminosity--mass relation.

\subsection{$r_{500}$ vs. global temperature}

As found in Evrard et al. (1996) the standard self-similar model
predicts $r_{500} \propto M^{1/3} \propto T^{0.5}$. The best fit for
the pilot LoCuSS sample confirms the standard self-similar model, and
gives $r_{500}=10^{-0.335 \pm 0.106} T_{0.2-0.5r_{500}}^{0.57\pm 0.12}
E(z)^{-1} (\Delta_{c,z}/\Delta_{c,0})^{-0.5}$~Mpc. It agrees with the
scaling relation, $r_{500} \propto T^{0.50 \pm 0.05}$ for 6 relaxed
nearby clusters ($z<0.15$, 2--9~keV, $T>3.5$~keV) in Arnaud et
al. (2005) also observed by XMM-Newton.

\bigskip

The above comparison of the X-ray scaling relations in this work with
the published results (Figs.~\ref{f:mt}--\ref{f:lm}) for the nearby
and more distant samples shows that the evolution of the scaling
relations can be accounted for by the redshift evolution given in
Sect.~\ref{s:prof}. As shown in Table~\ref{t:mtx_lite} the
normalization of the scaling relations agrees for the CCCs and
non-CCCs to better than 10\% using the temperature measured excluding
the cool core and the luminosity corrected for the cool core. The
X-ray quantities such as temperature and luminosity can be used as
reliable indicators of the cluster mass with the scatter less than
20\%. In general, the slopes of the scaling relations indicate the
need for non-gravitational processes. This fits into the general
opinion that galaxy clusters show a modified self-similarity up to
$z\sim 1$ (e.g. Arnaud 2005). As also demonstrated in simulations
(e.g. Poole et al. 2007), mergers only alter the structure of compact
cool cores, and the outer structure ($>0.1 r_{200}$ or $>0.2 r_{500}$)
is survived after the merger events. The scaling relations are thus
preserved for galaxy clusters.

\section{Discussion}
\label{s:discu}

\subsection{Comparison of the X-ray results}

The Chandra data of Abell~1689 (Xue \& Wu 2002), Abell~383 and
Abell~2390 (Vikhlinin et al. 2006a) were analyzed in details. Bauer et
al. (2005) present the cooling time of 8 clusters in the pilot LoCuSS
sample based on Chandra data. Smith et al. (2005) measured the
Chandra temperatures for 9 clusters in this sample. The published
Chandra results for these clusters agree with the XMM-Newton results
in this work. We also obtained an agreement with the published
results based on the same XMM-Newton data, e.g. Abell~1689 (Andersson
\& Madejski 2004), Abell~1835 (Majerowicz et al. 2002; Jia et
al. 2004) and Abell~2218 (Pratt et al. 2005).

\subsection{Gas profiles in the outskirts}
\label{s:infall}

The generally adopted $\beta$ model ($\beta=2/3$) gives $n_{\rm e}
\propto r^{-2}$. However, Vikhlinin et al. (1999) found a mild
trend for $\beta$ to increase as a function of cluster temperature,
which gives $\beta \sim 0.80$ and $n_{\rm e}
\propto r^{-2.4}$ for clusters around 10~keV. Bahcall (1999) also
found that the electron number density scales as $n_{\rm e} \propto
r^{-2.4}$ at large radii. Zhang et al. (2006) confirmed their
conclusion that $n_{\rm e} \propto r^{-2.42}$ for the REFLEX-DXL
clusters. Due to the gradual change in the slope, one should be
cautious to use a single slope double-$\beta$ model which might
introduce a systematic error in the cluster mass measurements (as also
described in e.g. Horner 2001). Similarly, we performed the power law
fit of the ICM density distributions at radii above $3^{\prime}$ for
the present sample and obtained an average of $n_{\rm e}
\propto r^{-2.2\pm 0.1}$.

\subsection{Lensing to X-ray mass ratios}
\label{s:mpro}

Smith et al. (2005) present a uniform strong lensing analysis of 9
clusters in this sample. We took these 9 overlapping clusters as a
subsample (hereafter the S05 subsample, X-ray selected) and compared
the strong lensing and X-ray masses. Ten clusters in the pilot LoCuSS
sample have CFH12k data. The detailed data reduction method can be
found in Bardeau et al. (2005) and the weak lensing results of the
sample in Bardeau et al. (2006), respectively. The CFH12k optical
luminosity weighted galaxy number density contours and weak lensing
masses are given in Bardeau et al. (2006). We took these 10 clusters
as a subsample (hereafter the B06 subsample, X-ray selected) and
compared the weak lensing and X-ray masses. In total eleven clusters
have both X-ray and weak/strong lensing masses.

\begin{figure*}
\begin{center}
\includegraphics[width=5cm,angle=270]{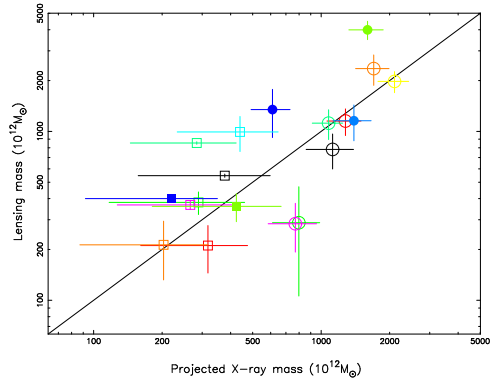}
\end{center}
\caption{Strong lensing mass vs. projected X-ray mass at $r_{2500}$ 
for the S05 subsample (boxes) and weak lensing mass vs. projected
X-ray mass at $r^{\rm wl}_{200}$ for the B06 subsample
(circles), respectively. The symbols for the CCCs are solid. The
colors have the same meaning as those in Fig.~\ref{f:scalet}.
\label{f:mslens_m2500} }
\end{figure*}

\begin{figure*}
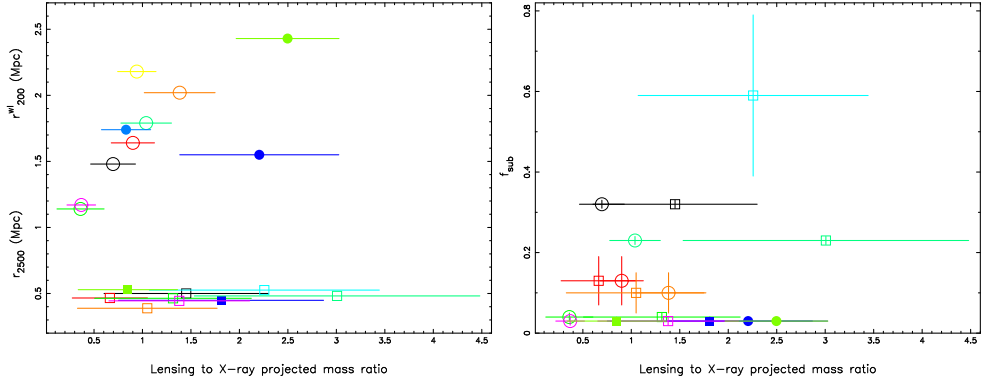

\begin{center}
\includegraphics[width=5cm,angle=270]{6567f15a.ps}
\includegraphics[width=5cm,angle=270]{6567f15b.ps}
\end{center}
\caption{{\it Left:} $r_{2500}$ vs. strong
lensing to X-ray projected mass ratio for the S05 subsample (boxes)
and $r^{\rm wl}_{200}$ vs. weak lensing to X-ray projected mass ratio
for the B06 subsample (circles), respectively. {\it Right:}
substructure fraction (defined in Smith et al. 2005) vs. strong
lensing to X-ray projected mass ratio at $r_{2500}$ for the S05
subsample (boxes) and weak lensing to X-ray projected mass ratio at
$r^{\rm wl}_{200}$ for the B06 subsample (circles), respectively. The
symbols for the CCCs are solid. The colors have the same meaning as
those in Fig.~\ref{f:scalet}. \label{f:mslens_m2500m200} }
\end{figure*}

To compare to the lensing mass we calculated the projected X-ray mass
distribution for each cluster using its observed radial mass
profile. We note that the clusters are detected up to the radii
between $r_{500}$ and $r_{200}$. Because the extrapolated virial radii
are about 2.2--2.6$r_{500}$, we projected the observational mass
profile using a truncation radius of $2.5 r_{500}$. We also performed
the projection to the weak lensing determined $r^{\rm wl}_{200}$. The
comparison of the projected X-ray masses using these 2 different
truncation radii, $2.5 r_{500}$ and $r^{\rm wl}_{200}$), shows that
the X-ray projected mass is insensitive to the truncation radius,
varying within a few per cent. The error introduced by the projection
to $2.5 r_{500}$ is relatively small compared to the projected mass
uncertainties, for example, $\sim 30$\% at $r^{\rm wl}_{200}$ for
Abell~68 using the uncertainties of the electron number density and
temperature measurements by Monte Carlo simulations. Therefore we
adopt the projected X-ray mass using the truncation radius of $2.5
r_{500}$ for further applications. We compare the X-ray masses to the
strong lensing masses in Smith et al. (2005) at the X-ray determined
$r_{2500}$ and to the weak lensing masses in Bardeau et al. (2006) at
the weak lensing determined $r^{\rm wl}_{200}$, respectively
(Fig.~\ref{f:mslens_m2500} and Fig.~\ref{f:mslens_m2500m200}). The
scatter in the strong/weak lensing to X-ray mass ratio is
comparable. 

Within the error, 6 out of 9 clusters show consistent mass
estimates between X-ray and strong lensing approaches. One third of
the clusters show higher strong lensing masses compared to the X-ray
masses up to a factor of 2.5. The mean of the strong lensing to X-ray
mass ratio is 1.53 with its scatter of 1.07. On average the strong
lensing approach tends to give larger masses than the X-ray approach
especially for dynamically active clusters.

Within the error, 4 out of 10 clusters show consistent mass estimates
between X-ray and weak lensing approaches. The mean of the weak
lensing to X-ray mass ratio is 1.12 with its scatter of 0.66. Among
the clusters showing inconsistent masses, half have higher weak
lensing masses compared to the X-ray masses. On average the X-ray and
weak lensing approaches agree better.

\subsection{Mass discrepancy} 

The mass discrepancy between X-ray and lensing can be a combination of
the measurement uncertainties and the physics in the individual
clusters. The following issues have to be considered in properly
comparing the lensing and X-ray masses. (1) Simulations in Poole et al
(2006) show that the temperature fluctuations $\Delta T/T \sim 20\%$
can pass the virialization point and persist in the compact cluster
cores. Therefore temperature fluctuations at the 20\% level do not
necessarily indicate a disturbed system, but introduce uncertainties
in the cluster mass estimates in the central region. (2) As
demonstrated in Puy et al. (2003), the relative error of the X-ray
surface brightness goes almost linearly as a function of the rotating
angle and axis ratio of the cluster, respectively. The projection of
the X-ray masses here is based on the assumption of spherical
symmetry, in which the total mass can be easily
overestimated/underestimated for aspherical clusters. (3) Specially
for the mass distribution in a CCC the cool core has to be well
resolved to guarantee the proper X-ray mass estimate in the central
region. (4) The line-of-sight merger can enhance the possibility to
observe the cluster by strong lensing. A strong lensing selected
sample can be boosted because of this. (5) The lack of reliable
redshift measurements of the faint galaxies in the weak lensing
analysis causes confusion of the lensed background galaxies and
unlensed cluster galaxies. This introduces a significant uncertainty
in the weak lensing mass estimate (Pedersen et al. 2006). (6) The lack
of reliable arc redshift measurements causes uncertainties in the
strong lensing mass estimate. (7) The mass distribution of the DM halo
can be more complex with multi-components which can significantly
affect the strong lensing analysis (e.g. Smith et al. 2005).

\begin{figure*}
\begin{center}
\includegraphics[width=5cm,angle=270]{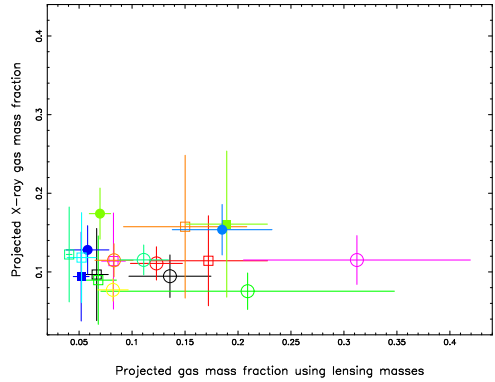}
\end{center}
\caption{Projected X-ray gas mass fraction vs. projected gas mass
fraction using strong lensing (boxes) and weak lensing (circles)
masses, for the S05 subsample and B06 subsample, respectively. The
symbols for the CCCs are solid. The colors have the same meaning as
those in Fig.~\ref{f:scalet}. \label{f:fg2500} }
\end{figure*}

The mass discrepancy between X-ray and lensing partially come from the
interpretation of galaxy dynamics and ICM structure in galaxy clusters
(e.g. Smith et al. 2005; Zhang et al. 2005a). The optical cluster
morphology is described in Bardeau et al. (2006). To reveal the link
between the X-ray lensing mass discrepancy and the characters of the
individual clusters, we investigated the strong/weak lensing to X-ray
mass ratio as a function of the X-ray properties such as temperature,
metallicity, central entropy, cooling radius, cooling time and X-ray
luminosity. However, no evident correlations are observed. This could
be a result of the low statistic of only 10 clusters in the
subsample. A large sample with better statistic could be helpful for
better understanding. In Fig.~\ref{f:fg2500} we present the projected
gas mass fractions at $r_{2500}$ using the strong lensing masses and
at $r^{\rm wl}_{200}$ using the weak lensing masses, which span a
broad range up to 0.3. The scatter of the projected gas mass
fractions using the strong (weak) lensing approach is larger by a
factor of 1.5 (2.5) than the scatter using the X-ray masses.

Combining data on the optical cluster morphology and the X-ray
appearance of the cluster may tell more about the cluster structure
and mass discrepancy. The soft band images are less temperature
dependent and thus better reflect the ICM density distribution. The
MOS1 data show less serious gaps which is important to check the
cluster morphology. To have a closer look at the cluster morphology we
created flat fielded point source included XMM-Newton MOS1 flux images
in the 0.7--2~keV band binned in $8^{\prime
\prime}\times 8^{\prime
\prime}$. The weighted Voronoi tesselation method (Cappellari \& Copin
2003; Diehl \& Statler 2006) was used to bin the image to $5\sigma$
significance. In Fig.~\ref{f:imga} ({\it also see
Figs.~\ref{f:imgb}--\ref{f:imgc} in the electronic edition of the
Journal for the whole sample}), we superposed the CFH12k optical
luminosity weighted galaxy density contours (Bardeau et al. 2006) on
the X-ray adaptively binned images.

\subsubsection{Strong lensing and X-ray masses}

The morphology of the X-ray images overlaid with optical contours can
be used to understand the mass discrepancy in the central regions
combined with the substructure fraction in the strong lensing analysis
(Smith et al. 2005). We thus present the strong/weak lensing to X-ray
mass ratio versus the substructure fraction
(Fig.~\ref{f:mslens_m2500m200}). Three clusters show discrepancies of
the masses using the strong lensing and X-ray approaches. Here we
discuss these 3 clusters, which show a significant link between the
substructure fraction and strong lensing to X-ray mass ratio.

Abell~383 has very low substructure fraction, appearing ``relaxed'' in
the strong lensing analysis. The X-ray temperature profile show a
strong decrease towards the cluster center, which can be even more
significant considering the PSF correction. This can add uncertainties
in the X-ray mass estimate.

Abell~773 and Abell~2218 have very high substructure fractions showing
multi-modal DM halos as found in Smith et al. (2005). In X-rays the
multi-clumps are smeared out and both clusters appear symmetric. This
may explain the large mass discrepancy between strong lensing and
X-ray for Abell~773 (by a factor of 2.3) and Abell~2218 (by a factor
of 3). Reliable arc redshift measurements have been obtained in the
strong lensing mass estimate for Abell~2218 (e.g. Smith et al. 2005
and the references in). The mass distribution of the DM halo in
Abell~2218 shows a complex structure with multi-components in the
central region (Smith et al. 2005). The pronounced mass discrepancy
between X-ray and strong lensing for Abell~2218 is mostly due to the
combination of (1), (2), (4) and (7) as partly indicated in Girardi et
al. (1997) and Pratt et al. (2005).

\subsubsection{Weak lensing and X-ray masses}

Among the 6 clusters with inconsistent weak lensing and X-ray masses,
the mass discrepancy is not significant for Abell~68 and Abell~1763.

The highest projected gas mass fractions were observed in Abell~963
($\sim 0.3$) and Abell~267 ($\sim 0.2$) using the weak lensing masses
(Fig.~\ref{f:fg2500}). The X-ray masses are likely reliable for these
2 clusters since both clusters appear regular in their X-ray flux
images. Therefore the high projected gas mass fractions using the weak
lensing masses may indicate that the weak lensing masses are
underestimated for these 2 clusters. This could explain the low weak
lensing to X-ray mass ratio ($\sim 0.4$) for Abell~267 and Abell~963.

Abell~1835 shows the most significant mass discrepancy (by a factor of
2.5) between X-ray and weak lensing. At $r^{wl,c02}_{200}=2.11$~Mpc
given by the weak lensing analysis in Clowe \& Schneider (2002), our
X-ray mass is $(10.4\pm 2.1) \times 10^{14} M_{\odot}$, in a good
agreement with their weak lensing mass $(8.5\pm0.8)\times 10^{14}
M_{\odot}$. The mass profile within $0.5 r_{500}$ agrees with the
strong lensing mass profile in Smith et al. (2005). The X-ray mass
($M_{500}$) of Abell~1835 also agrees with the previous X-ray results
in Majerowicz et al. (2002). The pronounced mass discrepancy between
X-ray and weak lensing using CFH12k data for Abell~1835 is mostly due
to the combination of (1), (2), (3) and (5). The mass discrepancy for
Abell~383 appears similar to Abell~1835.

As shown in Fig.~\ref{f:mslens_m2500m200}, there is a trend that the
weak lensing to X-ray mass ratio increases with the characteristic
cluster size (e.g. $r^{\rm wl}_{200}$). The disagreement between
the weak lensing and X-ray masses becomes worse with increasing
radius. This could be understood as follows. (i) There is little X-ray
emission in the outskirts beyond $r_{500}$. The extrapolation of the
X-ray mass profiles are too steep compared to the cluster potential in
the outskirts. The projected X-ray mass at $r^{\rm wl}_{200}$ could
thus be underestimated. (ii) The LSS component becomes important at
the boundary of the cluster (e.g. the REFLEX-DXL cluster
RXCJ0014.3$-$3022, Braglia et al. 2007) which can enhance the weak
lensing mass by including filaments in the projected cluster mass
(Pedersen \& Dahle 2006). (iii) Two colors were used to distinguish
the lensed background galaxies and un-lensed cluster galaxies. The
background confusion could also introduce some uncertainties, which
requires reliable photometric redshift measurements to improve the
situation (e.g. Gavazzi
\& Soucail 2007).

\subsection{Luminosity--mass relation using lensing masses}

The purpose to calibrate the $L$--$M$ relation is to seek the
representative $L$--$M$ scaling and to understand the scatter so that
the global luminosity can be used as a cluster mass indicator. The
global luminosity is thus used in the $L$--$M$ relations, but with
different independent cluster mass estimates. In such a way, the
difference of the scater in the $L$--$M$ relations using different
mass approaches can help us better understanding the scatter and thus
sources of systematics. The calibration of the $L$--$M$ relation could
thus provide the proper scatter and sources of systematic errors which
should be included in the cluster luminosity function for the cluster
cosmology.

We investigated the calibration of the luminosity--mass scaling
relation using independent approaches such as X-ray and gravitational
lensing for an X-ray selected flux-limited sample in a narrow redshift
bin ($ z\sim 0.2$). The X-ray cluster mass, based on the X-ray
selection criteria and calculated from the X-ray quantities, is
correlated with the X-ray luminosity. The gravitational lensing
approach provides a unique tool to measure the cluster mass without
assuming hydro-statistic equilibrium in the ICM required in the X-ray
approach. The lensing mass is independent of the X-ray properties such
as temperature, luminosity etc. Using the lensing masses instead of
the X-ray masses in the scaling relations can guarantee less
systematics and errors. The lensing and X-ray approaches can thus be
combined to understand the scaling relations and to reveal the physics
in the scatter of the scaling relations.

Smith et al.  (2005) performed such studies of the mass--temperature
relation on $r_{2500}$ scales combining the Chandra and HST
data. Petersen et al. (2006) performed similar studies for a collected
sample from archive. With the pilot LoCuSS sample, there are 3
advantages to the previous studies. (1) Lensing signals are sensitive
to the two-dimensional mass distribution and thus tend to pick up the
random structures in projection along the line-of-sight. The first
advantage is that the subsample of the overlapping clusters for
combining the X-ray and lensing studies is X-ray selected, no bias
towards the clusters showing line-of-sight mergers. (2) Both the
strong/weak lensing and X-ray data were uniformly observed and
consistently analyzed for the overlapping clusters. The second
advantage is that the systematics of the sample are better controlled
than the sample collected in the archive. (3) Both strong lensing and
weak lensing masses are combined with the high quality XMM-Newton data
covering the scales extending to radii larger than $r_{500}$.

\begin{figure*}
\begin{center}
\includegraphics[width=5cm,angle=270]{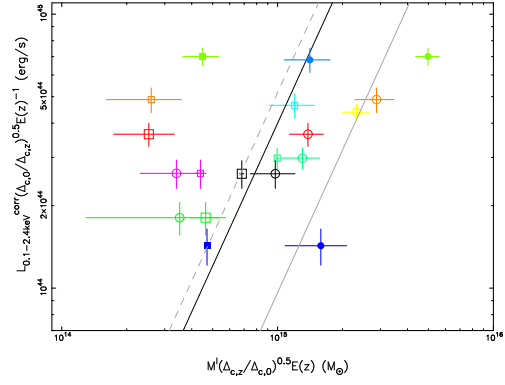}
\end{center}
\caption{X-ray luminosity in the 0.1-2.4~keV band vs.  strong
lensing mass within $r_{2500}$ for the S05 subsample (boxes), and
weak lensing mass within $r^{wl}_{200}$ for the B06 subsample
(circles), respectively. The symbols for the CCCs are solid. The
colors have the same meaning as those in Fig.~\ref{f:scalet}. The
slope of the fitting is set to 1.73. The grey lines denote the
best fit power law for the S05 subsample (dashed) and B06
subsample (solid). The black line denotes the X-ray
luminosity--mass relation for the pilot LoCuSS sample at $z \sim
0.2$. \label{f:mlens_lx} }
\end{figure*}

In Fig.~\ref{f:mlens_lx}, we show the luminosity--mass scaling
relations using X-ray and lensing masses. We observed the correlation
between the strong/weak lensing mass and X-ray luminosity. The scatter
using the lensing masses is about 40\%, larger than using the X-ray
masses ($<20\%$). The large scatter using the strong/weak lensing
masses can be due to the combination of (1) the dependence between the
X-ray luminosity and X-ray mass since the X-ray luminosity is measured
within the scale $2.5 r_{500}$ determined by the X-ray approach
instead of the scale $r_{200}^{\rm wl}$ determined by the lensing
approach, (2) the small size of the sample, (3) the possible
additional uncertainties in the X-ray and lensing mass estimates due
to the multi-component DM halo, (4) the use of a definition of the
global X-ray luminosity such as to reduce the scatter in the X-ray
scaling relations, and perhaps (5) the physics for the individual
clusters such as central AGNs.

We went through the radio archive, and found that there is potentially
one central AGN in each cluster for the LoCuSS sample. There is an
evidence of the link between their AGN activities and the scatter of
the scaling relation. Abell~1763 shows one twin-jet radio source in
the center (e.g. Hardcastle \& Sakelliou 2004), and one twin-lobe
radio source $\sim5^{\prime}$ off-center together with an extended low
surface brightness X-ray feature. The archive
SDSS\footnote{http://cas.sdss.org/dr4/en/tools/explore/obj.asp}
spectrum shows that the cD galaxy of Abell~1835 is a LINER. As 2
clusters with the most significant indication of AGN activities in the
sample, Abell~1763 and Abell~1835 coincidentally introduce the most
pronounced scatter in the $L$--$M$ scaling relations using both strong
lensing and weak lensing masses.

\section{Summary and conclusions}
\label{s:conclusion}

We performed a systematic analysis to measure the X-ray quantities
based on XMM-Newton observations for the pilot LoCuSS sample, a
flux-limited morphology-unbiased sample at $z \sim 0.2$ consisting of
12 X-ray luminous galaxy clusters. We investigated various X-ray
properties, the scaling relations and their scatter, and compared the
X-ray and gravitational lensing masses for this sample. We summarize
the main conclusions as follows.

\bigskip

(i) Self-similarity of the scaled profiles of the X-ray properties

\bigskip

An almost self-similar behavior of the scaled profiles of X-ray
properties, such as temperature, surface brightness, cooling time,
entropy, gravitational mass and gas mass, has been found for radii
above 0.2~$r_{500}$ for the sample.

\begin{itemize}

\item
Based on XMM-Newton observations, we obtained an average temperature
profile of the sample with a plateau at 0.2--0.5~$r_{500}$ and a drop
to 80\% of the maximum temperature at $0.5r_{500}$ with 20\%
scatter. For the 4 CCCs, we observed cool gas showing lower
temperatures than 70\% of the maximum temperatures in the cluster
cores. In the $r<0.2 r_{500}$ region, the average temperature profile
for the whole sample gives $T(r) \propto r^{0.21\pm 0.04}$, and for
the CCC subsample $T(r)
\propto r^{0.38 \pm 0.04}$, respectively. In the outskirts ($0.2
r_{500}<r<r_{500}$), it gives $T(r) \propto r^{-0.36
\pm 0.18}$ for the whole sample, without an evident difference between the
CCCs and non-CCCs at 1$\sigma$ significance level. The averaged
temperature profile for the pilot LoCuSS sample at $z \sim 0.2$ is
consistent with the nearby and more distant cluster samples within the
observational dispersion

\item
We determined the XMM-Newton surface brightness profiles up to at
least $r_{500}$. The surface brightness profiles of the non-CCCs show
flat cores populating a broad range of values up to 0.2~$r_{500}$. For
the CCCs, the cluster cores are resolved with current XMM-Newton
data. The surface brightness profiles are quite self-similar at $R>0.2
r_{500}$ for the sample.

\item
The scaled cooling time profiles show an almost self-similar behavior
above $0.2 r_{500}$. The best fit power law above $0.2 r_{500}$ gives
$t_{\rm c}(r) \propto r^{1.61\pm 0.01}$ for the whole sample, $t_{\rm
c}(r) \propto r^{1.70\pm 0.01}$ for the 4 CCCs, and $t_{\rm c}(r)
\propto r^{1.54\pm 0.01}$ for the 8 non-CCCs.

\item
We found an empirical scaling, $S_{0.2 r_{500}} \propto
T^{0.81\pm0.19}$. The $S$--$T$ relation for this sample agrees with
the REFLEX-DXL sample and the sample in Pratt et al. (2006). After the
redshift evolution correction, the entropy profiles for the pilot
LoCuSS sample agree with the nearby clusters in Ponman et al. (2003)
and Pratt et al. (2006) and the more distant clusters in Zhang et
al. (2006) within the observational dispersion. The entropy profiles
at $r>0.2r_{500}$ for the whole sample give $S(r)
\propto r^{1.01\pm 0.04}$. The non-CCCs show a combined 
entropy profile of $S(r) \propto r^{0.97\pm 0.05}$. The entropy
profiles for the 4 CCCs give, $S(r) \propto r^{1.10\pm 0.05}$,
consistent with the spherical accretion shock model prediction.

\end{itemize}

\bigskip

(ii) X-ray scaling relations

\bigskip

\begin{itemize}

\item
The scaling relations are sensitive to the exclusion of CCCs when the
X-ray luminosity and temperature are measured including the cluster
cores. The cluster cores ($<0.2 r_{500}$) contribute up to 70\% of the
bolometric X-ray luminosity. Using the X-ray luminosity corrected for
the cool core and temperature excluding the cool core, one not only
minimizes the scatter in the scaling relations but also obtains an
agreement better than 10\% between the normalization for the CCCs and
non-CCCs.

\item
For the pilot LoCuSS sample the mass scatter is less than 20\% in the
mass observable scaling relations. The X-ray scaling relations show no
evident evolution comparing the pilot LoCuSS sample to the nearby and
more distant samples within the observational dispersion after the
redshift evolution correction. This fits the general opinion
(e.g. Maughan et al. 2003; Arnaud 2005; Arnaud et al. 2005; Vikhlinin
et al. 2006a; Zhang et al. 2006) that the evolution of galaxy clusters
up to $z \sim 1$ is well described by a self-similar model for massive
clusters. With the current observations, the X-ray quantities such as
temperature and luminosity can be used as reliable indicators of the
cluster mass within 20\% scatter.

\end{itemize}

\bigskip

(iii) Mass calibration and the luminosity--mass relation

\bigskip

The discrepancies remain between the X-ray measured cluster masses and
the gravitational lensing masses. This can be due to a combination of
the measurement uncertainties and the physics in the individual
clusters. The scatter in the strong lensing to X-ray mass ratios and
weak lensing to X-ray mass ratios, respectively, is similar. We
observed the correlation between the X-ray luminosity and lensing
mass. The scatter using the gravitational lensing masses is
significant ($\sim 40$\%). This can be either due to the unknown
physics or due to that the sample is too small. A large X-ray selected
sample with high quality X-ray and gravitational lensing observations
should shed light on it.

\begin{acknowledgements}

The XMM-Newton project is an ESA Science Mission with instruments and
contributions directly funded by ESA Member States and the USA
(NASA). The XMM-Newton project is supported by the Bundesministerium
f\"ur Wirtschaft und Technologie/Deutsches Zentrum f\"ur Luft- und
Raumfahrt (BMWI/DLR, FKZ 50 OX 0001), the Max-Planck Society and the
Heidenhain-Stiftung. We acknowledge the anonymous referee for the
detailed comments improving the work. YYZ acknowledges discussions
with P. Schuecker, M. Arnaud, D. Pierini, M. Freyberg, S. Komossa,
Y. Chen, S.-M. Jia, J. Santos, G. Pratt and A. Simionescu. YYZ
acknowledges support from MPG. AF acknowledges support from BMBF/DLR
under grant No.\,50\,OR\,0207 and MPG. GPS acknowledges support from a
Royal Society University Research Fellowship.

\end{acknowledgements}

%===================begin{TABLES}=============================

\clearpage

\begin{table*} { \begin{center} \footnotesize
  {\renewcommand{\arraystretch}{1.3} \caption[]{ Primary
  parameters. Column~(1): cluster name; Col.~(2): optical redshift
  (e.g. Smith et al. 2005; Bardeau et al. 2006); Cols.~(3,4): sky
  coordinates in epoch J2000 of the cluster center; Col.~(5): hydrogen
  column density (Dickey \& Lockman 1990); Col.~(6): truncation radius
  corresponding to a $S/N$ of 3 of the observational surface
  brightness profile; Cols.~(7--9): bolometric luminosity including
  the $<0.2r_{500}$ region, excluding the $<0.2r_{500}$ region, and
  corrected for the $<0.2r_{500}$ region, respectively.}
  \label{t:global}}
\begin{tabular}{lcccccccccc}
\hline
\hline
Name & $z_{\rm opt}$ & \multicolumn{2}{c}{X-ray centroid} & $N_{\rm H}$ & $r_{\rm t}$ & $L_{\rm bol}^{\rm incc}$ & $L_{\rm bol}^{\rm excc}$ & $L_{\rm bol}^{\rm corr}$ \\
\hline
    &               &
R.A. & delc.    & $10^{22}$~cm$^{-2}$ & arcmin &
$10^{45}$erg/s & $10^{45}$erg/s & $10^{45}$erg/s \\
\hline
Abell~68    &  0.255 &    00   ~    37   ~  06.159 &$   +09  $~   09   ~   28.72   &  0.0493  &   6.1   & $  1.17  \pm  0.10 $ & $  0.63  \pm  0.07 $ & $  0.89  \pm  0.10  $\\
Abell~209   &  0.209 &    01   ~    31   ~  52.607 &$   -13  $~   36   ~   35.50   &  0.0164  &   7.4   & $  1.25  \pm  0.10 $ & $  0.80  \pm  0.07 $ & $  1.04  \pm  0.10  $\\
Abell~267   &  0.230 &    01   ~    52   ~  42.021 &$   +01  $~   00   ~   41.17   &  0.0280  &   5.3   & $  0.75  \pm  0.07 $ & $  0.35  \pm  0.04 $ & $  0.51  \pm  0.07  $\\
Abell~383   &  0.187 &    02   ~    48   ~  03.340 &$   -03  $~   31   ~   43.55   &  0.0392  &   7.7   & $  0.76  \pm  0.05 $ & $  0.25  \pm  0.04 $ & $  0.34  \pm  0.05  $\\
Abell~773   &  0.217 &    09   ~    17   ~  52.935 &$   +51  $~   43   ~   19.41   &  0.0144  &   7.7   & $  2.06  \pm  0.15 $ & $  1.05  \pm  0.11 $ & $  1.50  \pm  0.15  $\\
Abell~963   &  0.206 &    10   ~    17   ~  03.178 &$   +39  $~   02   ~   56.53   &  0.0140  &   6.1   & $  1.16  \pm  0.09 $ & $  0.56  \pm  0.07 $ & $  0.75  \pm  0.09  $\\
Abell~1689  &  0.184 &    13   ~    11   ~  29.330 &$   -01  $~   20   ~   26.66   &  0.0182  &   8.0   & $  2.94  \pm  0.10 $ & $  0.96  \pm  0.06 $ & $  1.49  \pm  0.10  $\\
Abell~1763  &  0.228 &    13   ~    35   ~  18.115 &$   +41  $~   00   ~   03.89   &  0.0936  &   7.5   & $  1.75  \pm  0.15 $ & $  1.21  \pm  0.12 $ & $  1.50  \pm  0.15  $\\
Abell~1835  &  0.253 &    14   ~    01   ~  01.865 &$   +02  $~   52   ~   35.48   &  0.0232  &   7.0   & $  5.18  \pm  0.17 $ & $  1.56  \pm  0.10 $ & $  2.29  \pm  0.17  $\\
Abell~2218  &  0.176 &    16   ~    35   ~  53.775 &$   +66  $~   12   ~   32.43   &  0.0324  &   7.7   & $  1.15  \pm  0.08 $ & $  0.70  \pm  0.06 $ & $  0.94  \pm  0.08  $\\
Abell~2219  &  0.226 &   16   ~   40   ~ 19.900 &  $+46$ ~   42   ~  41.00   &  0.0178  & ---     & ---                  & ---                   & --- \\
Abell~2390  &  0.233 &    21   ~    53   ~  37.115 &$   +17  $~   41   ~   46.41   &  0.0680  &   7.7   & $  3.90  \pm  0.26 $ & $  1.86  \pm  0.17 $ & $  2.54  \pm  0.26  $\\
Abell~2667  &  0.230 &    23   ~    51   ~  39.218 &$   -26  $~   05   ~   03.49   &  0.0165  &   7.6   & $  2.10  \pm  0.12 $ & $  0.81  \pm  0.08 $ & $  1.11  \pm  0.12  $\\
\hline
\hline
  \end{tabular}
  \end{center}
\hspace*{0.3cm}{\footnotesize The XMM-Newton observations of Abell~2219 are flared.
} }
\end{table*}

\begin{table*} { \begin{center} \footnotesize
  {\renewcommand{\arraystretch}{1.3} \caption[]{ Deduced properties of
  all 13 galaxy clusters. Column~(1): cluster name; Col.~(2): electron
  number density of the inner most radial bin; Col.~(3): entropy at
  $0.2 r_{500}$; Col.~(4): inner most radial bin of the surface
  brightness profile; Col.~(5): cooling time measured at $ r_{\rm
  cen}$; Col.~(6): cooling radius; Col.~(7): $r_{500}$; Cols.~(8--10):
  gas mass, total mass and gas mass fraction at $r_{500}$.}
  \label{t:catalog2}} \begin{tabular}{lrrrrrrrrr}
\hline
\hline
Name & $n_{e0}$        & $S_{0.2 r_{500}} $    & $ r_{\rm cen}$  & $t_{\rm c}
$ & $r_{\rm cool}$ & $r_{500}$ & $M_{\rm gas, 500}$ & $M_{500}$
& $f_{\rm gas, 500}$ \\
\hline
\multicolumn{2}{r}{$10^{-3}$~cm$^{-3}$} & keV~cm$^{2}$ & kpc &
Gyr & $r_{500}$ &  Mpc & $10^{14} M_{\odot}$ & $10^{14} M_{\odot}$
& \\
\hline
Abell~68            & $    6.4  \pm     0.3 $ & $  487  \pm    35 $ & $     3.55 $ & $      6.6  \pm       0.2 $ & $     0.10 $ & $     1.21 $ & $     0.68  \pm      0.07 $ & $     6.51  \pm      1.93 $ & $    0.105  \pm     0.060   $\\
Abell~209           & $    5.7  \pm     0.2 $ & $  370  \pm    26 $ & $     3.74 $ & $      6.4  \pm       0.3 $ & $     0.14 $ & $     1.15 $ & $     0.77  \pm      0.08 $ & $     5.33  \pm      1.71 $ & $    0.146  \pm     0.089   $\\
Abell~267           & $    7.4  \pm     0.4 $ & $  365  \pm    33 $ & $     3.29 $ & $      4.9  \pm       0.4 $ & $     0.15 $ & $     1.06 $ & $     0.47  \pm      0.04 $ & $     4.29  \pm      1.30 $ & $    0.109  \pm     0.060   $\\
Abell~383           & $   54.6  \pm     1.5 $ & $  303  \pm    14 $ & $     1.98 $ & $      0.5  \pm       0.2 $ & $     0.16 $ & $     0.98 $ & $     0.33  \pm      0.04 $ & $     3.17  \pm      0.94 $ & $    0.104  \pm     0.061   $\\
Abell~773           & $    7.4  \pm     0.4 $ & $  434  \pm    41 $ & $     3.15 $ & $      5.7  \pm       0.5 $ & $     0.14 $ & $     1.33 $ & $     1.05  \pm      0.12 $ & $     8.30  \pm      2.45 $ & $    0.126  \pm     0.070   $\\
Abell~963           & $   14.3  \pm     0.8 $ & $  393  \pm    18 $ & $     2.14 $ & $      2.6  \pm       0.2 $ & $     0.14 $ & $     1.14 $ & $     0.62  \pm      0.07 $ & $     5.19  \pm      1.52 $ & $    0.120  \pm     0.066   $\\
Abell~1689          & $   30.1  \pm     0.9 $ & $  505  \pm    25 $ & $     1.95 $ & $      1.5  \pm       0.2 $ & $     0.13 $ & $     1.44 $ & $     1.05  \pm      0.14 $ & $    10.26  \pm      2.96 $ & $    0.102  \pm     0.060   $\\
Abell~1763          & $    6.5  \pm     0.3 $ & $  402  \pm    47 $ & $     4.00 $ & $      6.1  \pm       0.5 $ & $     0.13 $ & $     1.12 $ & $     0.88  \pm      0.08 $ & $     4.96  \pm      1.46 $ & $    0.178  \pm     0.094   $\\
Abell~1835          & $   60.8  \pm     1.2 $ & $  334  \pm    20 $ & $     2.50 $ & $      0.6  \pm       0.3 $ & $     0.19 $ & $     1.22 $ & $     1.14  \pm      0.11 $ & $     6.62  \pm      1.94 $ & $    0.172  \pm     0.092   $\\
Abell~2218          & $    5.8  \pm     0.2 $ & $  444  \pm    51 $ & $     3.26 $ & $      8.0  \pm       0.5 $ & $     0.12 $ & $     1.07 $ & $     0.62  \pm      0.06 $ & $     4.18  \pm      1.27 $ & $    0.147  \pm     0.085   $\\
Abell~2390          & $   40.1  \pm     1.6 $ & $  486  \pm    34 $ & $     2.31 $ & $      0.9  \pm       0.3 $ & $     0.14 $ & $     1.29 $ & $     1.21  \pm      0.16 $ & $     7.67  \pm      2.28 $ & $    0.158  \pm     0.095   $\\
Abell~2667          & $   40.7  \pm     1.4 $ & $  399  \pm    27 $ & $     2.33 $ & $      0.8  \pm       0.3 $ & $     0.15 $ & $     1.19 $ & $     0.77  \pm      0.09 $ & $     6.02  \pm      1.74 $ & $    0.128  \pm     0.073   $\\
\hline
Mean             &  ---                   & ---           & ---      & ---                     & ---           & ---       & ---                     & ---     & $0.13 \pm 0.01$     \\
\hline
\hline
  \end{tabular}
  \end{center}
\hspace*{0.3cm}{\footnotesize   } }
\end{table*}

\begin{table*} { \begin{center} \footnotesize
  {\renewcommand{\arraystretch}{1.3} \caption[]{Generalized NFW model
  fit. Column~(1): cluster name; Cols.~(2,3): characteristic density
  and scale of the halo of the generalized NFW fit; Col.~(4): slope
  parameter of the generalized NFW fit.} \label{t:mcom}}
\begin{tabular}{lccc}
\hline
\hline
Name & $\rho_{\rm s}$ & $r_{\rm s}$ & $\alpha$
\\
\hline
     & $10^{14} M_{\odot}$~Mpc$^{-3}$ & Mpc &
\\
\hline
Abell~68      & $        82 \pm          3 $ & $     0.203 \pm      0.005 $ & $     0.000 \pm      0.100  $ \\
Abell~209     & $        38 \pm          5 $ & $     0.289 \pm      0.018 $ & $     0.000 \pm      0.063  $ \\
Abell~267     & $        82 \pm          5 $ & $     0.184 \pm      0.008 $ & $     0.000 \pm      0.100  $ \\
Abell~383     & $       108 \pm         15 $ & $     0.127 \pm      0.008 $ & $     0.518 \pm      0.044  $ \\
Abell~773     & $        53 \pm          2 $ & $     0.262 \pm      0.006 $ & $     0.000 \pm      0.100  $ \\
Abell~963     & $       173 \pm         33 $ & $     0.126 \pm      0.010 $ & $     0.000 \pm      0.090  $ \\
Abell~1689    & $        64 \pm         14 $ & $     0.220 \pm      0.021 $ & $     0.608 \pm      0.070  $ \\
Abell~1763    & $        69 \pm          6 $ & $     0.187 \pm      0.010 $ & $     0.000 \pm      1.000  $ \\
Abell~1835    & $       275 \pm         11 $ & $     0.117 \pm      0.002 $ & $     0.000 \pm      0.100  $ \\
Abell~2218    & $       123 \pm          4 $ & $     0.149 \pm      0.003 $ & $     0.000 \pm      1.000  $ \\
Abell~2390    & $       128 \pm         20 $ & $     0.168 \pm      0.011 $ & $     0.324 \pm      0.055  $ \\
Abell~2667    & $       165 \pm         14 $ & $     0.142 \pm      0.005 $ & $     0.037 \pm      0.035  $ \\
\hline
\hline
  \end{tabular}
  \end{center}
\hspace*{0.3cm}{}  }
\end{table*}

\begin{table*} { \begin{center} \footnotesize
  {\renewcommand{\arraystretch}{1.3} \caption[]{ Cluster global
  temperatures and metallicities. Column~(1): cluster name;
  Col.~(2,3): volume averaged radial temperature profile of
  0.1--0.5~$r_{500}$ and 0.2--0.5~$r_{500}$, respectively; Col.~(4-9):
  spectral measured temperatures and metallicities of the annuli of
  $R<2/3 r_{\rm t}$, $0.1r_{500}<R<0.5 r_{500}$ and $0.2r_{500}<R<0.5
  r_{500}$, respectively.}  \label{t:tglobal}}
  \begin{tabular}{lcccccccc}
\hline
\hline
Name & $T_{0.1-0.5r_{500}}$ & $T_{0.2-0.5r_{500}}$ & $T^{\rm spec}_{<2/3r_{\rm t}}$ & $T^{\rm spec}_{0.1-0.5r_{500}}$ & $T^{\rm spec}_{0.2-0.5r_{500}}$ & $Z_{<2/3r_{\rm t}}$ & $Z_{0.1-0.5r_{500}}$ & $Z_{0.2-0.5r_{500}}$\\
\hline
     &  keV                 & keV                  & keV                          & keV                             & keV                             & $Z_{\odot}$       & $Z_{\odot}$          & $Z_{\odot}$         \\
\hline
Abell~68       & $   8.2  \pm   0.3 $ & $   7.7  \pm   0.3 $ & $   7.9  \pm   0.3 $ & $   7.7  \pm   0.3 $ & $   7.3  \pm   0.3 $ & $  0.17  \pm  0.05 $ & $  0.15  \pm  0.04 $ & $  0.12  \pm  0.05   $\\
Abell~209      & $   6.4  \pm   0.2 $ & $   7.1  \pm   0.3 $ & $   6.9  \pm   0.2 $ & $   7.0  \pm   0.2 $ & $   7.1  \pm   0.3 $ & $  0.28  \pm  0.04 $ & $  0.23  \pm  0.03 $ & $  0.22  \pm  0.04   $\\
Abell~267      & $   6.1  \pm   0.2 $ & $   6.5  \pm   0.4 $ & $   6.1  \pm   0.2 $ & $   6.1  \pm   0.2 $ & $   6.2  \pm   0.4 $ & $  0.23  \pm  0.05 $ & $  0.23  \pm  0.05 $ & $  0.25  \pm  0.08   $\\
Abell~383      & $   4.6  \pm   0.1 $ & $   5.3  \pm   0.2 $ & $   4.5  \pm   0.1 $ & $   4.7  \pm   0.1 $ & $   4.7  \pm   0.2 $ & $  0.44  \pm  0.04 $ & $  0.29  \pm  0.04 $ & $  0.18  \pm  0.06   $\\
Abell~773      & $   7.8  \pm   0.3 $ & $   8.1  \pm   0.4 $ & $   7.9  \pm   0.2 $ & $   7.8  \pm   0.3 $ & $   8.3  \pm   0.4 $ & $  0.30  \pm  0.03 $ & $  0.26  \pm  0.04 $ & $  0.30  \pm  0.06   $\\
Abell~963      & $   6.5  \pm   0.2 $ & $   6.3  \pm   0.2 $ & $   6.4  \pm   0.2 $ & $   6.5  \pm   0.2 $ & $   6.5  \pm   0.2 $ & $  0.31  \pm  0.04 $ & $  0.30  \pm  0.03 $ & $  0.28  \pm  0.05   $\\
Abell~1689     & $   9.0  \pm   0.2 $ & $   8.4  \pm   0.2 $ & $   9.0  \pm   0.1 $ & $   9.0  \pm   0.2 $ & $   8.5  \pm   0.2 $ & $  0.26  \pm  0.02 $ & $  0.24  \pm  0.02 $ & $  0.26  \pm  0.04   $\\
Abell~1763     & $   7.1  \pm   0.2 $ & $   6.3  \pm   0.3 $ & $   6.3  \pm   0.2 $ & $   6.1  \pm   0.2 $ & $   5.8  \pm   0.3 $ & $  0.27  \pm  0.05 $ & $  0.23  \pm  0.04 $ & $  0.20  \pm  0.05   $\\
Abell~1835     & $   7.6  \pm   0.2 $ & $   8.0  \pm   0.3 $ & $   7.2  \pm   0.1 $ & $   8.1  \pm   0.2 $ & $   8.4  \pm   0.3 $ & $  0.29  \pm  0.02 $ & $  0.25  \pm  0.02 $ & $  0.23  \pm  0.04   $\\
Abell~2218     & $   8.0  \pm   0.2 $ & $   7.4  \pm   0.3 $ & $   6.9  \pm   0.2 $ & $   6.9  \pm   0.2 $ & $   6.6  \pm   0.3 $ & $  0.25  \pm  0.04 $ & $  0.18  \pm  0.21 $ & $  0.21  \pm  0.04   $\\
Abell~2390     & $   9.7  \pm   0.4 $ & $  10.6  \pm   0.6 $ & $  10.1  \pm   0.3 $ & $  10.6  \pm   0.4 $ & $  11.6  \pm   0.6 $ & $  0.35  \pm  0.04 $ & $  0.28  \pm  0.04 $ & $  0.24  \pm  0.06   $\\
Abell~2667     & $   7.0  \pm   0.2 $ & $   7.3  \pm   0.3 $ & $   6.6  \pm   0.1 $ & $   7.2  \pm   0.2 $ & $   7.0  \pm   0.3 $ & $  0.34  \pm  0.03 $ & $  0.30  \pm  0.03 $ & $  0.29  \pm  0.04   $\\
\hline
\hline
  \end{tabular}
  \end{center}
\hspace*{0.3cm}{\footnotesize   } }
\end{table*}

\begin{table*} { \begin{center} \footnotesize
  {\renewcommand{\arraystretch}{1.3} \caption[]{ Power law,
  $Y=Y_0\;X^{\gamma}$, parameterized X-ray scaling relations.}
  \label{t:mtx_lite}}
\begin{tabular}{lllllll}
\hline
\hline
$X$ & $Y$ & $Y_0$ & $\gamma$ & \multicolumn{2}{c}{Scatter} & Sample \\
\hline
    &     &       &           &       $X$  & $Y$ &                  \\
\hline
$\frac{T_{0.2-0.5r_{500}}}{\rm keV}$ &
$\frac{S_{0.2r_{500}}}{\rm keV\;cm^{2}} \; E(z)^{4/3} \; \left(\frac{\Delta_{c,z}}{\Delta_{c,0}}\right)^{2/3}$
   & $10^{2.08\pm0.26}$ & $0.90 \pm 0.30$ & --- & --- & REFLEX-DXL \\
&  & $10^{2.00\pm0.16}$ & $0.81 \pm 0.19$ & --- & --- & pilot LoCuSS sample \\
&  & $10^{1.96\pm0.22}$ & $0.89 \pm 0.25$ & --- & --- & non-CCC subsample \\
&  & $10^{2.04\pm0.18}$ & $0.71 \pm 0.21$ & --- & --- & CCC subsample \\
\hline
 $\frac{r}{r_{500}}$
   & $\frac{S}{\rm keV\;cm^{2}} \; E(z)^{4/3} \; \left(\frac{\Delta_{c,z}}{\Delta_{c,0}}\right)^{2/3} \left(\frac{T_{0.2-0.5r_{500}}}{10keV}\right)^{-0.65}$ 
   &$10^{3.52\pm0.03}$ & $1.00 \pm 0.07$ & --- & --- & REFLEX-DXL  \\
&  & $10^{3.50\pm0.02}$ & $1.01 \pm 0.04$ & --- & --- & pilot LoCuSS sample \\
&  & $10^{3.49\pm0.02}$ & $0.97 \pm 0.05$ & --- & --- & non-CCC subsample \\
&  & $10^{3.52\pm0.02}$ & $1.10 \pm 0.05$ & --- & --- & CCC subsample \\
\hline
$\frac{T_{0.2-0.5r_{500}}}{\rm keV}$
   & $\frac{M_{500}}{M_{\odot}} \; E(z) \; \left(\frac{\Delta_{c,z}}{\Delta_{c,0}}\right)^{0.5}$ 
   & $10^{13.69\pm0.03}h_{70}^{-1}$ & 1.5 (fixed) & 0.05 & 0.15 & pilot LoCuSS sample \\
&  & $10^{13.72\pm0.04}h_{70}^{-1}$ & 1.5 (fixed) & 0.05 & 0.20 & non-CCC subsample \\
&  & $10^{13.91\pm0.38}h_{70}^{-1}$ & $1.24\pm0.44$ & 0.06 & 0.15 & pilot LoCuSS sample \\
&  & $10^{13.41\pm0.53}h_{70}^{-1}$ & $1.87\pm0.63$ & 0.07 & 0.19 & non-CCC subsample \\
\hline
$\frac{T_{0.2-0.5r_{500}}}{\rm keV}$
   & $\frac{M_{\rm gas,500}}{M_{\odot}} \; E(z) \; \left(\frac{\Delta_{c,z}}{\Delta_{c,0}}\right)^{0.5}$ 
   & $10^{12.54\pm0.04}h_{70}^{-1}$ & 1.8 (fixed) & 0.06 & 0.10 & pilot LoCuSS sample \\
&  & $10^{12.54\pm0.05}h_{70}^{-1}$ & 1.8 (fixed) & 0.06 & 0.11 & non-CCC subsample \\
&  & $10^{11.72\pm0.70}h_{70}^{-1}$ & $2.76\pm 0.81$ & 0.05 & 0.13 & pilot LoCuSS sample \\
&  & $10^{11.65\pm0.98}h_{70}^{-1}$ & $2.86\pm 1.17$ & 0.05 & 0.13 & non-CCC subsample \\
\hline
$\frac{T_{0.2-0.5r_{500}}}{\rm keV}$
   &
$\frac{L_{\rm 0.1-2.4keV}^{\rm corr}}{\rm erg\;s^{-1}}\;E(z)^{-1} \; \left(\frac{\Delta_{c,z}}{\Delta_{c,0}}\right)^{-0.5}$
   & $10^{42.31\pm0.04}h_{70}^{-1}$ & 2.60 (fixed) & 0.05 & 0.12 & pilot LoCuSS sample  \\
&  & $10^{42.30\pm0.05}h_{70}^{-1}$ & 2.60 (fixed) & 0.05 & 0.13 & non-CCC subsample  \\
&  & $10^{41.87\pm0.71}h_{70}^{-1}$ & $3.11\pm 0.82$ & 0.04 & 0.13 &pilot LoCuSS sample  \\
&  & $10^{42.17\pm0.86}h_{70}^{-1}$ & $2.76\pm 1.01$ & 0.05 & 0.14 & non-CCC subsample  \\
\hline
$\frac{T_{0.2-0.5r_{500}}}{\rm keV}$
   & $\frac{L_{\rm bol}^{\rm corr}}{\rm erg\;s^{-1}}\;E(z)^{-1} \; \left(\frac{\Delta_{c,z}}{\Delta_{c,0}}\right)^{-0.5}$ 
   & $10^{42.39\pm0.04}h_{70}^{-1}$ & 2.98 (fixed) & 0.04 & 0.12 & pilot LoCuSS sample \\
&  & $10^{42.38\pm0.05}h_{70}^{-1}$ & 2.98 (fixed) & 0.05 & 0.13 & non-CCC subsample \\
&  & $10^{41.84\pm0.70}h_{70}^{-1}$ & $3.62 \pm 0.80$ & 0.04 & 0.13 &pilot LoCuSS sample \\
&  & $10^{41.95\pm0.90}h_{70}^{-1}$ & $3.49 \pm 1.06$ & 0.04 & 0.14 & non-CCC subsample \\
\hline
$\frac{M_{500}}{M_{\odot}} \; E(z) \; \left(\frac{\Delta_{c,z}}{\Delta_{c,0}}\right)^{0.5}$
   & $\frac{L_{\rm 0.1-2.4keV}^{\rm corr}}{\rm erg\;s^{-1}}\;E(z)^{-1} \; \left(\frac{\Delta_{c,z}}{\Delta_{c,0}}\right)^{-0.5}$
   & $10^{18.65\pm0.04}h_{70}^{-1}$ & 1.73 (fixed) & 0.14 & 0.11 & pilot LoCuSS sample \\
&  & $10^{18.61\pm0.04}h_{70}^{-1}$ & 1.73 (fixed) & 0.19 & 0.12 & non-CCC subsample \\
&  & $10^{18.12\pm4.74}h_{70}^{-1}$ & $ 1.77 \pm 0.32$ & 0.14 & 0.11 & pilot LoCuSS sample \\
&  & $10^{22.85\pm3.51}h_{70}^{-1}$ & $ 1.45 \pm 0.23$ & 0.19 & 0.12 & non-CCC subsample \\
\hline
$\frac{M_{500}}{M_{\odot}} \; E(z) \; \left(\frac{\Delta_{c,z}}{\Delta_{c,0}}\right)^{0.5}$
   & $\frac{L_{\rm bol}^{\rm corr}}{\rm erg\;s^{-1}}\;E(z)^{-1} \; \left(\frac{\Delta_{c,z}}{\Delta_{c,0}}\right)^{-0.5}$
   & $10^{15.16 \pm 0.04}h_{70}^{-1}$ & 1.99 (fixed)  & 0.14 & 0.12   & pilot LoCuSS sample  \\
&  & $10^{15.11 \pm 0.04}h_{70}^{-1}$ & 1.99 (fixed)  & 0.19 & 0.13   & non-CCC subsample  \\
&  & $10^{14.16 \pm 5.39}h_{70}^{-1}$ & $2.06 \pm 0.36$ & 0.14 & 0.12   & pilot LoCuSS sample  \\
&  & $10^{19.88 \pm 3.19}h_{70}^{-1}$ & $1.67 \pm 0.21$ & 0.19 & 0.13   & non-CCC subsample  \\
\hline
\hline
  \end{tabular}
  \end{center}
\hspace*{0.3cm}{}
  }
\end{table*}

%===================end{TABLES}=============================

\clearpage

%===================begin{FIGURES}=============================

%[h] means here; [t] means top; [b] means bottom

\clearpage

\clearpage

\clearpage 

\appendix

\section{XMM-Newton observational information}

\begin{table*} { \begin{center} \footnotesize
  {\renewcommand{\arraystretch}{1.3} \caption[]{Observational
  information of the XMM-Newton data. } \label{t:obs}}
  \begin{tabular}{lcccccccc}
\hline
\hline
  Name & Date &  Id & Filter & \multicolumn{2}{c}{Frame} & \multicolumn{3}{c}{Net exposure (ks)}   \\
\hline
       &      &     &        &    MOS & pn               &MOS1 & MOS2 & pn \\
\hline
  Abell~68 & 2006-03-14 & 0084230201 & Medium & FF & EFF & 24.94 & 23.81 & 18.15\\
 Abell~209 & 2001-01-15 & 0084230301 & Medium & FF & EFF & 17.33 & 16.09 & 12.80\\
 Abell~267 & 2002-01-02 & 0084230401 & Medium & FF & EFF & 16.06 & 15.74 & 10.96\\
 Abell~383 & 2002-08-17 & 0084230501 & Medium & FF & EFF & 28.07 & 28.05 & 21.50\\
 Abell~773 & 2001-04-26 & 0084230601 & Medium & FF & EFF & 13.01 & 14.68 & 15.90\\
 Abell~963 & 2001-11-02 & 0084230701 & Medium & FF & EFF & 23.82 & 24.69 & 17.92\\
Abell~1689 & 2001-12-24 & 0093030101 & Thin   & FF & EFF & 33.67 & 33.34 & 29.24\\
Abell~1763 & 2002-12-13 & 0084230901 & Medium & FF & EFF & 12.32 & 12.01 &  9.34\\
Abell~1835 & 2000-06-28 & 0098010101 & Thin   & FF & FF  & ---   & 25.27 & 24.73\\
Abell~2218 & 2002-09-28 & 0112980101 & Thin   & FF & EFF & 16.69 & 16.91 & 13.79\\
Abell~2219 & 2002-06-14 & 0112231801 & Medium & FF & EFF & --- & --- & ---\\
           & 2002-06-24 & 0112231901 & Medium & FF & EFF & --- & --- & ---\\
Abell~2390 & 2001-06-19 & 0111270101 & Thin   & FF & FF  & 10.27 & 10.03 &  8.85\\
Abell~2667 & 2003-06-21 & 0148990101 & Medium & FF & FF  & 22.21 & 22.87 & 14.10\\
\hline
\hline
  \end{tabular}
  \end{center}
\hspace*{0.3cm}{\footnotesize The MOS1 data of Abell~1835 are in window mode,
which cannot be used for this work. The XMM-Newton observations of
Abell~2219 are flared.  } }
\end{table*}

\clearpage
\section{Individual cluster profiles}
\label{apx:tprofs}

\clearpage

\clearpage

\clearpage

\clearpage

\clearpage

\clearpage

%===================end{FIGURES}=============================
\end{document}